\documentclass{lmcs}
\pdfoutput=1

\usepackage{lastpage}
\lmcsdoi{15}{1}{18}
\lmcsheading{}{\pageref{LastPage}}{}{}%
{Sep.~01,~2017}{Mar.~04,~2019}{}

\usepackage{thmtools}
\usepackage{thm-restate}
\usepackage{graphicx}
\usepackage{mathrsfs}
\usepackage{caption}
\usepackage{subcaption}
\usepackage{mathpartir}
\usepackage{amsmath}
\usepackage{cmll}
\usepackage[all]{xy}
\usepackage{tikz}
\usepackage{hyperref}
\usepackage{amsthm}
\usepackage{stmaryrd}
\usepackage{todonotes}
\usepackage{MnSymbol}

\usepackage{bm}
\usepackage[refpage]{nomencl}

\makeatletter

\def\@@@nomenclature[#1]#2#3{%
\def\@tempa{#2}\def\@tempb{#3}%
\protected@write\@nomenclaturefile{}%
{\string\nomenclatureentry{#1\nom@verb\@tempa @[{\nom@verb\@tempa}]%
|nompageref{\begingroup\nom@verb\@tempb\protect\nomeqref
{\theequation}}}%
{\thepage}}
\endgroup
\@esphack}

\def\nompageref#1#2{%
\if@printpageref\pagedeclaration{#2}\else\null\fi
\linebreak#1\nomentryend\endgroup}

\def\pagedeclaration#1{\dotfill\nobreakspace#1}
\def\nomentryend{.}

\makeatother

\makenomenclature

\makeindex

\newcommand{\Con}{\mathrm{Con}}
\newcommand{\conf}[1]{\mathscr{C}(#1)}
\newcommand{\cov}{{{\mathrel-\joinrel\subset}}}
\newcommand{\longcov}[1]{{\stackrel{#1}{\mathrel-\joinrel\relbar\joinrel\subset\,}}}

\newcommand{\ttrue}{\mathbf{t\!t}}
\newcommand{\imc}{\rightarrowtriangle}

\newcommand{\tfalse}{\mathbf{f\!f}}
\newcommand{\tbool}{\mathbb{B}}

\newdir{C}{%
!/4.5pt/@{(}*:(1,-.2)@{}*:(1,+.2)@_{}}

\newdir{|>}{%
!/4.5pt/@{|}*:(1,-.2)@{>}*:(1,+.2)@_{>}}

\renewcommand{\tilde}[1]{\widetilde{#1}}

\newcommand{\minconflict}[2]{\xymatrix { #1 \ar@{~}[r]& #2}}
\newdir{pb}{:(1,-1)@^{|-}}
\def\pb#1{\save[]+<16 pt,0 pt>:a(#1)\ar@{pb{}}[]\restore}

\newcommand{\CC}{{\rm C\!\!\!C}}
\renewcommand{\cc}{\mathrm{\,c\!\!\!\!c\,}}

\newcommand{\change}[1]{{#1}}
\newcommand{\coin}{\mathbf{coin}}

\newcommand{\tensor}{\otimes}
\newcommand{\ES}{\mathcal{E}}
\newcommand{\ESS}{\mathcal{E}_\sim}
\newcommand{\ESSP}{\mathcal{EP}_\sim}

\newcommand{\proj}{\downarrow}
\newcommand{\secbij}[1]{\mathscr{B}^\text{sec}_{#1}}
\def\profto{\!\!\!\xymatrix@C-.75pc{\ar[r]|-{\! +\!} &}\!\!\! }
\newcommand{\CG}{\mathrm{CG}}

\newcommand{\iso}{\cong}

\newcommand{\inter}{\circledast}
\newcommand{\ointer}{\circledast'}

\newcommand{\lift}[1]{\overline{#1}}

\newcommand{\conflict}{\mathrel{\#}}    

\newcommand{\tto}{\Rightarrow}
\newcommand{\Qu}{\mathcal{Q}}
\newcommand{\An}{\mathcal{A}}
\newcommand{\A}{\mathcal{A}}
\newcommand{\D}{\mathcal{D}}
\newcommand{\C}{\mathcal{C}}
\newcommand{\B}{\mathcal{B}}
\newcommand{\E}{\mathcal{E}}
\newcommand{\X}{\mathcal{X}}
\newcommand{\Y}{\mathcal{Y}}
\renewcommand{\S}{\mathcal{S}}
\newcommand{\T}{\mathcal{T}}
\newcommand{\ifpcf}{\mathbf{if}}
\renewcommand{\com}{\mathbf{com}}
\newcommand{\run}{\mathbf{run}}
\newcommand{\done}{\mathbf{done}}
\newcommand{\q}{\mathbf{q}}
\renewcommand{\b}{\mathbf{b}}

\newcommand{\mconflict}{\!\!\xymatrix@C=15pt{\, \ar@{~}[r]& \,}\!\!\!\!\!}
\newcommand{\intr}[1]{\llbracket #1 \rrbracket}
\newcommand{\lbl}{\mathrm{lbl}}
\newcommand{\ind}{\mathrm{ind}}
\newcommand{\pol}{\mathrm{pol}}
\newcommand{\just}{\mathrm{just}}
\newcommand{\bij}{\simeq}
\newcommand{\tuple}[1]{\langle #1 \rangle}
\newcommand{\id}{\mathrm{id}}
\DeclareMathOperator{\dom}{dom}

\DeclareMathOperator{\codom}{codom}
\newcommand{\TCG}{\mathbf{Tcg}}

\newcommand{\CHO}{\mathbf{Cho}}
\newcommand{\tnat}{\mathbb{N}}
\newcommand{\tskip}{\mathbf{skip}}
\newcommand{\tsucc}{\mathbf{succ}}
\newcommand{\iszero}{\mathbf{iszero}}
\newcommand{\var}{\mathbf{ref}}

\newcommand{\pred}{\mathbf{pred}}
\newcommand{\newref}{\mathbf{newref}}

\newcommand{\tin}{\mathbf{in}}
\newcommand{\mkvar}{\mathbf{mkvar}}

\newcommand{\tx}{\mathbb{X}}
\newcommand{\e}{\mathbf{e}}
\newcommand{\emptyar}{\text{\boldmath$1$}}
\newcommand{\wiso}{\approx}
\newcommand{\sym}[1]{\widetilde{#1}}
\newcommand{\init}{\mathrm{init}}
\renewcommand{\P}{\mathcal{P}}
\newcommand{\choto}{\stackrel{\CHO}{\profto}}
\newcommand{\ttuple}[1]{\llangle #1 \rrangle}

\newcommand{\tcgto}{\stackrel{\TCG}{\profto}}
\newcommand{\inc}{\trianglelefteq}

\newcommand{\tread}{\mathbf{R}}
\newcommand{\twrite}{\mathbf{W}}
\newcommand{\tok}{\mathbf{ok}}
\newcommand{\cell}{\mathrm{cell}}
\newcommand{\Cell}{\mathrm{Cell}}
\newcommand{\newcell}{\mathrm{newcell}}
\newcommand{\restrict}{\upharpoonright}
\newcommand{\ev}{\mathrm{ev}}
\newcommand{\siso}{\cong}
\newcommand{\sbij}{\simeq}
\newcommand{\isf}{\cong}
\newcommand{\refl}{\mathrm{refl}}

\newcommand{\ocajm}{\oc_{\text{\tiny{AJM}}}}

\usepackage{color}
\newcommand{\red}[1]{#1}

\title{\red{Thin Games with Symmetry and} Concurrent Hyland-Ong Games}
\author{Simon Castellan}
\address{Univ Lyon, CNRS, ENS de Lyon, UCB Lyon 1, LIP}
\email{simon.castellan@ens-lyon.fr}
\author{Pierre Clairambault}
\address{Univ Lyon, CNRS, ENS de Lyon, UCB Lyon 1, LIP}
\email{pierre.clairambault@ens-lyon.fr}
\author{Glynn Winskel}
\address{Computer Laboratory, University of Cambridge}
\email{glynn.winskel@cl.cam.ac.uk}

\begin{document}


\maketitle

\begin{abstract}
We build a cartesian closed category, called $\CHO$, based on event structures. It allows an interpretation of
higher-order stateful concurrent programs that is 
refined and precise: on the one hand it is
conservative with respect to standard Hyland-Ong games when interpreting purely functional programs
as innocent strategies, 
while on the other hand it is 
much more expressive. The interpretation of
programs constructs compositionally a representation of their execution that 
exhibits causal dependencies 
and remembers the points of non-deterministic branching.

The construction is in two stages. First, we build a compact closed category $\TCG$. It is a variant
of Rideau and Winskel's category $\CG$, with the 
difference that games and strategies in $\TCG$ are
equipped 
with \emph{symmetry} 
to express that certain events are essentially the same. This
is analogous to the underlying category of AJM games enriching simple games with an equivalence
relations on plays. Building on this category, we construct the cartesian closed category $\CHO$ as
having as objects the standard arenas of Hyland-Ong games, 
with 
strategies, 
represented by certain events structures,
playing on games with symmetry obtained as expanded forms of these arenas.

To illustrate and 
give an operational light 
on these constructions, we interpret (a
close variant of) Idealized Parallel Algol in $\CHO$.
\end{abstract}


\section{Introduction}
In game semantics, \emph{computation} is represented within a two-player game
played between the \emph{program} and its \emph{execution environment} -- the
program is often 
considered to be \emph{Player} and the execution environment
\emph{Opponent}. The  two players make \emph{moves} corresponding to
computational events: the program calling an external function is a Player move,
and this function returning is an Opponent move. Originally motivated by the very
foundational quest of understanding higher-order sequentiality
\cite{ho,ajm}, game semantics developed into a rich
subject, with a wide scope spanning logical aspects of computation, through the
Curry-Howard correspondence, to the conception of new decision algorithms for
equivalence or verification problems on higher-order programs.

Game semantics often plays one of the two following roles in the literature.

\emph{(1) A syntax-free, compositional operational semantics.} The strategy
interpreting a program is a syntax-free object -- in essence it is a
representation of the \emph{behaviour} of the program, with no information as to
\emph{how} this behaviour is written down in the syntax. In particular, it
abstracts cleanly from the bureaucratic aspects of the syntax and reduction of
the language under examination. It is by nature \emph{compositional}, because the
strategy for a term is calculated by induction on its syntax tree, following the
methodology of denotational semantics. In particular, the application of one term
to another is interpreted as the composition of the corresponding strategies.

A compositional fine-grained description of the execution of higher-order
programs is a useful tool -- for instance, it provides methodologies to study
problems such as termination or complexity in the abstract, in a syntax-free
manner
\cite{DBLP:journals/apal/ClairambaultH10,DBLP:journals/corr/Clairambault15}.
Such a representation is also key to further program analysis. It can provide an
invariant for compilation
\cite{DBLP:conf/popl/Ghica07,DBLP:journals/corr/Schopp14}, or a compositional
model construction on which to perform model-checking
\cite{DBLP:conf/tacas/AbramskyGMO04}.  Even in the purely functional case, it was
recently proposed by Jones \emph{et al} \cite{jones} as a convenient closure-free
way to compute the partial evaluation of a term.

Although historically the focus on game semantics has often been on the second
role mentioned below, a good part of its recent developments have been indeed as
a syntax-free, compositional operational semantics. In this direction, it is to
be compared with various similar frameworks. \emph{Normal form bisimulations}
\cite{DBLP:conf/csl/LassenL07} are close relatives, as recently emphasized by
Levy and Staton \cite{DBLP:conf/csl/LevyS14}.  Recent developments of the
\emph{geometry of interaction} also pursue similar methods and objectives
\cite{DBLP:conf/csl/HoshinoMH14,DBLP:journals/corr/LagoFVY15}. Finally,
Hirschowitz and collaborators have provided a very general framework in which one
can give syntax-free descriptions of different kinds of programs
\cite{DBLP:journals/cuza/HirschowitzP12,DBLP:conf/calco/EberhartHS15}.

\emph{(2) An observational classification of effects.} Beyond the use of game
semantics as an operational semantics, the separation of the \emph{observable
behaviour} of a term from its syntax allowed researchers to study computational
features of programs in terms of the observations that they permitted.  One of
the most acclaimed achievements of early game semantics is the identification of
conditions on strategies (visibility, innocence, well-bracketing) in the context
of Hyland-Ong games, that characterize, not syntactically but
\emph{observationally}, the behaviour of programs having access to certain
effects. Indeed, \emph{innocent well-bracketed} strategies are essentially purely
functional programs \cite{ho}: relaxing innocence to \emph{visibility} captures
the use of ground state \cite{am} while removing well-bracketing captures control
operators \cite{DBLP:conf/lics/Laird97}.  Finally, removing visibility captures
terms that have access to higher-order references
\cite{DBLP:conf/lics/AbramskyHM98}.  This is known as the \emph{semantic cube} or
\emph{Abramsky cube}.

Each combination of these conditions corresponds to a certain programming
language, for which the strategies have exactly the same observation power as
programs. In many cases the resulting model is \emph{fully abstract}, without the
need for a quotient. This allowed researchers in game semantics, starting with
Ghica and McCusker \cite{DBLP:journals/tcs/GhicaM03}, to give decision procedures
for observational equivalence of programs in certain programming languages where
the fully abstract games model is algorithmically presented
\cite{DBLP:journals/tcs/MurawskiW08,DBLP:conf/fossacs/Cotton-BarrattH15}.

Despite this impressive flexibility, each game semantics model comes with its
limitations. The notion of \emph{play}, which is at the heart of any game
semantics model, specifies the observational power of the execution environment.
Whereas the capabilities of Player can be adjusted to a certain extent via
conditions on strategies, this cannot be pushed further than what is wired into
the model construction. For instance in Hyland and Ong's original model, plays
are well-bracketed and visible for both players -- it follows that we only
observe parts of programs visible to an evaluation context with no access to
higher-order state and control operators. Clearly that can be solved; for
instance by allowing more general plays as in \cite{DBLP:conf/lics/AbramskyHM98}.
Then again, the whole model has to be rethought if one wishes to allow Opponent
to be concurrent. The same line of thought led Ghica and Tzevelekos to define
\emph{system-level game semantics} \cite{DBLP:journals/entcs/GhicaT12}, in an
effort to take as few assumptions as possible as to the power of the execution
environment. We advocate here another option, namely 
to use a
\emph{causal} model.

\subsubsection*{Causality.} In \emph{causal} game semantics, a program is not
represented by an enumeration of all its possible interactions with an Opponent
of observational strength wired into the model. Instead, it is represented by an
abstract structure displaying information about the \emph{causal choices} behind
the program's actions.  On the one hand, this means that the model is more
intensional and most likely further away from quotient-free full abstraction
results. On the other, the representation makes few
assumption as to the computational features available to the execution
environment. This makes the model more modular, and  a finer
representation: from the causal game semantics of a program, it is
always possible to recover -- in an effective manner -- the set of
plays corresponding to an observation by a certain kind of environment.
The causal representation has other advantages. For instance, as long
advocated by Melli\`es, it allows us to get rid of artificialities in
the standard play-based composition mechanism for innocent strategies,
making more explicit the fact that it is relational (this is key to the
full completeness results for fragments of linear logic
\cite{DBLP:conf/lics/AbramskyM99,DBLP:conf/lics/Mellies05}).
Furthermore, the importance of causal representations for programs has
been advocated in the past for various purposes, ranging from error
diagnostics \cite{DBLP:journals/tac/BenvenisteFHJ03} to the study of
reversible aspects of computation \cite{DBLP:conf/lics/CristescuKV13}.
Last but not least, causal representations display the evolution of a
concurrent system with partial orders rather than interleavings (it is
\emph{``truly concurrent''}). Such representations avoid the
\emph{state explosion} problem of interleaving-based ones
\cite{DBLP:books/sp/Godefroid96}, leading to potential applications to
the verification of concurrent systems. 

\subsubsection*{Contributions.} Giving causal representations of the execution of
programs is not a new problem. Such models have been developed for various process
languages, from CCS \cite{DBLP:conf/ac/Winskel86}, up to (recently) the
full $\pi$-calculus
\cite{DBLP:conf/calco/EberhartHS15,DBLP:conf/ictac/CristescuKV15}. There seem to
exist few developments on truly concurrent semantics of concurrent languages with
shared state, with the notable exception of a Petri net semantics for a simple
imperative programming language \cite{DBLP:conf/lics/HaymanW06}. 

In this paper, we give a general framework in which one can define such truly
concurrent models for higher-order concurrent languages, with various
synchronization primitives. This has the form of a cartesian closed category of
arenas and concurrent strategies, which are certain event structures.  The
approach is conservative over the category of Hyland-Ong innocent strategies for
PCF (and over the more recent work \cite{DBLP:conf/lics/TsukadaO15} in the
non-deterministic case), in the sense that a pure term is interpreted as its
forest of \emph{P-views}. In this paper, we develop this category in detail, and
illustrate it by spelling out the interpretation of Idealized Parallel Algol
(IPA), a higher-order concurrent programming language with shared state and
semaphores as synchronization primitives. The methodology is that of game
semantics, which provides a well-rooted hope that any of the many languages that
one can model in game semantics could be given a truly concurrent representation
in this framework as well.

\red{To achieve that, we build on the category $\CG$ of \emph{concurrent games on
event structures} introduced by Rideau and Winskel \cite{lics11,cg1}, which
itself belongs to a family of game semantics focusing on positionality rather
than sequentiality initiated by Abramsky and Melli\`es in
\cite{DBLP:conf/lics/AbramskyM99} and subsequently pushed by Melli\`es and others
\cite{DBLP:conf/lics/Mellies05,DBLP:conf/concur/MelliesM07}. Relative to $\CG$, 
the main technical difficulty is to handle \emph{replication}
in programming languages: the same resource may be accessed many times, but
subsequent behaviour cannot depend on the low-level details of how replication
is implemented in the model (it is the \emph{uniformity} problem solved with
\emph{equivalence relations} in AJM games). The first layer of our framework
addresses this difficulty: we first build a compact closed category $\TCG$,
designed to play a similar role as $\CG$ in that it offers the basic
compositional mechanisms on which further semantic constructions rely, while
handling uniformity. Then, relying in $\TCG$ we build the announced cartesian closed
category $\CHO$, designed as a conservative extension of usual Hyland-Ong games.
We consider the main contributions of this paper to be these two categories of
strategies $\TCG$ and $\CHO$; the interpretation of IPA in $\CHO$ is mostly there
for motivations and illustrations purposes. We insist that the intermediate
$\TCG$ is a contribution in itself rather than a means to an end for the
construction of $\CHO$; and indeed in some further work we have found it
convenient to build on $\TCG$ directly rather than through $\CHO$.} 

\red{
\subsubsection*{Other work. 
} Here, we find it 
appropriate to give some further
insight on the broader context of the developments presented in this paper.
The bulk of the present paper is the detailed construction of the games model
used in the conference paper \cite{lics15} to build a fully abstract games model
of PCF based on parallel evaluation. The result of \cite{lics15} also
necessitates the development of a concurrent notion of innocence \cite{ho}, not
covered here. The approach to uniformity adopted here is that given in
\cite{lics15}, rather than the earlier approach of \cite{lics14} (which we will
however mention in the course of the paper).

Our aim for the current paper is that it serves as a reference for the
construction of $\TCG$ and $\CHO$. In that, besides the illustrative
interpretation of IPA, it contains no striking application of the framework.
Hence, to help motivate the rather lengthy and technically involved model
construction, we feel 
it is helpful to add some further perspective on the use of
this model. This framework is a cornerstone for a number of subsequent
developments, obtaining achievements that were not within reach
with the previously existing tools of game semantics. In \cite{lics15}, pairing
$\CHO$ with a notion of concurrent innocence we have given a fully abstract model
of PCF based on parallel evaluation. This was extended to non-deterministic PCF
in the first author's PhD thesis \cite{DBLP:phd/hal/Castellan17}. The category
$\TCG$ was found to accommodate transparently probabilities, supporting the
first notion of probabilistic innocence. This was applied successfully to
probabilistic PCF \cite{lics18} and the probabilistic $\lambda$-calculus
\cite{csl18}. Besides those a number of developments are currently under way, in
which the constructions detailed here play a crucial role. A common trend in all
these developments is that further structure sits on $\TCG$ and $\CHO$ in a
high-level modular way, and do not interact with the details of the construction,
making those convenient for further semantic constructions despite their
intricacies. }

\subsubsection*{Outline.} In Section \ref{sec:replication}, we give some basic
ideas behind the formalization of game semantics on top of event structures, and
introduce the key issues that we will have to solve in order to push these ideas
to a fully-fledged games model. In particular, we will show that we need to move
to a setting of event structures \emph{with symmetry}, in order to handle
uniformity of strategies with respect to replicated resources. In Section
\ref{sec:gameswithsymmetry}, we give the first main contribution of this paper:
the compact closed category $\TCG$. In Section \ref{sec:cho}, we rely on it to
construct our cartesian closed category $\CHO$. Finally in Section
\ref{sec:interpretation}, we illustrate $\CHO$ by describing the interpretation
of IPA.


\section{Arenas, concurrent strategies, and uniformity}
\label{sec:replication}
\red{This first section has several purposes: firstly, it aims to introduce the
basic ideas behind our concurrent formulation of Hyland-Ong games. Secondly, it
recalls from \cite{cg1} the required preliminaries on games on event structures.
Finally, it introduces the main difficulties encountered in trying to obtain a cartesian closed
category based on this, motivating the definitions of Section
\ref{sec:gameswithsymmetry}.}

\subsection{Preliminaries on Idealized Parallel Algol}

Before presenting our game semantics, we fix a syntax (inspired 
by \cite{gm}) for Idealized
Parallel Algol (IPA). It will not be exactly the same language as in \cite{gm} -- notably, it lacks
semaphores. 
We omit them in order to keep the language simple, but they can be interpreted
with the same methodology than shared variables.
Note that the language is mostly here to fix
notations and for providing
examples and illustrations. Indeed, the focus on the paper is on the model construction rather than
its applications, which will come later in companion papers.

The \textbf{types} of IPA are the following.
\[
A, B ::= \com \mid \tbool \mid \tnat \mid A \to B \mid \var
\]

The type $\com$ is a type of \emph{commands}, which returns no useful value (if it returns at all, it returns $\tskip$),
but may perform read/write operations on the memory. 
The types $\tnat$ and $\tbool$ are types for expressions
that (if they return) return respectively a natural number or a boolean. 
Finally, $\var$ is the type for integer variables.
Note that we consider \emph{active expressions}, \emph{i.e.} 
the evaluation of a term of type $\tbool$ or $\tnat$ can trigger side effects.

Raw \textbf{terms} of IPA are described as follows.

\[
\begin{array}{rcl}
M, N &::=& x \mid M\,N \mid \lambda x.\,M \mid Y \mid\\
&& \ttrue \mid \tfalse \mid \ifpcf\,M\,N_1\,N_2 \mid\\
&& n \mid \tsucc\,M \mid \pred\,M \mid \iszero\,M \mid\\
&& \tskip \mid M;N \mid M\parallel N \mid\\
&& \newref\,r\,\tin\,M \mid M:=N \mid~!M \mid \mkvar
\end{array}
\]

The first three lines describe the syntax of PCF \cite{plotkin}. The fourth line describes \emph{commands}
and combinators for them. Finally, the fifth line gives the primitives for manipulating variables.
We include the so-called \emph{bad variable} constructor \cite{am}, but it will only play a very
minor role in our development.

These terms are subject to mostly standard typing rules. We give most of them in Figure \ref{fig:typing_ipa}, 
omitting the standard rules for the $\lambda$-calculus, the fixpoint combinator, and constants.
By convention, we use $\tx$ to range over ground types: $\com, \tbool, \tnat$. By abuse of notation, 
we will also use $\tnat$ and $\tbool$ respectively for the sets of (total) natural numbers and booleans.

\begin{figure}
\begin{mathpar}
\inferrule
	{ \Gamma \vdash M : \com \\ \Gamma \vdash N : \com }
	{ \Gamma \vdash M \parallel N : \com }
\and
\inferrule
	{ \Gamma \vdash M : \com \\ \Gamma \vdash N : \tx }
	{ \Gamma \vdash M;N : \tx }
\and
\inferrule
	{ \Gamma \vdash M : \var \\ \Gamma \vdash N : \tnat }
	{ \Gamma \vdash M:=N : \com }
\and
\inferrule* 
	{ \Gamma, x: \var \vdash M:\tx }
	{ \Gamma \vdash \newref\,x\,\mathbf{in}\,M : \tx }
\and
\inferrule
	{ \Gamma \vdash M : \tbool \\ \Gamma \vdash N_i : \tx }
	{ \Gamma \vdash \ifpcf\,M\,N_1\,N_2 : \tx }
\and
\inferrule
	{ \Gamma \vdash M : \tnat \to \com \\ \Gamma \vdash N : \tnat }
	{ \Gamma \vdash \mkvar\,M\,N : \var }
\end{mathpar}
\caption{Typing rules of IPA}
\label{fig:typing_ipa}
\end{figure}

Although some of our typing rules seem restricted to output ground types, the full rules can be derived
as syntactic sugar. For instance, a version of $\ifpcf$ that eliminates to $\var$ can be obtained as:
\begin{mathpar}
\inferrule
	{ \Gamma \vdash M : \tbool \\ \Gamma \vdash N_i : \var }
	{ \Gamma \vdash \mkvar\,(\lambda x.\,\ifpcf\,M\,(N_1:=x)\,(N_2:=x))\,(\ifpcf\,M\,!N_1\,!N_2) : \var }
\end{mathpar}

It is an easy verification that the other rules can be generalized similarly.

The language can be equipped with standard (small-step) operational semantics, see \cite{gm} for
details. We omit it here since it will play no role in the technical development.

%
%
%

\subsection{Partial orders and conflicts for strategies}
\label{subsec:dialogues}

We now start introducing our semantics. In the remainder of this section we introduce gradually the main 
ideas behind our model, relying as much as possible on examples. Our starting
point will be
the standard Hyland-Ong innocent semantics for PCF, which we will use to motivate
concurrent games on event structures. This section will culminate on the issues of replication and
uniformity, which will prompt the developments of Section
\ref{sec:gameswithsymmetry}.

\subsubsection{Dialogue games.}
\emph{Hyland-Ong games} formalize the intuition that a program is a strategy having a dialogue with its execution
environment. A possible dialogue on the type $\tbool \to \tbool \to \tbool \to \tbool$ appears in
Figure \ref{fig:dialogue}.
\begin{figure}[h!]
\[
\xymatrix@R=0pt{
\tbool \ar@{}[r]|{\to} & \tbool \ar@{}[r]|{\to}& \tbool \ar@{}[r]|{\to}& \tbool\\
&&&\q&(-,\Qu)\\
\q\ar@{--}@/^/[urrr]&&&&(+,\Qu)\\
\ttrue  \ar@{--}@/^/[u]&&&&(-,\An)\\
&\q\ar@{--}@/^/[uuurr]&&&(+,\Qu)\\
&\ttrue \ar@{--}@/^/[u]&&&(-,\An)\\
&&&\ttrue       \ar@{--}@/_/[uuuuu]&(+,\An)
}
\]
\caption{A dialogue on $\tbool \to \tbool \to \tbool \to \tbool$}
\label{fig:dialogue}
\end{figure}
The diagram is read from top to bottom.
Each move is either by Player or Opponent, and is either a Question or an Answer. Questions correspond to variable
calls, whereas Answers indicate a call terminating. The dashed lines between moves (traditionally called \emph{justification
pointers}) convey information about
thread indexing; in this example they are redundant but become required at higher types -- we will see more on them later.

In natural language, this diagram would read: \emph{``The environment starts the evaluation of a term of type $\tbool \to \tbool \to \tbool \to \tbool$
by interrogating its return type $(-, \Qu)$. The evaluation requires information on the first argument, so the term
triggers its evaluation by playing under the corresponding component of the type $(+, \Qu)$. The evaluation
of the argument terminates with value $\ttrue$ $(-, \An)$. Knowing that its first argument is $\ttrue$, the term
now needs information on its second argument $(+, \Qu)$. This argument returns $\ttrue$ $(-, \An)$, and 
now the term computes and returns $\ttrue$ at toplevel $(+, \An)$.''} The reader should recognize here
a description of an execution of $\ifpcf$.

%
In Hyland-Ong games, \textbf{sequential innocent strategies} consist of sets of dialogues as above,
where Opponent moves are \emph{justified} by the preceding one
-- such dialogues are known as \emph{$P$-views}.
A strategy for a term of PCF contains several such dialogues, specifying the term entirely.
The full strategy for $\ifpcf$ contains in total four maximal P-views, displayed
in Figure \ref{fig:dialogues_if}.
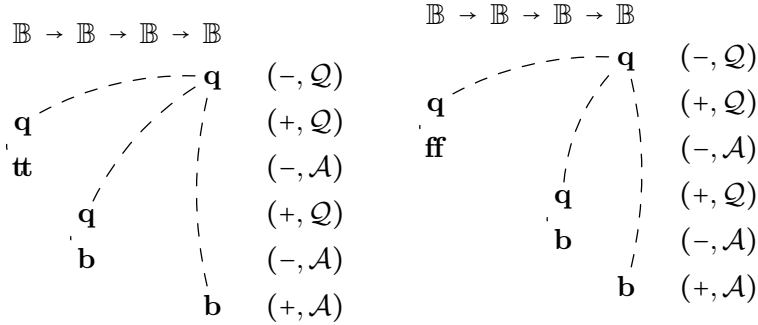
\begin{figure}[h]
\[
\xymatrix@R=0pt@C=10pt{
\tbool \ar@{}[r]|{\to} & \tbool \ar@{}[r]|{\to}& \tbool\ar@{}[r]|{\to}&\tbool\\
&&&\q&(-,\Qu)\\
\q\ar@{--}@/^/[urrr]&&&&(+,\Qu)\\
\ttrue  \ar@{--}@/^/[u]&&&&(-,\An)\\
&\q\ar@{--}@/^/[uuurr]&&&(+,\Qu)\\
&\b\ar@{--}@/^/[u]&&&(-,\An)\\
&&&\b\ar@{--}@/^/[uuuuu]&(+,\An)
}\hspace{20pt}
\raisebox{10pt}{
\xymatrix@R=0pt@C=10pt{
\tbool \ar@{}[r]|{\to} & \tbool \ar@{}[r]|{\to}& \tbool\ar@{}[r]|{\to}&\tbool\\
&&&\q&(-,\Qu)\\
\q\ar@{--}@/^/[urrr]&&&&(+,\Qu)\\
\tfalse \ar@{--}@/^/[u]&&&&(-,\An)\\
&&\q\ar@{--}@/^/[uuur]&&(+,\Qu)\\
&&\b        \ar@{--}@/^/[u]&&(-,\An)\\
&&&\b       \ar@{--}@/_/[uuuuu]&(+,\An)
}}
\]
\caption{Maximal $P$-views for $\ifpcf$}
\label{fig:dialogues_if}
\end{figure}
Such non-empty sets of P-views (satisfying further conditions: determinism and well-bracketing) are
usually called \emph{well-bracketed innocent strategies}. Because of their correspondence with
certain normal forms (\emph{PCF B\"ohm trees} \cite{curien}), they are the cornerstone of Hyland-Ong games and 
of the full abstraction results they allowed.

\subsubsection{Partially ordered dialogues.}
In Hyland-Ong games, every P-view is a total order, meaning that the whole sequential innocent strategy is a tree. In
our framework, we question this premise. For instance, informally, the intuitive behaviour of the parallel composition operation 
$\parallel:\com \to \com \to \com$ of IPA is most elegantly represented as in Figure \ref{fig:po_dialogue}.

\begin{figure}[h!]
\[
\xymatrix@R=10pt{
\com	\ar@{}[r]|\to&
\com	\ar@{}[r]|\to&
\com\\
&&\run	\ar@{-|>}[dl]\ar@{-|>}[dll]&(-,\Qu)\\
\run	\ar@{--}@/^/[urr]
	\ar@{-|>}[d]&
\run	\ar@{--}@/^/[ur]
	\ar@{-|>}[d]&&(+,\Qu)\\
\done	\ar@{-|>}[drr]
	\ar@{--}@/^/[u]&
\done	\ar@{-|>}[dr]
	\ar@{--}@/^/[u]&&(-,\An)\\
&&\done	\ar@{--}@/_/[uuu]&(+,\An)
}
\]
\caption{A partially ordered dialogue}
\label{fig:po_dialogue}
\end{figure}
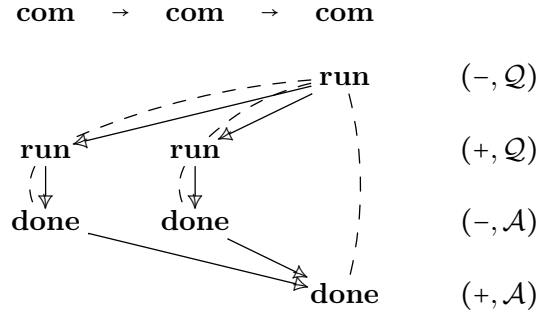

The diagram of Figure \ref{fig:po_dialogue} is analogous to the previous ones, but is now partially 
ordered rather than totally ordered. The relation $\imc$ denotes immediate causality;
it was unnecessary before, as it coincided with chronological contiguity. The justification
pointers remain -- we will see more on their precise nature later. Note that
in the standard game semantics for IPA \cite{gm}, this
partial order would be only implicit; and given by all the possible linear orderings of
the partial order above. Here instead, the partial order will be the first-class object of
interest. Strategies, in particular, will be partially ordered.

\subsubsection{Non-determinism.} Concurrent languages are, in general, non-deterministic. Note
that we do not mean non-deterministic in the sense that, as above, the execution admits
multiple distinct linear orderings. For us, non-determinism means that execution takes 
irreconciliable routes, even up to permutation of independent events. We illustrate that 
by the following two examples.

\begin{figure}[h!]
\[
\xymatrix@R=10pt@C=5pt{
&\tbool\\
&\q	\ar@{-|>}[dl]
	\ar@{-|>}[dr]&&(-,\Qu)\\
\ttrue	\ar@{~}[rr]
	\ar@{--}@/^/[ur]&&
\tfalse	\ar@{--}@/_/[ul]&(+,\An)
}
\hspace{40pt}
\xymatrix@R=10pt@C=0pt{
&\com\!\!\!\!\!\!\!\!	\ar@{}[rrrrrr]|\to&&~~~~~&~~~~~&&&\com\\
&&&&&&& \run
	\ar@{-|>}[dlllllll]
	\ar@{-|>}[dlllll]&(-,\Qu)\\
\run	\ar@{-|>}[d]
	\ar@{--}@/^/[urrrrrrr]
	\ar@{~}[rr]&&
\run	\ar@{-|>}[d]
	\ar@{--}@/^/[urrrrr]&&&&&&(+,\Qu)\\
\done	\ar@{--}@/^/[u]&&
\done	\ar@{--}@/^/[u]
	\ar@{-|>}[drrrrr]&&&&&&(-,\An)\\
&&&&&&& \done
	\ar@{--}@/_/[uuu]&(+,\An)
}
\]
\caption{Non-deterministic dialogues}
\label{fig:nd_dialogues}
\end{figure}
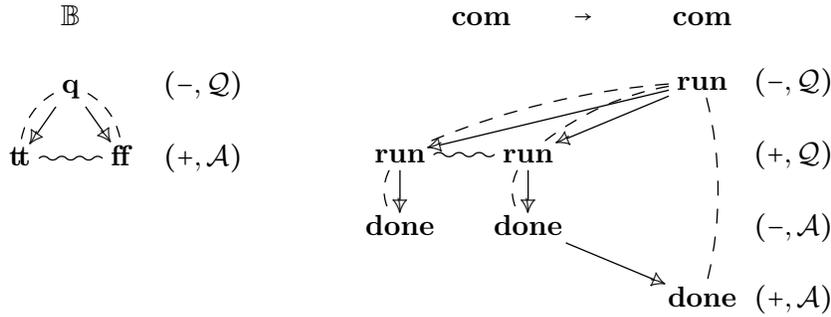

In the above two diagrams, the wiggly line $\mconflict\ \,$ indicates \emph{immediate conflict}. Two 
moves/events related by immediate conflict are incompatible, and can never occur together in an execution. Accordingly,
the first example is a representation of our strategy for the non-deterministic boolean, which answers either
true or false. The second example illustrates another key aspect of our model: we remember explicitly the point
of non-deterministic choice. Here, Player silently flips a coin. If the result is heads, they evaluate the argument,
then terminate. However, if the result is tails, they evaluate the argument, then diverge. Typical play-based game 
semantics would forget the halting branch, which is contained in the other. Instead, our model represents the two branches
explicitly.

We now show how to make such diagrams formal.

\pagebreak
\subsection{Prestrategies on arenas}

\subsubsection{Event structures.} Such a combination of causality and non-determinism is elegantly expressed 
via Winskel's \emph{event structures} \cite{DBLP:conf/ac/Winskel86}. 

\begin{defi}
An \textbf{event structure} \index{Event structure} (\emph{es} for short) is a tuple $(E, \leq_E, \Con_E)$ where $E$ is a set of \emph{events},
$\leq_E$ is a partial order indicating \emph{causality} and $\Con_E$ is a 
nonempty set of finite subsets of
$E$, satisfying:
\[
\begin{array}{l}
\forall e\in E,~[e]_E = \{e' \in E \mid e' \leq_E e\}\text{ is finite}\\
\forall e\in E,~\{e\}\in \Con_E\\
\forall X \in \Con_E,~\forall Y \subseteq X,~Y\in \Con_E\\
\forall X \in \Con_E,~\forall e\in E,~\forall e'\in X,~e\leq_E e' \implies X\cup \{e\} \in \Con_E
\end{array}
\]%
\nomenclature[aa]{$\leq_E$}{Causal order on event structure $E$}%
\nomenclature[ab]{$\Con_E$}{Consistent subsets of event structure $E$}%
\end{defi}

The set $\Con_E$ of \emph{consistent subsets} specifies which events can occur together in an 
execution of the system. 
The \emph{states} of an event structure $E$, called the (finite)
\textbf{configurations}\index{Configurations}, are those finite sets $x \subseteq E$ that are%
\nomenclature[ac]{$\conf{E}$}{Configurations of event structure $E$}%
both consistent and \textbf{down-closed} (\emph{i.e.}  for all
$e\in x$, for all $e'\leq e$, then $e'\in x$) -- the set of
configurations on $E$ is written $\conf{E}$, and is partially ordered
by inclusion. Configurations with a maximal element are called
\textbf{prime configurations}\index{Prime configurations}, they are those of the
form $[e]$ for%
\nomenclature[ad]{$[X]_E$}{Down-closure in $E$ of $X\subseteq E$, written $[e]$ for a singleton
$\{e\}$}%
$e\in E$ (note that we drop the $E$ in $[e]_E$ whenever, as above, this is clear
from the context) -- more generally, we will write $[X]_E$ for the down-closure
of a set of events $X$ in $E$.
We will also use the notation $[e) = [e] \setminus \{e\}$.%
\nomenclature[ae]{$[e)$}{Dependencies of $e$, $[e) = [e] \setminus \{e\}$}%
We write $x \longcov{e}$ to mean that $e \not \in x$ and%
\nomenclature[af]{$x\longcov{e}$}{Configuration $x$ extends with event $e$}%
$x \cup \{e\} \in \conf{E}$. Finally, when drawing event structures as
above, we do not represent the full partial order $\leq$ but the
\textbf{immediate causality} generating it, defined as $e \imc e'$%
\nomenclature[ag]{$\imc$}{Immediate causality}%
whenever $e < e'$ and for any $e \leq e'' \leq e'$, either $e = e''$
or $e'' = e'$.
We will often omit the subscripts in $\leq$ or $\imc$ when they are clear from the context.

In this paper, most of the event structures we consider (such as those in the previous subsection)
have a simpler consistency structure.

\begin{defi}
An \textbf{event structure with binary conflict}\index{Binary conflict} is a triple $(E, \leq_E, \conflict_E)$, where
$\leq_E$ is a partial order and $\conflict_E$ is an irreflexive symmetric binary
relation on $E$,%
\nomenclature[ah]{$e_1 \conflict e_2$}{Binary conflict between events $e_1$ and
$e_2$}%
such that:
\[
\begin{array}{l}
\forall e\in E,~[e]_E~\text{is finite}\\
\forall e_1 \conflict_E e_2,~\forall e_2 \leq_E e'_2,~e_1 \conflict_E e'_2
\end{array}
\]
\end{defi}

It is easy to check that an event structure with binary conflict is an event structure, with 
$\Con_E = \{X \in \mathcal{P}_f(E) \mid \forall e_1, e_2 \in X,~\neg(e_1 \conflict_E e_2)\}$. On the
other hand, not every event structure can be described via a binary conflict (take \emph{e.g.} three
events with any subset of cardinal less than two being consistent).
The strategies in the cartesian closed category we aim to build will only have binary
conflict, and accordingly in Section \ref{sec:cho} we will restrict to event structures with binary
conflict. In the meantime, some aspects of the theoretical development are smoother when carried out
with arbirary consistency.

In an event structure with binary conflict, we can trace back conflicts to their original cause.
For $e_1 \conflict_E e_2$ we say that the conflict is
\textbf{minimal}\index{Minimal conflict}, written%
\nomenclature[ai]{$e_1 \mconflict~e_2$}{Immediate conflict between events $e_1$
and $e_2$}%
$e \mconflict_E~e'$, iff for all $e''\leq e$
we have $\neg (e'' \conflict_E e')$ and for all $e'' \leq e'$ we have
$\neg (e \conflict_E e')$.
As above, we will often drop the subscripts in $\conflict$ or $\mconflict$ when they are clear from the context.
Following this notation, all the diagrams of the previous subsection can be 
regarded as representations of event structures with binary conflict (ignoring the dashed lines).

\subsubsection{Games and arenas.} In game semantics, dialogues as in Subsection \ref{subsec:dialogues}
obey the rules of a \emph{game} inherited from the type. In order to define it, let us first recall
the following notion from \cite{cg1}.

\begin{defi}%
\nomenclature[aj]{$\pol_A$}{Polarity function on esp $A$}%
An \textbf{event structure with polarities (esp)}\index{Event structure!with polarities (esp)} is an event structure $A$ along with a \textbf{polarity function}
\[
\pol_A : A \to \{-, +\}
\]
associating to any event a polarity, that is either $-$ for Opponent or $+$ for Player.

By a \textbf{game}\index{Game}, we simply mean an event structure with polarities.
\end{defi}

Those games form the objects of the category $\CG$ of concurrent games of
\cite{cg1}. \red{As one key aim of the paper is to reconstruct a version of
Hyland-Ong games, certain games called \emph{arenas} will play a particular role
in our development (being the objects of our cartesian closed category $\CHO$).}

\begin{defi}\label{def:arena}
An \textbf{arena}\index{Arena} is a \textbf{conflict-free} (all finite sets consistent) esp/game satisfying:
\begin{itemize}
\item \emph{Forest.} if $a_1, a_2 \leq a \in A$, then either
$a_1 \leq a_2$ or $a_2 \leq a_1$.
\item \emph{Alternation.} if $a_1 \imc a_2$, then $\pol(a_1) \neq \pol(a_2)$.
\end{itemize}

An arena $A$ is \textbf{negative}\index{Negative!arena} if all its minimal events have negative polarity.
\end{defi}

Arenas are close representations of types.
Although formulated a bit differently, our arenas are the same as in
\cite{ho} (with the exception of the absence of the Question/Answer
labeling, \red{which we do not require: as we do not aim for a full abstraction
result, no notion of well-bracketing is needed}).

\begin{exa}\label{ex:arenas}
Leaving for later the general interpretation of types, we have:
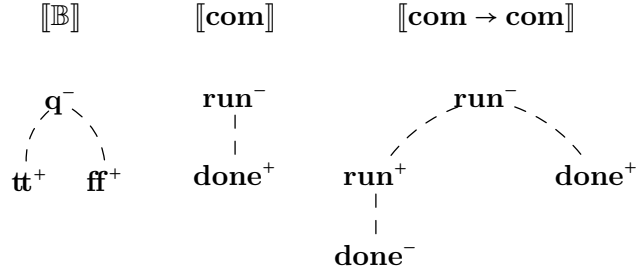
\[
\xymatrix@C=0pt@R=15pt{
&\!\!\!\intr{\tbool}\!\!\!\\
&\!\!\!\q^{-}\!\!\!	\ar@{--}@/_/[dl]
	\ar@{--}@/^/[dr]\\
\!\!\!\ttrue^{+}\!\!\!&&\tfalse^{+}
}
\hspace{20pt}
\xymatrix@C=0pt@R=15pt{
\intr{\com}\\
\run^{-}	\ar@{--}[d]\\
\done^{+}
}
\hspace{20pt}
\xymatrix@C=0pt@R=15pt{
&\!\!\!\!\!\!\intr{\com \to \com}\!\!\!\!\!\!\\
&\!\!\!\run^{-}\!\!\!	\ar@{--}@/_/[dl]
	\ar@{--}@/^/[dr]\\
\!\!\!\run^{+}\!\!\!	\ar@{--}[d]&&\!\!\!\done^{+}\!\!\!\\
\!\!\!\done^{-}\!\!\!
}
\]

Throughout this paper, we will often omit the semantic brackets on types when this causes
no confusion and simply refer to these arenas as $\tbool, \com,$ \emph{etc}.
\end{exa}

By convention, we represent immediate causality in arenas by dashed lines $\xymatrix{\,\!\ar@{--}[r]&\,\!}$ 
rather than $\imc$. Events are ordered from top to bottom, and are annotated with their polarity. We observe in 
the third example -- and it will be a general fact once we give the formal definitions -- that each move in
the arena $\intr{\com \to \com}$ \emph{comes from} a well-defined occurrence of a base type $\com$ in 
$\com \to \com$: $\run^{-}$ and $\done^{+}$ come from the output $\com$, and 
$\run^{+}$ and $\done^{-}$ come from the input $\com$.
As usual in game semantics, this is used in the representation of dialogues (as in Subsection \ref{subsec:dialogues}):
whenever possible, moves are placed under the corresponding base type occurrence.

\subsubsection{Prestrategies.} 
The dialogues of Subsection \ref{subsec:dialogues}, and our notion of strategies (called \emph{prestrategies} for now -- 
more conditions are to come), will be event structures
\emph{labeled by a game}. In other words, a prestrategy will be an event stucture $S$ along with
a labeling function $\sigma : S \to A$ associating to each event in $S$ an image in the game. These
labeling functions need to satisfy conditions corresponding to the notion of \emph{map of event
structures}.

\begin{defi}\label{def:map_es}%
\nomenclature[aja]{$\ES$}{Category of event structures and maps of event
structures}%
Let $E, F$ be event structures. A \textbf{morphism (map) of event structures}\index{Morphisms!of event structures} $f : E \to F$ is a function, satisfying:
\begin{itemize}
\item \emph{Preservation of configurations.} For all $x\in \conf{E}$, $f x \in \conf{F}$,
\item \emph{Local injectivity.} For all $e, e'\in x \in \conf{E}$, if $f e = f e'$ then $e = e'$.
\end{itemize}
Event structures and maps between then form a category $\ES$.

A \textbf{prestrategy}\index{Prestrategy} on a game $A$ is a map of event structures $\sigma : S \to A$.
\end{defi}

So a prestrategy $\sigma : S \to A$ must only reach valid states of $A$, and behaves \emph{linearly}:
in a configuration, each event of the game appears at most once. We note in passing that for non-linear
languages, this linearity assumption will be circumvented by creating duplicates of events -- more on
that later.

If $\sigma : S \to A$ is a prestrategy, then $S$ automatically inherits from $A$ a polarity
function that we write $\pol_S : S \to \{-, +\}$, 
leaving the dependency on $\sigma$ implicit. Of course, it is equivalent 
to require $S$ to be explicitely equipped with polarities, in a way preserved by
$\sigma$ \red{-- we then say that $\sigma$ is a \textbf{map of esps}.}

\subsubsection{Representations of prestrategies.}
We will only draw prestrategies with binary conflict. When drawing 
such a $\sigma : S \to A$ as in Subsection \ref{subsec:dialogues}, we only draw $S$ (more precisely, with immediate causality
$\imc$ and immediate conflict $\mconflict$), where each event is presented as its image through $A$, and 
placed under the corresponding ground type occurrence in the type. We use
the dashed lines $\xymatrix{\,\!\ar@{--}[r]&\,\!}$ to represent the relation on $S$ induced by immediate causality
on $A$. For instance, the
second diagram of Figure \ref{fig:nd_dialogues} is a representation of the map of event structures below.

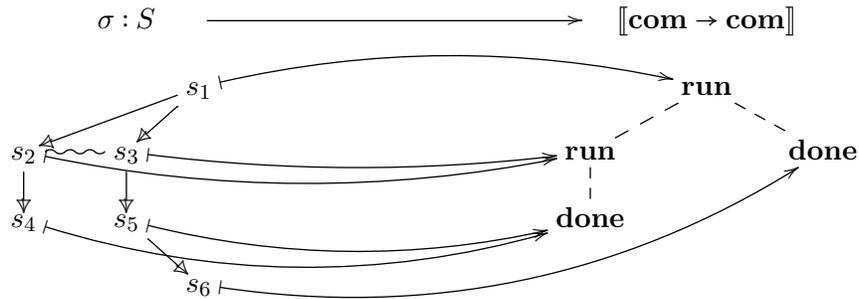
\[
\xymatrix@R=10pt@C=5pt{
&&\sigma:S&	\ar[rrrrrrrrrrr]&&&&&&&&&&&&\!\!\!\!\!\!\!\!\!\intr{\com \to \com}\!\!\!\!\!\!\!\!\!\\
&&& s_1
	\ar@{|->}@/^1pc/[rrrrrrrrrrrr]
        \ar@{-|>}[dlll]
        \ar@{-|>}[dl]			&&&&&&&&~~~~~~~~~~&&&&\run	\ar@{--}[dl]
								\ar@{--}[dr]\\
s_2     \ar@{|->}@/_1pc/[rrrrrrrrrrrrrr]
	\ar@{-|>}[d]
        \ar@{~}[rr]&&
s_3     \ar@{|->}@/_/[rrrrrrrrrrrr]
	\ar@{-|>}[d]&&&&&&&&&&&&		\run	\ar@{--}[d]&&\done\\
s_4	\ar@{|->}@/_1.5pc/[rrrrrrrrrrrrrr]&&
s_5     \ar@{|->}@/_1pc/[rrrrrrrrrrrr]
	\ar@{-|>}[dr]&&&&&&&&&&&&\done\\
&&& s_6
	\ar@{|->}@/_2pc/[uurrrrrrrrrrrrr]\\
~
}
\]

As the reader can see, this explicit map notation is a bit cumbersome. Its representation as in Figure \ref{fig:nd_dialogues}
conveys the relevant information -- the only thing lost is the ``name'' ($s_1, \dots, s_6$) of moves in $S$.
More formally, it should be clear to the reader that such a representation displays a finite
prestrategy $\sigma : S \to A$ adequately
up to \textbf{isomorphism} of prestrategies:

\begin{defi}\label{def:2cell_strat}
Let $\sigma : S \to A$ and $\tau : T \to A$ be prestrategies. A \textbf{morphism}\index{Morphisms!of prestrategies} from $\sigma$ to
$\tau$ is
a map of event structures $f : S \to T$ such that $\tau \circ f = \sigma$.

Accordingly, an \textbf{isomorphism}\index{Isomorphism!of prestrategies} between $\sigma$ and $\tau$ is given by $(f, g)$, where $f : S
\to T$ and $g : T \to S$ are maps between $\sigma$ and $\tau$ such that $g \circ f = \id_S$ and $f
\circ g = \id_T$. We write $\sigma \siso \tau$ to mean that $\sigma$ and $\tau$
are isomorphic -- in%
\nomenclature[ak]{$\sigma \siso \tau$}{(Strong) isomorphism between
prestrategies}%
that case we might sometimes say that $\sigma$ and $\tau$ are \textbf{strongly isomorphic} to
emphasize the distinction with \emph{weak isomorphisms}, to be defined in Definition
\ref{def:weak_mor}.
\end{defi}

\subsection{Compositional structure}
\label{sec:plain_comp}

In order to obtain such representations of programs compositionally, the standard methodology of denotational semantics
suggests to organize them as a category. Rideau and Winskel \cite{lics11} give the basic ingredients for the
construction of a (bi)category of games on event structures. We give here the main ideas and definitions, but refer
the reader to \cite{cg1} for a more in-depth construction with proofs.

We start with the following simple definition.

\begin{defi}%
\nomenclature[al]{$A\parallel B$}{Simple parallel composition}%
The \textbf{simple parallel composition}\index{Simple parallel composition!of event structures} $E_1 \parallel E_2$ of two event structures $E_1$ and $E_2$ has:
\begin{itemize}
\item \emph{Events.} The disjoint union $\{1\} \times E_1 \cup \{2\} \times E_2$,
\item \emph{Causality.} We have $(i, e) \leq_{E_1 \parallel E_2} (j, e')$ iff $i = j$ and $e \leq_{E_i} e'$,
\item \emph{Consistency.} For $X = \{1\} \times X_1 \uplus \{2\} \times X_2$ (often written simply $X_1
\parallel X_2$) a finite subset of $E_1 \parallel E_2$, we have $X \in \Con_{E_1 \parallel E_2}$
iff $X_1 \in \Con_{E_1}$ and $X_2 \in \Con_{E_2}$.
\end{itemize}

The simple parallel composition of event structures with binary conflict still has binary
conflict. Simple parallel composition can also be applied to maps $f : E \to E'$
and $g : F \to F'$ to form $f \parallel g : E \parallel F \to E' \parallel F'$ in
the obvious way; as usual we also write $f \parallel F$ for $f \parallel \id_F$.
If $A$ and $B$ have polarities, so is $A\parallel B$, with $\pol_{A\parallel B}((1, a)) = \pol_A(a)$ and
$\pol_{A\parallel B}((2, b)) = \pol_B(b)$.
\end{defi}

In other words, the two event structures are put side by side, without any interaction. If $A$ is a
game, then there is also its \textbf{dual}\index{Dual!of an esp} $A^\perp$,
defined as having the same events, causality,%
\nomenclature[am]{$A^\perp$}{Dual}%
consistency as $A$, but reversed polarity: $\pol_{A^\perp}(a) = -\pol_A(a)$. Both operations $- \parallel -$ and
$(-)^\perp$ are defined on all esps/games, but preserve arenas.

\subsubsection{Morphisms.} Given games $A$ and $B$, a \textbf{prestrategy from $A$ to
$B$}\index{Prestrategy!from $A$ to $B$} is a prestrategy:
\[
\sigma : S \to A^\perp \parallel B
\]
We will sometimes write $\sigma : A \profto B$ to keep the $S$ anonymous.%
\nomenclature[ama]{$\sigma : A \profto B$}{Prestrategy $\sigma : S \to A^\perp \parallel B$, with
$S$ anonymous}%
The basic example of a prestrategy from $A$ to $A$ is the \emph{copycat strategy}. 

\begin{defi}\label{def:copycat_str}%
\nomenclature[an]{$\cc_A : \CC_A \to A^\perp \parallel A$}{Copycat strategy}%
Let $A$ be a game. We define an event structure $\CC_A$ as having:
\begin{itemize}
\item \emph{Events.} Those of $A^\perp \parallel A$,
\item \emph{Causality.} The transitive closure of the relation:
\[
\begin{array}{l}
\{((1, a), (1, a'))\mid a \leq_{A^\perp} a'\} \cup \{((2, a), (2, a'))\mid a \leq_A a'\} \cup\\
\{((1, a), (2, a))\mid \pol_A(a) = +\}\cup \{((2, a), (1, a))\mid \pol_A(a) = -\}
\end{array}
\]
\item \emph{Consistency.} For $X$ a finite subset of $\CC_A$, we have 
$X\in \Con_{\CC_A}$ iff $[X]_{\CC_A} \in \Con_{A^\perp \parallel A}$.
\end{itemize}

In particular, if $A$ is an arena, then $\CC_A$ is conflict-free. 

The \textbf{copycat prestrategy}\index{Copycat strategy} is the identity function, which is a map of es:
\[
\cc_A : \CC_A \to A^\perp \parallel A
\]
\end{defi}

\begin{exa}\label{ex:copycat}
The copycat prestrategy from $\intr{\com \to \com}$ to itself is:
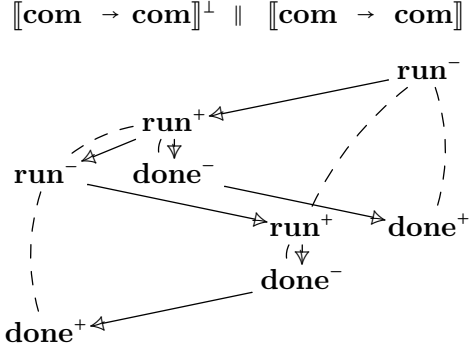
\[
\xymatrix@R=5pt@C=10pt{
\llbracket\com	\ar@{}[r]|\to&
\com\rrbracket^\perp	\ar@{}[r]|{\parallel}&
\llbracket\com	\ar@{}[r]|\to&
\com\rrbracket\\
&&&\run^-	\ar@{-|>}[dll]\\
&\run^+		\ar@{-|>}[dl]
		\ar@{-|>}[d]\\
\run^-		\ar@{-|>}[drr]
		\ar@{--}@/^/[ur]&
\done^-		\ar@{-|>}[drr]
		\ar@{--}@/^/[u]\\
&&\run^+	\ar@{-|>}[d]
		\ar@{--}@/^/[uuur]&
\done^+		\ar@{--}@/_/[uuu]\\
&&\done^-	\ar@{-|>}[dll]
		\ar@{--}@/^/[u]\\
\done^+		\ar@{--}@/^/[uuu]
}
\]
\end{exa}

Note that the partial order above is a tree, whose branches are exactly the P-views of the usual corresponding copycat strategy
in Hyland-Ong games.

\subsubsection{Interaction.} \label{subsubsec:rep_interaction}%
As usual in game semantics, composition is obtained by a two-step process: parallel interaction,
plus hiding. The main difficulty in defining the composition of prestrategies is parallel interaction -- we first explain how it
is done on a closed interaction between $\sigma : S \to A$ and $\tau : T \to A^\perp$.
The \emph{interaction}%
\nomenclature[ao]{$\sigma \wedge \tau : S \wedge T \to A$}{Interaction of $\sigma : S \to A$ and
$\tau : T \to A^\perp$}%
of $\sigma$ and $\tau$, written $\sigma \wedge \tau : S \wedge T \to A$, will be
a labeled event structure (by which we mean an event structure $S \wedge T$ with
a map $\sigma \wedge \tau$, the \emph{labelling function}) describing the behaviours
accepted by both $\sigma$ and $\tau$.

Its construction is done in several stages. Firstly, its \emph{states} should correspond to certain pairings between matching states of
$\sigma$ and states of $\tau$, \emph{i.e.} pairs $(x, y) \in \conf{S} \times \conf{T}$ such that $\sigma x = \tau y$. Note that in such a
case, the local injectivity assumption on $\sigma$ and $\tau$ induces a bijection $\varphi_{x, y}$ between $x$ and $y$ -- in 
fact matching pairs $(x, y)$ are in one-to-one correspondence with bijections $\varphi : x \sbij y$ between configurations
of $S$ and $T$ such that for all $s \in x$, $\tau\,(\varphi\,s) = \sigma\,s$, indicating which events \emph{synchronise} with 
each other. However, not all such bijections represent valid
states of the interaction, as $\sigma$ and $\tau$ might not agree on the \emph{order} in which to play events in $x, y$.
This is addressed by requiring bijections to be \emph{secured}, as below.

\begin{defi}\label{def:secbij}%
\nomenclature[ap]{$\varphi : x \sbij y$}{Secured bijection between $x$ and $y$}%
\nomenclature[aq]{$\secbij{\sigma, \tau}$}{Set of secured bijections between
(configurations of) $\sigma$ and $\tau$}%
Let $\sigma : S \to A$ and $\tau : T \to A^\perp$ be prestrategies. A \textbf{secured
bijection}\index{Secured bijection} between $x\in \conf{S}$
and $y \in \conf{T}$ is a bijection
\[
\varphi : x \sbij y
\]
such that for all $s\in x$ we have $\tau\,(\varphi\,s) = \sigma\,s$, and which is \textbf{secured}, in the sense that
the reflexive transitive closure of
\[
(s, t) \vartriangleleft (s', t') \Leftrightarrow s <_S s' \vee t <_T t'
\]
is a partial order written $\leq_\varphi$ on (the graph of) $\varphi$, making $(\varphi, \leq_\varphi)$ a poset.
We write $\secbij{\sigma, \tau}$ the set of secured bijections between $\sigma$ and $\tau$.
\end{defi}

\begin{exa}
In the diagram below are represented two prestrategies $\sigma$ on $\com \parallel \com$, and $\tau$ on $(\com \parallel \com)^\perp$.
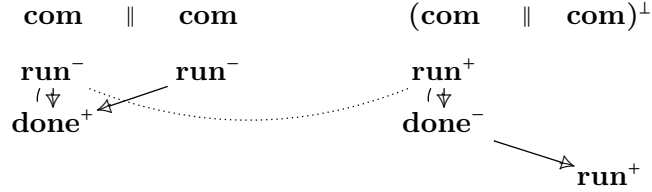
\[
\xymatrix@R=5pt{
\com	\ar@{}[r]|\parallel &\com&&
(\com	\ar@{}[r]|\parallel &\com)^\perp\\
\run^-  \ar@{-|>}[d]
	\ar@{.}@/_1.5pc/[rrr]&
\run^-  \ar@{-|>}[dl]&&
\run^+  \ar@{-|>}[d]\\
\done^+ \ar@{--}@/^/[u]&&&
\done^- \ar@{--}@/^/[u]
        \ar@{-|>}[dr]\\
&&&&\run^+
}
\]
The dotted line is the only pair in the unique maximal secured bijection in $\secbij{\sigma, \tau}$. The maximum
configurations of $\sigma$ and $\tau$ are matching, but not in a secured way.
\end{exa}

This gives a notion of state of the interaction, but we expected to build a labeled event structure. Hence we wish to
present $\secbij{\sigma, \tau}$ (up to isomorphism) as the set of configurations of an event structure $S \wedge T$. This is done
via the \emph{prime construction}. Say that a secured bijection $(\varphi, \leq_\varphi)$ is a
\textbf{prime}\index{Prime secured bijection} when
it has exactly one maximal element $(s_\varphi, t_\varphi)$. In other words, a prime secured bijection is the 
data of one synchronisation $(s_\varphi, t_\varphi)$, plus a causally valid history for it. We now form:

\begin{defi}\label{def:es_pb}
The event structure $S \wedge T$ is obtained as follows:
\begin{itemize}
\item \emph{Events.} Prime secured bijections $\varphi \in \secbij{\sigma, \tau}$.
\item \emph{Causal order.} Inclusion of secured bijections.
\item \emph{Consistency.} For $X$ a finite subset of $\secbij{\sigma, \tau}$ of
prime secured bijections, we have $X \in \Con_{S\wedge T}$ iff $\cup X \in \secbij{\sigma, \tau}$.
\end{itemize}
There is a map of es $\sigma \wedge \tau : S\wedge T \to A$, given by
$(\sigma \wedge \tau)\,\varphi = \sigma\,s_\varphi = \tau\,t_\varphi$.
\end{defi}

In passing, we note that 
if $\sigma : S \to A$ and $\tau : T \to A^\perp$ have binary conflict (meaning $S$ and $T$ have
binary conflict), then so does $\sigma \wedge \tau : S \wedge T \to A$, with 
$\neg (\varphi_1 \conflict_{S\wedge T} \varphi_2)$ iff $\varphi_1 \cup \varphi_2 \in \secbij{\sigma,
\tau}$ -- this easily boils down to the lemma below.

\begin{lem}\label{lem:bin_conf}
Assume $\sigma, \tau$ have binary conflict, and 
let $X$ be a finite subset of $\secbij{\sigma, \tau}$. Then the two following statements are
equivalent.
\begin{itemize}
\item[(1)] We have $\cup X \in \secbij{\sigma, \tau}$,
\item[(2)] For all $\varphi, \psi \in X$, $\neg(\varphi \conflict \psi)$.
\end{itemize}
\end{lem}
\begin{proof}
\emph{(1) $\Rightarrow$ (2).} Obvious, since $\varphi \cup \psi$ is a down-closed subset of $\cup X$.

\emph{(2) $\Rightarrow$ (1).} First we show that $\cup X$ is a bijection from a
configuration of $S$ to a configuration of $T$. Indeed if $(s,t_1), (s, t_2) \in
\cup X$ then $(s, t_1) \in \varphi$ and $(s, t_2) \in \psi$ for some $\varphi,
\psi \in X$. But then $\varphi \cup \psi \in \secbij{\sigma, \tau}$ by
hypothesis, so in particular it is a bijection, so $t_1 = t_2$. Hence $\cup X$ is
a bijection. Furthermore its projections are configurations. Indeed, 
since $X$ is pairwise compatible, $\pi_1 X$ is a pairwise compatible set of
configurations of $S$, hence $\cup (\pi_1 X) \in \conf{S}$ since $S$ has binary
conflict -- and likewise for $\cup (\pi_2 X)$.
It remains to prove that $\cup X$ is secured. But all $\varphi \in X$ are
down-closed partial orders induced as subsets of a common preorder of
synchronized events. Taking their union cannot introduce cycles, as these would
already occur in the separate components by down-closure. Hence $\cup X \in
\secbij{\sigma, \tau}$.
%
%
%
%
\end{proof}

It will also be useful later to have a concrete understanding of how minimal conflict arises in an
interaction; hence we prove the following lemma.

\begin{lem}\label{lem:pb_mconf}
Let $\sigma : S \to A$ and $\tau : T \to A^\perp$ be prestrategies, and $\varphi \in \secbij{\sigma,
\tau}$. Then if $\varphi : x \sbij y$ extends in $\secbij{\sigma, \tau}$ with $(s_1, t_1)$ and $(s_2, t_2)$ but
$\varphi \cup \{(s_1, t_1), (s_2, t_2)\} \not \in \secbij{\sigma, \tau}$. Then, either $x \cup
\{s_1, s_2\} \not \in \conf{S}$ or $y \cup \{t_1, t_2\} \not \in \conf{T}$.

In particular, if $S, T$ have binary conflict and $\varphi, \psi$ are events in $S\wedge T$
such that $\varphi\!\mconflict~\psi$, then $s_\varphi\!\mconflict~s_\psi$ or $t_\varphi\!\mconflict~t_\psi$.
\end{lem}
\begin{proof}
If $x \cup \{s_1, s_2\} \in \conf{S}$, $y \cup \{t_1, t_2\} \in \conf{T}$, and $s_1, s_2$ and $t_1,
t_2$ are distinct, then clearly $\varphi \cup \{(s_1, t_1),(s_2, t_2)\} \in \secbij{\sigma, \tau}$.
However if \emph{e.g.} $s_1 = s_2$, then $\tau\,t_1 = \tau\,t_2$, hence $y \cup \{t_1, t_2\} \not
\in \conf{T}$ by local injectivity -- and similarly if $t_1 = t_2$.

The second part of the statement follows easily.
%
%
\end{proof}

In fact, what we have described above is the \emph{pullback} construction in
$\ES$. There are maps of event structures:%
\nomenclature[aq]{$\Pi_i : S_1 \wedge S_2 \to S_i$}{Projections of an interaction
pullback}%
\[
\begin{array}{rcrclcrcrcl}
\Pi_1 &:& S\wedge T &\to& S &~~~~~~~~~& \Pi_2 &:& S\wedge T &\to& T\\
&&         \varphi     &\mapsto& s_\varphi&&&&\varphi &\mapsto& t_\varphi
\end{array}
\]
making the following square commute, and a pullback (from Lemma 2.11 of \cite{cg1}):
\[
\xymatrix@R=10pt@C=10pt{
&S\wedge T
	\ar[dl]_{\Pi_1}
	\ar[dr]^{\Pi_2}
	\pb{270}\\
S	\ar[dr]_\sigma&&
T	\ar[dl]^\tau\\
&A
}
\]

We motivated the pullback by asking for an es whose configurations are secured bijections. And indeed,
those are in a very close correspondence.

\begin{prop}\label{prop:rep_int}
For any $x\in \conf{S\wedge T}$, we have $\varphi_x = \cup x : \Pi_1 x \sbij \Pi_2 x$. Moreover, the assignment:
\[
\begin{array}{rcl}
\conf{S\wedge T} &\to& \secbij{\sigma, \tau}\\
x &\mapsto& \varphi_x
\end{array}
\]
is an order-isormorhism (with both sets ordered by inclusion). Finally, there is a family of order-isomorphisms:
\[
\begin{array}{rcrcl}
\nu_x &:& x &\iso& \varphi_x\\
&& \psi &\mapsto& (s_\psi, t_\psi)
\end{array}
\]
that is natural in $x$, \emph{i.e.} for $x\subseteq y$ and $\varphi \in x$ we
have $\nu_x \varphi  = \nu_y \varphi$.
\end{prop}
\begin{proof}
Direct extension of Lemma 2.9 in \cite{cg1}.
\end{proof}

This allows us, when reasoning on configurations of a pullback, to manipulate directly secured
bijections rather than compatible
sets of prime secured bijections. Likewise, when reasoning on events of the pullback in an ambiant configuration, we can directly
apply $\nu$ and reason on synchronised pairs. In the proofs, we will often use this proposition implicitely and transfer
silently between the different representations.

Finally, we are in position to define the parallel interaction of two prestrategies $\sigma : S \to A^\perp \parallel B$ and
$\tau : T \to B^\perp \parallel C$. We simply form the pullback:
\[
\xymatrix@R=15pt@C=0pt{
&(\sigma \parallel C)\wedge (A\parallel \tau)
	\ar[dl]_{\Pi_1}
	\ar[dr]^{\Pi_2}
	\pb{270}\\
S\parallel C
	\ar[dr]_{\sigma \parallel C}&&
A\parallel T
	\ar[dl]^{A\parallel \tau}\\
&A\parallel B \parallel C
}
\]

We write $T\circledast S = (\sigma \parallel C)\wedge (A\parallel \tau)$ for the
interaction, and $\tau \circledast \sigma : T\circledast S \to A\parallel B
\parallel C$%
\nomenclature[ar]{$\tau \circledast \sigma : T \circledast S \to A\parallel B \parallel
C$}{Interaction pullback of $\sigma : S \to A^\perp \parallel B$ and $\tau : T \to B^\perp \parallel
C$}%
for either side of the pullback square. Hence we get the interaction of $\sigma$ and $\tau$, $\tau \circledast \sigma$, as a labeled event structure.

\begin{exa}\label{ex:interaction}
Consider the following prestrategies $\sigma : S \to \intr{\com \to \com}$ and $\tau : T \to \intr{\com\to
\com}^\perp \parallel \intr{\com}$ (note that to match the definition of interaction above, we can
consider $\sigma : S \to \emptyar^\perp \parallel \intr{\com \to \com}$, where
$\emptyar$ is the empty arena).%
\nomenclature[as]{$\emptyar$}{Empty game/arena}%
\[
\xymatrix@C=0pt@R=5pt{
\sigma:& \llbracket\com    \ar@{}[rr]|\to&&
\com\rrbracket&&~~~~~~~~~~~~~~~&\tau:&
\llbracket\com   \ar@{}[rr]|\to&&
\com\rrbracket^\perp  \ar@{}[rr]|\parallel&&
{\com}\\
&&&{\run}^{-}
        \ar@{-|>}[dll]&&&&&&&&
{\run}^{-}
        \ar@{-|>}[dll]\\
&{\run}^{+}
	\ar@{--}@/^/[urr]
        \ar@{-|>}[d]&&&&&&&&
{\run}^{+}
        \ar@{--}@/^/[urr]
        \ar@{-|>}[dll]
        \ar@{-|>}[d]\\
&{\done}^{-}
        \ar@{--}@/^/[u]
        \ar@{-|>}[drr]&&&&&&
{\run}^{-}
        \ar@{--}@/^/[urr]
        \ar@{-|>}[dl]
        \ar@{-|>}[dr]&&
{\done}^{-}
        \ar@{--}@/_/[u]
        \ar@{-|>}[drr]\\
&&&{\done}^{+}
        \ar@{--}@/_/[uuu]&&&
{\done}^{+}\!\!
        \ar@{--}@/^/[ur]
        \ar@{~}[rr]&&
{\done}^{+}\!\!\!\!\!\!\!\!\!\!
        \ar@{--}@/_/[ul]&&&
{\done}^{+}
        \ar@{--}@/_/[uuu]
}
\]

We display below a representation of the interaction:
\[
\tau \circledast \sigma : T \circledast S \to (\com \to \com) \parallel \com
\]

We only display polarities for the moves in the right hand side $\com$. Indeed events on the left hand side
(synchronised) part of the interaction have no well-defined polarity, as the two strategies disagree
on them.

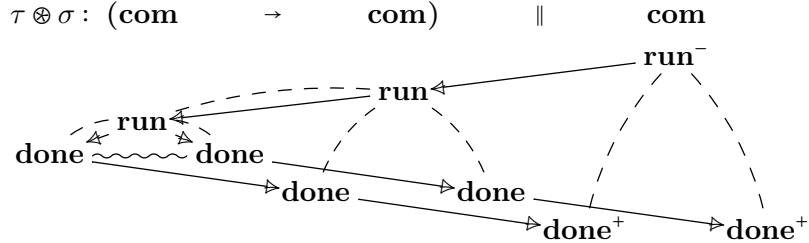
\[
\xymatrix@R=0pt@C=0pt{
\tau \circledast \sigma :&
(\com   \ar@{}[rrr]|\to&&&
\com)   \ar@{}[rrr]|\parallel&&&
\com\\
&&&&&&&\run^{-}
        \ar@{-|>}[dlll]\\
&&&&{\run}
        \ar@{-|>}[dlll]\\
&{\run}
        \ar@{--}@/^/[urrr]
        \ar@{-|>}[dl]
        \ar@{-|>}[dr]\\
{\done}
        \ar@{--}@/^/[ur]
        \ar@{~}[rr]
        \ar@{-|>}[drrr]&&
{\done}
        \ar@{--}@/_/[ul]
        \ar@{-|>}[drrr]\\
&&&{\done}
        \ar@{--}@/^/[uuur]
        \ar@{-|>}[drrr]&&
{\done}
        \ar@{--}@/_/[uuul]
        \ar@{-|>}[drrr]\\
&&&&&&
{\done}^{+}
        \ar@{--}@/^/[uuuuur]&&
{\done}^{+}
        \ar@{--}@/_/[uuuuul]
}
\]

We leave it to the reader to check that each event in this diagram corresponds uniquely to a
configuration in $S \parallel \com$ and a matching configuration in $T$ such that the induced bijection
is secured.
\end{exa}

\subsubsection{Hiding.} \label{subsubsec:hiding}
Once we have performed the interaction, it is fairly simple to obtain the composition by ignoring the \emph{synchronised events}, \emph{i.e.} 
those that map to $B$. This is an instance of the following \emph{projection} operation.

\begin{defi}\label{def:proj}%
\nomenclature[at]{$E\proj V$}{Projection of an event structure (hiding)}%
Let $E$ be an event structure, and $V \subseteq E$ a set of events of $E$. The \textbf{projection of
$E$ to $V$}\index{Projection of an event structure}, written $E\proj V$,
has components:
\begin{itemize}
\item \emph{Events.} $V$.
\item \emph{Causality.} The order $\leq_E$ restricted to $V$.
\item \emph{Consistency.} The sets $X \in \Con_E$ such that $X \subseteq V$.
\end{itemize}
\end{defi}

This gives an event structure -- it is clear that a hiding of an event structure with binary
conflict still has binary conflict. Note as well the \emph{unique witness} property reminiscent of
that used in studying the composition of deterministic strategies in standard game semantics: for any $x\in \conf{E\proj V}$, there exists 
a \emph{unique} $y \in \conf{E}$
such that $y \cap V = x$, and whose maximal
events are those of $x$; obtained as $y = 
[x]_E = \{e \in E \mid \exists e'\in x,~e\leq_E e'\} \in \conf{E}$.

Finally, we define composition. From $\sigma : S \to A^\perp \parallel B$ and $\tau : T \to B^\perp \parallel C$, first 
compute the interaction $\tau \circledast \sigma : T \circledast S \to A\parallel B \parallel C$. Then, set $V \subseteq T\circledast S$
to comprise all $\varphi \in T\circledast S$ such that $(\tau \circledast
\sigma)\,\varphi \not\in B$. Writing $T\odot S = T\circledast S \proj V$,%
\nomenclature[ata]{$\tau \odot \sigma : T\odot S \to A^\perp \parallel C$}{Composition of $\sigma :
S \to A^\perp \parallel B$ and $\tau : T \to B^\perp \parallel C$}%
the composition of $\sigma$ and $\tau$ is:
\[
\begin{array}{rcrcl}
\tau \odot \sigma &:& T\odot S &\to& A^\perp \parallel C\\
&& \varphi &\mapsto& (\tau \circledast \sigma)\,\varphi
\end{array}
\]

From the fact that interaction and hiding preserve binary conflict, it follows that for $\sigma : S
\to A^\perp \parallel B$ and $\tau : T \to B^\perp \parallel C$, if $S, T, A, C$ have binary
conflict, then so does $T\odot S$.

\begin{exa}
Consider the interaction of Example \ref{ex:interaction}. After hiding, the resulting composition
is:
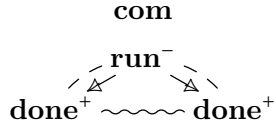
\[
\xymatrix@R=5pt@C=0pt{
&\com\\
&\run^-
	\ar@{-|>}[dl]
	\ar@{-|>}[dr]\\
\done^+	\ar@{--}@/^/[ur]
	\ar@{~}[rr]&&
\done^+	\ar@{--}@/_/[ul]
}
\]
Note that the conflict between the two maximal events, although it was inherited in the interaction,
becomes minimal after projection as its original cause has been hidden away.
\end{exa}

Composition is associative up to isomorphism \cite{cg1}. However, copycat is not neutral for composition with respect to prestrategies -- it is 
only the case for \emph{strategies} (see \cite{cg1} for details):

\begin{defi}\label{def:strat}
A prestrategy $\sigma : S \to A$ on a game $A$ is a \textbf{strategy}\index{Strategy} if it is:
\begin{itemize}
\item \emph{Receptive.}\index{Receptive} For all $x\in \conf{S}$, if $\sigma x\longcov{a^-}$, then there exists a unique $s\in S$ such that 
$x\longcov{s}$ and $\sigma\,s = a$.
\item \emph{Courteous.}\index{Courteous} If $s_1 \imc_S s_2$ and $\pol(s_1) = +$ or $\pol(s_2) = -$, then $\sigma\, s_1 \imc_A \sigma\,s_2$.
\end{itemize}
\end{defi}

Putting everything together, we get \cite{cg1}:

\begin{thm}
There is a compact closed category $\CG$ of games and strategies up to isomorphism.
\end{thm}

\subsection{Interpreting programs and replication}
\label{subsec:replication}

The category $\CG$ is a general framework for composing concurrent strategies. We can rely on it to build a model of an affine 
variant of IPA: that involves restricting to negative arenas, and interpreting function space in IPA via the usual arrow arena construction.
We refrain from giving the details here, since we will give them in the non-affine case later on.
However, before going on to handling replication, we will give examples in the affine case and try
to convey further intuition as to what the model computes. 
Then we will present the expanded arenas used to handle replication, and we will introduce the issue of uniformity.

\subsubsection{Concurrent strategies and view functions.} As the reader familiar with Hyland-Ong games may have noticed, our examples before showed
how to represent as concurrent strategies, \emph{view functions} rather than expanded strategies -- or, in Curien's terminology \cite{curien},
\emph{meager} rather than \emph{fat} innocent strategies. And indeed, in our framework it is the case that a pure program
will be interpreted directly as its view function, never constructing the full set of plays. For illustration,
the interpretation of the affine pure program:
\[
\lambda f^{\tbool \to \com}. f~\ttrue : (\tbool \to \com) \to \com
\]
will be the strategy:
\[
\xymatrix@R=5pt@C=10pt{
(\tbool	\ar@{}[r]|\to&
\com)	\ar@{}[r]|\to&
\com\\
&&\run^-	\ar@{-|>}[dl]\\
&\run^+	\ar@{--}@/^/[ur]
	\ar@{-|>}[dl]
	\ar@{-|>}[d]\\
\q^-	\ar@{--}@/^/[ur]
	\ar@{-|>}[d]&
\done^-	\ar@{-|>}[dr]
	\ar@{--}@/^/[u]\\
\ttrue^+\ar@{--}@/^/[u]
&&\done^+	\ar@{--}@/_/[uuu]
}
\]
which the reader can match against the tree of $P$-views for the corresponding Hyland-Ong innocent strategy. The composition of
such strategies is computed directly using pullbacks in $\ES$, never constructing the expanded plays. In other words, we never
work with full Hyland-Ong strategies, but always with their causal representations: the view functions.

But the usual strategies for stateful programs \cite{am} are not generalizations of \emph{meager} innocent strategies, but
of \emph{fat} ones: the behaviour of programs must be observed not only on P-views but
on general plays. Hence, the reader may wonder if evading them causes us to lose that ability. Fortunately it is not the case, 
and strategies for stateful programs can be represented causally just as innocent strategies. For instance, consider the
following example.

\begin{exa}\label{ex:state1}
Consider the following term of IA.
\[
\newref~b~\tin~\lambda f^{\com\to \com}.~f~(b:=\ttrue);~!b : (\com \to \com) \to \tbool
\]
Following (the affine variant of) the interpretation of Section
\ref{sec:interpretation}, it yields the strategy:
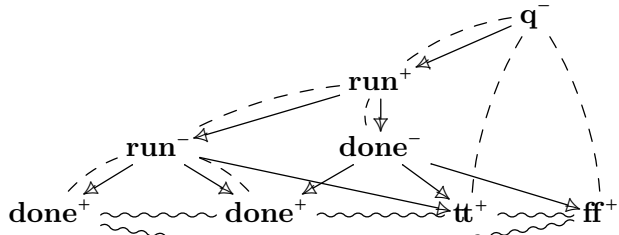
\[
\xymatrix@R=10pt@C=5pt{
&(\com \ar@{}[rr]|\to&&
\com)   \ar@{}[rr]|\to&&
\tbool\\
&&&&&\q^-	\ar@{-|>}[dll]\\
&&&\run^+	\ar@{-|>}[d]
		\ar@{--}@/^/[urr]
		\ar@{-|>}[dll]\\
&\run^-		\ar@{-|>}[dl]
		\ar@{-|>}[dr]
		\ar@{-|>}[drrr]
		\ar@{--}@/^/[urr]
&&\done^-	\ar@{-|>}[dr]
		\ar@{-|>}[drrr]
		\ar@{--}@/^/[u]
		\ar@{-|>}[dl]\\
\done^+		\ar@{~}[rr]
		\ar@{--}@/^/[ur]
		\ar@{~}@/_1pc/[rrrrrr]
&&\done^+	\ar@{~}[rr]
		\ar@{--}@/_/[ul]
&&\ttrue^+	\ar@{~}[rr]
		\ar@{--}@/^/[uuur]&&
\tfalse^+	\ar@{--}@/_/[uuul]
}
\]
\end{exa}

The $\done^+$ to the left is duplicated, witnessing the two outcomes of the race in the memory that happens
if the argument does not respect the evaluation stack, and concurrently returns $\done^-$ and asks for its argument.
The reader familiar with the game semantics for Idealized Algol \cite{am} can check that taking the set of
(well-bracketed) alternating linear orderings of configurations of this event structure yields the expected set of plays.

\subsubsection{Replication.} \label{subsubsec:replication}%
But so far, we have only seen \emph{affine} programs and strategies, \emph{i.e.} that call
each resource at most once. As it stands, the local injectivity condition in Definition \ref{def:map_es} forbids us 
from having two compatible events corresponding to the same move in the game. It is natural to consider dropping it, 
but then we lose access to the nice structural properties of $\ES$ (such as
pullbacks). It is unclear to us how one would go about defining composition
of strategies in such a setting, let alone proving that it forms a category; in particular if one
insists on remembering the point of non-deterministic branching.

\red{Instead, the solution behind our category $\CHO$}
takes inspiration from AJM games \cite{ajm} and from the reconstruction of HO
games in \cite{hhm} : we explicitly duplicate moves in arenas. Rather
than playing directly on an arena $A$, our strategies play in $\oc A$, a variant
of $A$ with events duplicated countably many times, in depth. More formally:

\begin{defi}%
\nomenclature[au]{$\lbl\,\alpha$}{Label of an indexing function $\alpha$}%
\nomenclature[av]{$\ind\,\alpha$}{Copy index of an indexing function $\alpha$}%
Let $A$ be an arena, and $a\in A$. An \textbf{indexing function}\index{Indexing function} for $a$ is a function:
\[
\alpha : [a] \to \omega
\]
which associates, to $a$ and its dependencies, a \textbf{copy index}\index{Copy index}.
From $\alpha : [a] \to \omega$, we write $\lbl\,\alpha = a$ for its \textbf{label}\index{Label}, and $\ind\,\alpha = \alpha\,(\lbl\,\alpha)$ for
the copy index of $a$.
\end{defi}

Indexing functions will be the \emph{events} of $\oc A$. Its full structure will be:

\begin{defi}\label{def:oc}%
\nomenclature[aw]{$\oc A$}{Expanded arena}%
From an arena $A$, we build a new arena $\oc A$, comprising:
\begin{itemize}
\item \emph{Events.} indexing functions $\alpha : [a] \to \omega$,
\item \emph{Causal order.} for $\alpha : [a] \to \omega$ and $\beta : [b] \to \omega$, we have
$\alpha \leq_{\oc A} \beta$ iff $a \leq_A b$ and for all $a'\leq_A a,~\alpha(a') = \beta(a')$.
\item \emph{Polarity.} For $\alpha \in \oc A$, $\pol_{\oc A}(\alpha) = \pol_A(\lbl\,\alpha)$.
\end{itemize}
\end{defi}

Moves in $\oc A$ have a rather complex structure. However, note that just as $A$ -- which was
required to be an arena rather than a general game --  $\oc A$ is
a forest. For each $\alpha : [a] \to \omega$, either $a$ is minimal and then so
is $\alpha$, or there is a unique $a' \imc_A a$ -- in which case the restriction of 
$\alpha$ to $[a']$ gives a unique $\alpha'$ such that $\alpha' \imc_{\oc A} \alpha$. In other
words, $\alpha$ is entirely determined by the data of $\lbl\,\alpha = a$, $\ind\,\alpha = \alpha(a)$, and its
immediate predecessor $\alpha'$, called its \textbf{justifier}\index{Justifier}
$\just(\alpha)$. Using this%
\nomenclature[ax]{$\just(\alpha)$}{Justifier (unique immediate dependency) of
$\alpha$ in $\oc A$}%
decomposition inductively, we can unambiguously draw configurations of $\oc A$ by annotating
each event by its copy index, and its justifier.

\begin{exa}\label{ex:hooc}
The following is a representation of a configuration of $\oc \tbool$:
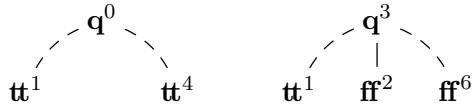
\[
\xymatrix@C=10pt@R=10pt{
&\q^0 &&&&\q^3\\
\ttrue^1	\ar@{--}@/^/[ur]&&
\ttrue^4	\ar@{--}@/_/[ul]&&
\ttrue^1	\ar@{--}@/^/[ur]&
\tfalse^2	\ar@{--}[u]&
\tfalse^6	\ar@{--}@/_/[ul]
}
\]
where, \emph{e.g.} the events labeled $\ttrue^1$ respectively denote
$\{\q\mapsto 0, \ttrue \mapsto 1\}$ and $\{\q\mapsto 3, \ttrue \mapsto 1\}$.
\end{exa}

Using the additional space granted by $\oc A$, we can now represent programs 
evaluating their arguments multiple times. For instance, a valid strategy
$\sigma : S \to \oc \intr{(\tbool \to \tbool) \to \tbool \to \tbool}$
for the term $\lambda f^{\tbool \to \tbool}x^\tbool.\,f\,(f\,x)$ could contain, for 
$i, j\in \omega$ and injective function $\tuple{-, -} : \omega^2 \to \omega$,%
\nomenclature[axa]{$\tuple{\__1, \dots, \__n}$}{Any injective function $\omega^n
\to \omega$}%
a configuration:

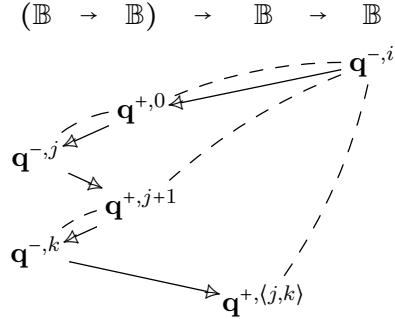
\[
\xymatrix@R=0pt@C=10pt{
(\tbool	\ar@{}[r]|\to&
\tbool)	\ar@{}[r]|\to&
\tbool	\ar@{}[r]|\to&
\tbool\\
&&&\q^{-,i}	\ar@{-|>}[dll]\\
&\q^{+,0}	\ar@{--}@/^/[urr]
		\ar@{-|>}[dl]\\
\q^{-,j}	\ar@{--}@/^/[ur]
		\ar@{-|>}[dr]\\
&\q^{+,j+1}	\ar@{--}@/^/[uuurr]
		\ar@{-|>}[dl]\\
\q^{-,k}	\ar@{--}@/^/[ur]
		\ar@{-|>}[drr]\\
&&\q^{+,\tuple{j,k}}
		\ar@{--}@/_/[uuuuur]
}
\]

This diagram exploits the representation introduced just above for configurations of $\oc A$: each
event is specified through its label and copy index. The full strategy $\sigma$ would comprise
such configurations for all $i, j\in \omega$. For each positive event, $\sigma$ must provide
a copy index -- this choice must be made globally, in a way that avoids collisions to maintain
local injectivity of $\sigma$.

\subsubsection{\change{Strategies up to copy indices}}\label{subsubsec:weakiso}
Using the composition mechanism introduced before, one may define an interpretation
of terms of IPA as concurrent strategies on expanded arenas. However, as observed above, such strategies
not only carry information about the events they play and their causal history, but also the data of
specific \emph{copy indices} that seem largely irrelevant -- \emph{e.g.}, as above, the choice of
an injection $\tuple{-, -}$. In fact, for reasons familiar from AJM games \cite{ajm}, strategies will
not satisfy the laws of cartesian closed categories unless we consider them 
\emph{up to their specific choice of copy indices}. Let us observe that on an example.

\begin{exa}\label{ex:wiso}
Consider the term $M$ to be $f : \tbool \to \tbool \to \tbool \vdash \lambda x^\tbool.\,f\,x\,x : \tbool \to \tbool$.
Its interpretation could contain:
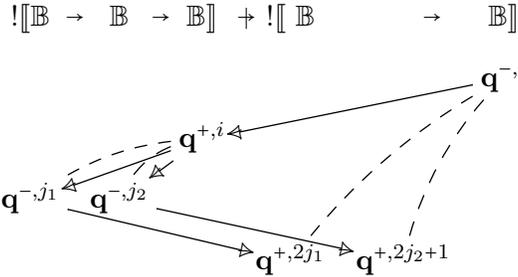
\[
\xymatrix@C=5pt@R=5pt{
\oc \llbracket\tbool
	\ar@{}[r]|\to&
\tbool	\ar@{}[r]|\to&
\tbool\rrbracket
	\ar@{}[r]|\to||&
\oc\llbracket ~~~ \tbool&
	\ar@{}[r]|{\!\!\!\!\!\!\!\!\!\!\to~~}&
\tbool\rrbracket\\
&&&&&\q^{-,i}
	\ar@{-|>}[llld]\\
&&\q^{+,i}
	\ar@{-|>}[dll]
	\ar@{-|>}[dl]\\
\q^{-,j_1}
	\ar@{--}@/^/[urr]
	\ar@{-|>}[drrr]&
\q^{-, j_2}
	\ar@{--}@/^/[ur]
	\ar@{-|>}[drrr]\\
&&&\q^{+,2j_1}
	\ar@{--}@/^/[uuurr]&
\q^{+, 2j_2+1}
	\ar@{--}@/^/[uuur]
}
\]
Because of the contraction, the Opponent events of indices $j_1$ and $j_2$ corresponding to different events in the 
arena trigger Player events corresponding to the same event in the arena. Local
injectivity is ensured by the functions $2n$ and $2n+1$ having disjoint codomain.

Likewise, consider two terms, with chosen configurations of their strategies:
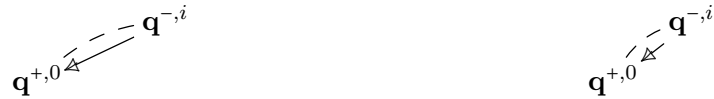
\[
\xymatrix@R=5pt@C=5pt{
\intr{\lambda x^\tbool y^\tbool.\,x} &:& \oc \llbracket\tbool
	\ar@{}[r]|\to&
\tbool	\ar@{}[r]|\to&
\tbool\rrbracket\\
&&&&\q^{-,i}
	\ar@{-|>}[dll]\\
&&\q^{+, 0}
	\ar@{--}@/^/[urr]
}
\qquad\qquad
\xymatrix@R=5pt@C=5pt{
\intr{\lambda x^\tbool y^\tbool.\,y} &:& \oc \llbracket\tbool
	\ar@{}[r]|\to&
\tbool	\ar@{}[r]|\to&
\tbool\rrbracket\\
&&&&\q^{-,i}
	\ar@{-|>}[dl]\\
&&&\q^{+, 0}
	\ar@{--}@/^/[ur]
}
\]

Then we have:
\[
\xymatrix@R=5pt@C=5pt{
\intr{M}\odot\intr{\lambda x^\tbool y^\tbool.\,x} &:& \oc \llbracket\tbool
        \ar@{}[r]|\to&
\tbool\rrbracket\\
&&&\q^{-,i}
        \ar@{-|>}[dl]\\
&&\q^{+, 0}
        \ar@{--}@/^/[ur]
}
\qquad\qquad
\xymatrix@R=5pt@C=5pt{
\intr{M}\odot \intr{\lambda x^\tbool y^\tbool.\,y} &:& \oc \llbracket\tbool
        \ar@{}[r]|\to&
\tbool\rrbracket\\
&&&\q^{-,i}
        \ar@{-|>}[dl]\\
&&\q^{+, 1}
        \ar@{--}@/^/[ur]
}
\]

These are required to be the same by the laws of cartesian closed categories,
since $(\lambda f.\,M)\,(\lambda xy.\,x) =_\beta (\lambda f.\,M)\,(\lambda xy.\,y)$.
However, they are not isomorphic as strategies.
\end{exa}

In order to solve this mismatch, we first need to formalize what it means for two configurations of $\oc A$ to 
be the same up to the choice of copy indices.

\begin{defi}\label{def:index_isos}
Let $x, y \in \conf{\oc A}$. A \textbf{reindexing iso}\index{Reindexing
isomorphism} between $x$ and $y$ is an order-isomorphism:%
\nomenclature[ay]{$x\iso y$}{Order-isomorphism between (possibly implicitly)
ordered sets}%
\[
\theta : x \iso y
\]
which \emph{preserves labels}: for all $\alpha \in x$, $\lbl\,\alpha = \lbl\,(\theta\,\alpha)$.

A reindexing iso $\theta : x \iso y$ is \textbf{positive}\index{Reindexing isomorphism!positive} iff it preserves the copy index of negative events, \emph{i.e.}
for all $\alpha^- \in x$, $\ind\,\alpha = \ind\,(\theta\,\alpha)$.
\textbf{Negative} \index{Reindexing isomorphism!negative}reindexing isos
are defined dually.
\end{defi}

Intuitively, two configurations of $\conf{\oc A}$ related by a reindexing iso are distinct representations of 
one \emph{thick subtree} of $A$ in the sense of Boudes \cite{boudes}, \emph{i.e.} a subtree of the arena with duplicated
sub-arenas. Two strategies are to be identified iff they are isomorphic, with the commuting triangle to $\oc A$ being weakened
to a commutation \emph{up to reindexing iso} -- in fact, it turns out to be simpler to strengthen that to \emph{positive reindexing isos}.
Altogether:

\begin{defi}\label{def:weak_mor}%
\nomenclature[az]{$\sigma \wiso \tau$}{Weak isomorphism}%
Let $\sigma : S \to \oc A$, $\tau : T \to \oc A$ be two strategies. A \textbf{weak morphism from
$\sigma$ to $\tau$} \index{Weak morphism!between strategies on $\oc A$} is 
$f : S \to T$, such that the triangle 
\[
\xymatrix@C=5pt@R=5pt{
S	\ar@/^/[rr]^f
	\ar[dr]_{\sigma}&&
T	\ar[dl]^{\tau}\\
&\oc A
}
\]
commutes \emph{up to positive symmetry}, in the sense that for all $x \in \conf{S}$, the set:
\[
\{(\sigma\,s, \tau\,(f\,s))\mid s \in x\}
\]
is a positive reindexing iso. If $f : S \to T$, $g: T \to S$ are two weak morphisms such that
$g\circ f = \id_S$ and $f \circ g = \id_T$, we say that $(f, g)$ is a \textbf{weak
isomorphism}\index{Weak isomorphism!between strategies on $\oc A$}, and
write $\sigma \wiso \tau$ to mean that $\sigma$ and $\tau$ are weakly isomorphic.
\end{defi}

\change{\subsubsection{Uniformity}} The two strategies of Example \ref{ex:wiso} are weakly isomorphic. And in fact, a consequence of
the developments of this paper is that the natural interpretation of terms $\vdash M : A$ of PCF
as strategies on $\oc A$ hinted at here \emph{is} sound and computationally adequate
(it is reasonable to expect the same statement for IPA to be true as well, but it does not follow
from the results in this paper). However,
proving it bumps into a significant difficulty: without further contraints on strategies, 
weak isomorphism is \emph{not} a congruence. Indeed, strategies can behave differently depending
on Opponent's choice of copy indices. For instance, composing the two weakly isomorphic 
$\intr{M}\odot \intr{\lambda x^\tbool y^\tbool.\,x}$ and $\intr{M}\odot \intr{\lambda x^\tbool y^\tbool.\,y}$ 
of Example \ref{ex:wiso} with
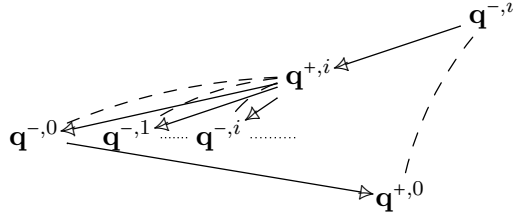
\[
\xymatrix@R=5pt@C=10pt{
\llbracket&\tbool	\ar@{}[rr]|\to&&
\tbool\rrbracket	\ar@{}[r]|\to||&
\llbracket\tbool	\ar@{}[r]|\to&
\tbool\rrbracket\\
&&&&&\q^{-, i}
	\ar@{-|>}[dll]\\
&&&\q^{+, i}
	\ar@{-|>}[dlll]
	\ar@{-|>}[dll]
	\ar@{-|>}[dl]\\
\q^{-, 0}
	\ar@{-|>}[drrrr]
	\ar@{--}@/^/[urrr]&
\q^{-,1}\ar@{--}@/^/[urr]
	\ar@{.}[r]&
\q^{-,i}\ar@{--}@/^/[ur]
	\ar@{.}[r]&~\\
&&&&\q^{+,0}
	\ar@{--}@/^/[uuur]
}
\]
yields, in the one hand, a strategy that calls its argument, and on the other, one that does not. Clearly, they are
not weakly isomorphic. This is because the strategy above is not \emph{uniform}: its behaviour not
only depends on
Opponent's moves, but also on their copy index. A useful analogy is that of a program that looks up the address where it is
loaded in memory, and uses that information to specify its behaviour.	

In AJM games \cite{ajm}, uniformity is ensured by equipping games with
an equivalence relation on plays not unlike our reindexing
isomorphisms, and then requiring strategies to satisfy closure
properties with respect to it. Our approach to uniformity bears some
resemblance with that, but the richer structure of our plays prevents us from
following 
the AJM recipe  directly. 
\red{To conclude this section and transition to the
next, we give a few ideas motivating our approach to uniformity as detailed in
the next section.

In traditional game semantics, plays are sequences of moves, \emph{i.e.} total
orders. Hence, the mere fact that two plays $s_1$ and $s_2$ (necessarily of the
same length) are in relation in AJM games informs a one-to-one correspondence
between the moves appearing in $s_1$ and those appearing in $s_2$. In contrast,
in our present setting the states (in the game) are partially ordered \emph{configurations}
$x\in \conf{\oc A}$, so an equivalence $x \sim_{\oc A} y$ between $x,
y\in \conf{\oc A}$ no longer informs such a one-to-one correspondence. This
correspondence will then have to become primitive: the equivalence between
two configurations must be \emph{witnessed} by a specific isomorphism,
such as a reindexing iso. We like indeed to
think of the set of reindexing isomorphisms as forming a ``proof-relevant''
equivalence relation between configurations of $\oc A$ -- it expresses not only
which configurations are in relation, but also \emph{how}.

How, then, to express that a strategy $\sigma : S \to \oc A$ is uniform? It is
tempting, for $x, y \in \conf{S}$, to simply set $x \sim_S y$ when $\sigma x
\sim_{\oc A} \sigma y$ -- the witness isomorphism $\sigma x \iso \sigma y$ then
informing an isomorphism $x \iso y$. This yields a set of isomorphisms between
configurations of $S$, which then may be asked to satisfy closure properties
ensuring uniformity. However, it turns out that this is too much to ask: such a
simple definition rejects perfectly uniform strategies, some of which are
definable through IPA. Intuitively (and unlike for the deterministic sequential
strategies of AJM games), a term/strategy may \emph{play symmetric
moves for non-symmetric reasons}. So we opt instead for a more intensional option:
we endow strategies with their own ``proof-relevant equivalence relation''
expressing which configurations they \emph{considers} equivalent, and how. This
``uniformity witness'' is part of strategies with symmetry, and propagated
through composition and other constructions on strategies. It cannot be
recovered uniquely (except for \emph{innocent} strategies \cite{lics15}, though
this is out of scope for this paper) -- Appendix \ref{app:unif_wit} contains a more
detailed discussion along the lines of this paragraph, with examples.

We now go on to the formalization of these ideas.}

\section{Thin concurrent games}
\label{sec:gameswithsymmetry}
Dealing with \emph{uniformity} requires us to
replicate the construction of concurrent games in a more expressive setting,
capable to express that certain events or configurations might be
\emph{symmetric}, \emph{i.e.} interchangeable; making their indistinguishability
part of the structure. 
\emph{Event structures with symmetry} \cite{symmetry} were designed precisely to
cope with such situations. In this section, we construct a replacement for $\CG$
based on those, enforcing uniformity of strategies over symmetric events, and
hence supporting uniform replication.

In the previous section, a game was an event structure structure $A$, while a
strategy was an event structure $S$ labeled by $A$, \emph{i.e.} a map $\sigma : S
\to A$ in the category $\E$ of event structures and maps between them --
furthermore, composition of strategies was obtained by leveraging universal
constructions in $\E$, \emph{e.g.} pullback for interaction. In order to build
games with symmetry, it is mathematically appealing to replicate the
definitions and constructions, but this time based on the category $\ESS$
of event structures with symmetry and maps between them (to be defined in
Definition \ref{def:ess}). It is also as economical as we
can do with the generality we wish to give the model (and in particular, so that
it supports IPA) -- we have explored a number of simpler alternatives, which fail
in various ways. 

In Section \ref{subsec:sym_sat}, we first recall event structures with symmetry,
and expand on the above methodology for enriching concurrent games with symmetry.
Though mathematically appealing, pushing these guidelines bumps against
significant obstacles, allowing for two main solutions: the \emph{saturated} (or
\emph{fat}) and \emph{thin} approaches (respectively appearing in conference
papers \cite{lics14} and \cite{lics15}). After briefly reviewing the fat case,
the rest of the section commits to the thin. In Section \ref{subsec:thin}, we
develop thin concurrent games, focusing on the problem of uniformity, which
imposes the most constraints on their design. Finally in Section
\ref{subsec:compact}, we show that thin games with symmetry form a compact closed
category. Then, before going further along the main narrative of the paper and
constructing the cartesian closed category of Concurrent Hyland-Ong games in
Section \ref{sec:cho}, we show that $\TCG$ also supports the construction of an
AJM-style exponential modality.

\subsection{Symmetry on event structures and games} 
\label{subsec:sym_sat} 
In this first part, we review the main technical tool -- event
structures with symmetry -- and introduce the main challenges in
constructing games based on those. 

\subsubsection{Event structures with symmetry} \label{subsubsec:essp}
Intuitively, symmetry on event structures should behave as an equivalence
relation, but also satisfy bisimulation-like properties in order to ensure that
symmetric states have the same futures (up to symmetry). Looking for a notion of
symmetry on event structures satisfying these two aspects, a natural methodology
is to instantiate known categorical constructions.

In their seminar paper \cite{DBLP:conf/lics/JoyalNW93}, Joyal,
Nielsen and Winskel gave a categorical notion of bisimulation between
objects $E$ and $F$ as a \emph{span}
\[
\xymatrix@R=10pt@C=10pt{
&B
	\ar[dl]_{l}
	\ar[dr]^{r}\\
E&&F
}
\]
where $l$ and $r$ are \emph{open maps}, \emph{i.e.} satisfy a path lifting
condition formulated as a factorisation property. Winskel later
defined an \emph{event structure with symmetry} in \cite{symmetry} as an event
structure $E$ with a span of open maps as above (with $E = F$), additionally satisfying
categorical formulations of the laws of equivalence relations -- we omit
details, opting below for a more concrete equivalent definition. 
Besides their algebraic genesis,
event structures with symmetry have already proved adequate as a modeling
framework: most notably, Hayman and Winskel proved in
\cite{DBLP:conf/fsttcs/HaymanW08} that the universal characterization of the
unfolding of \emph{safe} Petri nets as event structures could be extended to
\emph{general} Petri nets, provided one unfolds to \emph{event structures with
symmetry} -- the multiple tokens in one place yielding distinct yet
\emph{symmetric} copies of enabled transitions.

Symmetry on event structures can be also defined via \emph{isomorphism families}
\cite{symmetry}.

\begin{defi}[Isomorphism families and event structures with symmetry]%
\nomenclature[ba]{$\A$}{Event structure with symmetry $\A = (A, \tilde{A})$}%
\nomenclature[bb]{$\tilde{A}$}{Isomorphism family of an ess $\A$}%
\nomenclature[bc]{$\theta \restrict x$}{Restriction of $\theta : x'\isf y' \in
\tilde{A}$ to
subconfiguration $x \subseteq x'$}%
  Let $A$ be an event structure and $\tilde A$ be a set of bijections
  between configurations of $A$. Then, $\tilde A$ is an
  \textbf{isomorphism family}\index{Isomorphism family} on $A$ if it satisfies:

  \begin{itemize}
  \item \emph{(Groupoid)} The set $\tilde A$ contains all identity
    bijections, and is stable under composition and inverse of
    bijections.
  \item \emph{(Restriction)} For every bijection
    $ \theta : x   \bij y  \in  \tilde A$ and
    $x' \in \mathscr{C} (A)$ such that $x' \subseteq x$, then the
    restriction $\theta \restrict x'$ of $ \theta $ to $x'$ is in $\tilde A$. In
    particular, $\theta\,x' \in \conf{A}$.
  \item \emph{(Extension)} For every
    $ \theta : x  \bij  y  \in  \tilde A$ and extension
    $x \subseteq x' \in \mathscr{C} (A)$, there exists a
    (non-necessarily unique) $y \subseteq y' \in \mathscr{C} (A)$ and
    an extension $ \theta \subseteq \theta '$ such that
    $ \theta ': x'  \bij  y'  \in  \tilde A$.
  \end{itemize}

  In this case the pair $\A = (A, \tilde A)$ is called an \textbf{event
    structure with symmetry (ess)}\index{Event structure!with symmetry (ess)}. We
will use $\S, \T, \A, \B,  \ldots $ to range
  over event structures with symmetry. If $A$ additionally has polarities and
bijections in
$\tilde{A}$ preserve them, we say that $\A$ is an \textbf{event structure with
symmetry and
polarities (essp)}\index{Event structure!with symmetry and polarities (essp)}.
%
\end{defi}

One natural way to think of an isomorphism family is as a
\emph{proof-relevant history-preserving bisimulation and equivalence relation} between
configurations: the equivalence between configurations is witnessed by
precise bijections between their events. 

The following represents a rather trivial example of an event
structure with symmetry:
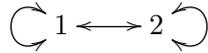
\[
\xymatrix{
1	\ar@(dl,ul)@{<->}[]
	\ar@{<->}[r]&
2	\ar@(ur,dr)@{<->}[]
}
\]
with events $\{1, 2\}$, no non-trivial causality and all finite sets
consistent, and symmetry comprising all bijections between
configurations meaning that all events are interchangeable. In this
diagram we use an arrow $\xymatrix@R=5pt@C=15pt{~\ar@{<->}[r]&~}$ to convey the information
that events are symmetric. In general however, the information of
symmetry is more contextual and cannot be represented that easily. For
instance, one may consider an event structure with symmetry with events
$\{1, 2, 1', 2'\}$, trivial causality and all finite sets consistents,
and symmetry comprising all bijections between configurations that are
subsets of two maximal bijections:
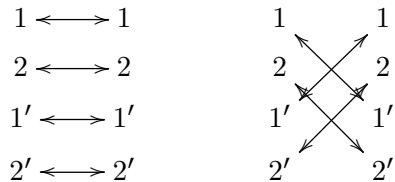
\[
\xymatrix@R=5pt{
1	\ar@{<->}[r]&
 1\\
2	\ar@{<->}[r]&
 2\\
1'	\ar@{<->}[r]&
 1'\\
2'	\ar@{<->}[r]&
 2'
}
\qquad\qquad
\xymatrix@R=5pt{
1       \ar@{<->}[ddr]&
 1\\
2       \ar@{<->}[ddr]&
 2\\
1'      \ar@{<->}[uur]&
 1'\\
2'      \ar@{<->}[uur]&
 2'
}
\]

It is a good exercise to verify that this yields an event
structure with symmetry. There one can observe that from the empty
bijection one can put $2$ and $2'$ in correspondance, however if $(1,
1)$ is already in the bijection, then its extension with $(2, 2')$
forms the bijection between configurations $\{(1, 1), (2, 2')\}$ which
is \emph{not} in the isomorphism family, not being a subset of either
of the bijections above. In that sense symmetry is
\emph{contextual}: whether two events can be interchanged depends on which
events have already been exchanged. Being sets of bijections, symmetries on event
structures are rather hard to picture. In this paper, most diagrams representing
event structures with symmetry will only display the event structure (the
observable actions) and keep the symmetry (witnessing uniformity) implicit.

The following lemma, easy consequence of the \emph{(Restriction)}
axiom, is important to keep in mind when manipulating symmetry:
\begin{lem}
  Let $\A$ be an ess and   $ \theta : x \bij y \in \tilde{A}$. Then, $
\theta $ is an
  order-isomorphism.
\end{lem}

Hence, if $\theta \in \tilde{A}$, we write 
$\theta : x \iso y$ (rather than just $x \bij y$) to indicate that
$\theta$ preserves and reflects the (implicit, inherited from $\leq_A$)
ordering on $x$ and $y$.
Instead of $ \theta : x \iso y \in \tilde A$, we will also often use
the more compact notation
$\theta : x \isf_{\tilde A} y$; and we will refer to $\theta$ as
\textbf{a symmetry}\index{Symmetry} between $x$ and
$y$. Given a symmetry $ \theta $, we write $\dom \theta $ and
$\codom \theta $ for its domain and codomain respectively.%
\nomenclature[bd]{$\theta : x \iso y$}{Order-preserving bijection
between $x$ and $y$}%
\nomenclature[be]{$\theta : x \isf_{\tilde{A}} y$}{$\theta : x \iso y
\in \tilde{A}$}%
\nomenclature[bf]{$\dom\,\theta$}{Domain of $\theta$}%
\nomenclature[bg]{$\codom\,\theta$}{Codomain of $\theta$}%

\subsubsection{Constructions on ess.}
In the interpretation of games, as discussed in Section
\ref{subsubsec:weakiso} all non-trivial symmetries come from
\emph{replication}. Accordingly our key construction of ess comes
from the discussion of Section \ref{subsubsec:replication} -- the
following is established by a direct verification.

\begin{prop}\label{prop:bang_isos}%
\nomenclature[bh]{$\tilde{\oc A}_-$}{Set of negative reindexing isos on
$\oc A$}%
\nomenclature[bi]{$\tilde{\oc A}_+$}{Set of positive reindexing isos on
$\oc A$}%
  Let $A$ be an arena. Recall $\oc A$ (Definition
  \ref{def:oc}) whose events are functions
  $ \alpha : [a] \rightarrow \mathbb{N} $ with $a \in A $. The sets
  $\tilde{\oc A}$ of \emph{reindexing isos} (see Definition
\ref{def:index_isos}), $\tilde{\oc A}_-$
  of \emph{negative reindexing isos}, and $\tilde{\oc A}_+$ of
\emph{positive reindexing isos}, are
  isomorphism families on $\oc A$.
\end{prop}

This will be at the heart of our construction of the
cartesian closed category of concurrent Hyland-Ong games, in Section
\ref{sec:cho}. We mention in passing another construction on event
structures with symmetry, more reminiscent of \emph{AJM games}
\cite{ajm}. 

\begin{defi}\label{def:bang_ajm}
If $\A$ is an ess, then $\ocajm \A$ has events, causality, consistency
that of $\parallel_\omega A$. Its isomorphism family comprises those
bijections $\theta :\,\parallel_{i\in I} x_i \bij\,\parallel_{j\in J}
y_j$ such that there exists a permutation $\pi : I \bij J$, and for all
$i\in I$ a symmetry $\theta_i : x_i \isf_\A y_{\pi(i)}$ such that 
for all $(i, a) \in\,\parallel_{i\in I} x_i$, we have $\theta(i, a) =
(\pi(i), \theta_i(a))$.
\end{defi}

The reader familiar with AJM games will recognize the similarity with
the definition of the equivalence relation between plays in an
exponential games in the AJM setting. In this paper we focus on
HO-style games, but it has become crucial in further work that our setting with
symmetry supports AJM-style games as well -- we will come back to that in Section
\ref{subsubsec:ajm}. Throughout the paper we reserve the notation $\oc$ for
HO-style replication, \emph{i.e.} the ess $\oc A$ of Proposition
\ref{prop:bang_isos} from an arena $A$. In contrast, the AJM-style exponential
will always be denoted by $\ocajm$ to reflect its lesser importance in the
present development.

Besides the above, ess support all the basic constructions on event structures.
\begin{defi}\label{def:op_sym}%
\nomenclature[bj]{$\theta_1 \parallel \theta_2$}{Parallel composition
$\theta_1 \parallel \theta_2 :
x \parallel y \bij x' \parallel y'$ of bijections $\theta_1 : x \bij
x'$ and
$\theta_2 : y \bij y'$}%
  Let $\A$ and $\B$ be ess. We build their \textbf{simple parallel
composition}\index{Simple
parallel composition!of event structures with symmetry (ess)} as
  $(A  \parallel  B, \tilde A  \parallel  \tilde B)$ where $\tilde A
\parallel  \tilde B$ is the
  set of bijections of the form $ \theta _1  \parallel   \theta _2 : x
\parallel  y  \bij  x'
\parallel  y'$ where
  $x, x'  \in   \mathscr{C} (A), y, y'  \in   \mathscr{C} (B)$, $
\theta _1  \in  \tilde A,  \theta
_2  \in  \tilde B$ and
  $ \theta _1  \parallel   \theta _2$ is defined as $(i, a) \mapsto (i,
\theta _i(a))$.

If $\A$ is a game with symmetry, its
\textbf{dual}\index{Dual!of event structures with
symmetry (ess)} $\A^\perp$ has the same isomorphism
family $\tilde{A}$ on $A^\perp$.
\end{defi}

\subsubsection{Categorical structure.} Just as maps of event structures
were at the heart of the construction of concurrent games, games with
symmetry will make use of maps between ess.

\begin{defi}\label{def:ess}%
\nomenclature[bk]{$f\,\theta$}{Forward image $f\,\theta = \{(f\,a_1,
f\,a_2)\mid (a_1, a_2) \in
\theta\}$ of a bijection}%
  Let ${\A}, \B$ be event structures with symmetry. A map of event
structures
  $f : {A}  \rightarrow  B$ \textbf{preserves
symmetry}\index{Morphisms!preserving symmetry} iff for all
  $ \theta : x  \isf _{{\tilde A}} y$, the bijection
  $f \theta  = \{(fa, fa') \mid (a, a')  \in   \theta \}$ is in $
\tilde B$.
  In that case, $f$ is a \textbf{map of event stuctures with
symmetry}\index{Morphisms!of
event structures with symmetry (ess)}, written $f : \A \to
  \B$.
Event structures with symmetry and their maps form a category written%
\nomenclature[bka]{$\ESS$}{Category of event structures with symmetry
and
morphisms preserving symmetry}%
$\ESS$.

Finally, if $\A, \B$ have polarities, maps are required to preserve
those as well.
\end{defi}


In $\ESS$, morphisms can be compared \emph{up to symmetry}, abstracting
away from
the comparison of morphisms \emph{up to the choice of copy indices} of
the previous section.

\begin{defi}\label{def:sym_maps}%
\nomenclature[bl]{$f \sim_{\tilde{B}} g$}{Maps $f, g : \A \to \B$ in
$\ESS$ are
symmetric}%
  Let $f, g : \A  \rightarrow  \B$ be maps of event
  structures with symmetry. They are \textbf{symmetric}\index{Symmetric
morphisms of ess} (written
  $f \sim_{\tilde B} g$) when for all $x  \in   \mathscr{C} (A)$, the
bijection
  $\{(fs,gs)\mid s  \in  x\}$ is in $\tilde B$.
\end{defi}

As usual, we write $f \sim g$ instead of $f \sim_{\tilde B} g$ when $\tilde{B}$
is clear from the context. 

With this definition, we will be able to reformulate Definition \ref{def:weak_mor}
by requiring the triangle to commute up to symmetry, \emph{i.e.} $\tau \circ f
\sim \sigma$ (postponed for now).

\subsubsection{Pullbacks and pseudo-pullbacks.} The construction of 
$\CG$ relies crucially on properties of the category $\ES$ of event
structures and their maps. In order to construct games with symmetry,
it is appealing to attempt replicating the same constructions, but
building on $\ESS$ rather than $\ES$. In other words, a \emph{game with
symmetry} would be an ess $\A$ with polarities, and a strategy on $\A$
would be a map $\sigma : \S \to \A$ between ess. A strategy \emph{from
$\A$ to $\B$} would be $\sigma : \S \to \A^\perp \parallel \B$, and
those would be composed by pullback as in Section
\ref{sec:replication}, and so on.

A first obstacle in replicating those constructions is that unlike
$\ES$, the category $\ESS$ does not have all pullbacks (a proof of that
appears in Appendix \ref{app:no_pb}) -- one may understand that by the
fact that $\ESS$ really may be more adequately regarded as enriched
over equivalence relations: indeed, maps should be compared up to
symmetry rather than on the nose. Hence, one is tempted to have the
role of pullbacks played by a universal construction taking symmetry
into account, such as \emph{pseudo-pullbacks} or \emph{bi-pullbacks}.

\begin{defi}\label{def:pseudo_pb}
Let $f : \A \to \C$ and $g : \B \to \C$ in $\ESS$. A
\textbf{pseudo-pullback}\index{Pullback!pseudo-} of $f$ and $g$ is
\[
\xymatrix@R=10pt@C=10pt{
&\P	\ar[dl]_{\Pi_1}
	\ar[dr]^{\Pi_2}
	\pb{270}\\
\A	\ar@{}[rr]|\sim
	\ar[dr]_f&&
\B	\ar[dl]^g\\
&\C
}
\]
commuting \emph{up to symmetry}, and such that for all $f' : \X \to
\A$, $g' : \X \to \B$ such that $f\circ f' \sim g \circ g'$ there
exists a \emph{unique} map $h : \X \to \P$ such that $\Pi_1 \circ f =
f'$ and $\Pi_2 \circ g = g'$.

It is a \textbf{bi-pullback}\index{Pullback!bi-} iff for
all $f' : \X \to \A$, $g' : \X \to \B$ such that $f\circ f' \sim g
\circ g'$ there is $h : \X \to \P$, \emph{unique up to symmetry}, such that
$\Pi_1 \circ f \sim f'$ and $\Pi_2 \circ g \sim g'$. 
\end{defi}

In particular, any pseudo-pullback is a bi-pullback. It turns out that
$\ESS$ has all pseudo-pullbacks \cite{symmetry}, so one may opt to use
those for composing strategies. Though sensible, this choice has technically 
heavy
consequences -- we will come back to them later. However, there is
another possibility to compose strategies with symmetry: it may be that
though $\ESS$ does not have \emph{all} pullbacks, it has all those
required to compute the interaction of strategies. And indeed, using
polarity it is easy to capture maps that interact well in the presence
of symmetry: just as
receptivity prevents a strategy from refusing an Opponent move,
\emph{$\sim$-receptivity} prevents it from refusing to consider two
Opponent moves to be symmetric.

\begin{defi}
If $\A$ is a essp, a map of ess $\sigma : \S \rightarrow {\A}$ is 
  \textbf{$\sim$-receptive} \index{$\sim$-receptive} iff for all
  $ \theta :x_1 \isf _{{\tilde S}} x_2$, for all $x_1 \longcov{s_1^-}$
and
$\sigma\,x_2 \longcov{a_2^-}$ such that $\sigma\,\theta \cup
\{(\sigma\,s_1, a_2)\} \in \tilde{A}$,
there is a unique $s_2$ such that $\sigma\,s_2 = a_2$, and we have
$\theta \cup \{(s_1, s_2)\} \in \tilde{S}$.
%

Moreover, $\sigma$ as above is \textbf{strong-receptive} if it is both
receptive and $\sim$-receptive.
\end{defi}

Dual $\sim$-receptive maps always have pullbacks in
$\ESS$: intuitively, the polarity helps determining whose
responsibility it is to set two events as symmetric. However,
before we prove that, let us include an example illustrating the
following fact: besides ensuring pullbacks, $\sim$-receptivity is key to
ensure that strategies based on ess are indeed uniform.

\begin{exa}
  Recall the non-uniform strategy from Section \ref{sec:replication}:
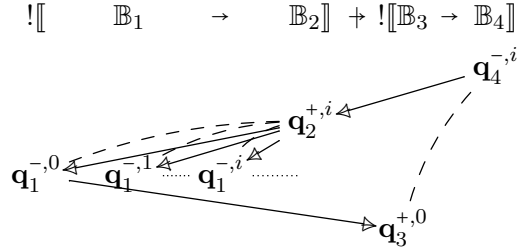
\[\xymatrix@R=0pt@C=10pt{
\oc\llbracket&\tbool_1  \ar@{}[rr]|\to&&
\tbool_2\rrbracket      \ar@{}[r]|\to||&
\oc\llbracket\tbool_3   \ar@{}[r]|\to&
\tbool_4\rrbracket\\
&&&&&\q_4^{-, i}
        \ar@{-|>}[dll]\\
&&&\q_2^{+, i}
        \ar@{-|>}[dlll]
        \ar@{-|>}[dll]
        \ar@{-|>}[dl]\\
\q_1^{-, 0}
        \ar@{-|>}[drrrr]
        \ar@{--}@/^/[urrr]&
\q_1^{-,1}\ar@{--}@/^/[urr]
        \ar@{.}[r]&
\q_1^{-,i}\ar@{--}@/^/[ur]
        \ar@{.}[r]&~\\
&&&&\q_3^{+,0}
        \ar@{--}@/^/[uuur]
}
\]
Assume now there is an isomorphism family $\tilde S$ on this event
structure $S$ such that the labelling map
$ \sigma  : \S  \rightarrow  \oc  \llbracket \tbool  \to  \tbool
\rrbracket^\perp  \parallel  \oc
\llbracket \tbool  \to  \tbool \rrbracket$ is $\sim$-receptive.

By $\sim$-receptivity (since the identity on $\{\q_4^{-, 0},
\q_2^{+,0}\}$
must be in $\tilde S$), we must have that the bijection
$\{\q_4^{-, 0}, \q_2^{+, 0}, \q_1^{-, 0}\} \isf \{\q_4^{-, 0}, \q_2^{+,
0},
\q_1^{-, 1}\}$ is in $\tilde S$. However, only the left hand side part
can be
extended by $\q_3^{+, 0}$, absurd.
\end{exa}

As $\sim$-receptivity is crucial to ensure uniformity, it will be required
whether we wish to compute interaction via pullback or pseudo-pullback.

We now prove that as claimed above, pullbacks exist along dual
$\sim$-receptive maps of ess. To define the isomorphism family for the
pullback, we first notice that bijections on configurations of the (plain)
pullback induce bijections on their projections:

\begin{lem}
  Let $\sigma : S \to A$ and $\tau : T \to A$ be maps of event
structures.
  Let $ \theta  : w  \bij  z$ be a bijection, where $w, z \in
\conf{S\wedge T}$.
  There are (unique) bijections
  $ \theta _S :  \Pi _1\,w  \bij   \Pi _1\,z$ and $ \theta _T :  \Pi
_2\,w  \bij   \Pi _2\,z$ satisfying
  $ \Pi _1 \circ  \theta  =  \theta _S \circ  \Pi _1$ and $ \Pi _2
\circ  \theta  =  \theta _T \circ
   \Pi _2$.
  Moreover, the mapping $ \theta  \mapsto ( \theta _S,  \theta _T)$ is
monotonic \emph{w.r.t.}
  inclusion.
\end{lem}
\begin{proof}
  By local injectivity, $ \Pi _1$ defines a bijection
  $w \bij \Pi _1 w$ and $z \bij \Pi _1 z$.  With this remark,
  $ \theta _S$ is simply defined as
  $ \Pi _1 \circ \theta \circ \Pi _1^{-1}$. The equation and uniqueness
are by definition, and
  monotonicity is obvious. The definition of $\theta_T$ is symmetric.
\end{proof}

For $\sigma : \S \to \A$ and $\tau : \T \to \A$ maps of ess, define
$\tilde{S} \wedge \tilde T$ to contain those bijections $ \theta : w
\isf z$ such that $ \theta _S :  \Pi _1 w  \isf   \Pi _1 z \in  \tilde
S$ and $ \theta _T :  \Pi _2 w  \isf   \Pi _2 z  \in  \tilde T$.
Bearing in mind the correspondence between configurations of $S \wedge
T$ and secured bijections $x \sbij y$, there is an order-isomorphism
between those bijections $\theta \in \tilde{S} \wedge \tilde T$ and
commutative squares between secured bijections $x \sbij y$ and $x'
\sbij y'$ (ordered by componentwise union):

\[
\xymatrix@R=15pt@C=15pt{
\ar@{}[d]|{\theta_S \rotatebox{90}{$\isf$}_{\tilde S}}
x \ar@{}[r]|{\sbij}^ \sigma  &  \sigma x \ar@{}[r]|= &  \tau y
\ar@{}[r]|{\sbij}^ \tau  & y
\ar@{}[d]|{\theta_T \rotatebox{90}{$\isf$}_{\tilde T}}\\
x' \ar@{}[r]|{\sbij}^ \sigma  &  \sigma x' \ar@{}[r]|= &  \tau y'
\ar@{}[r]|{\sbij}^ \tau  & y' \\
}
\]

This definition indeed yields a pullback in $\ESS$:

\begin{restatable}{lemma}{pbsym}
\label{lem:pb_sym}
  Let $\sigma : \S \to \A$ and $\tau : \T \to \A^\perp$ be
$\sim$-receptive maps of ess.
  The set $\tilde{S} \wedge \tilde {T}$ is an isomorphism family on
  $S \wedge T$ and the ess
  $(S \wedge T, \tilde{S} \wedge \tilde T)$ is a pullback in $\ESS$ of
$ \sigma $ and $ \tau $,
  written $\S \wedge \T$.
\end{restatable}
\begin{proof}[Proof idea]
The only difficulty is in proving the \emph{(Extension)} condition.
The argument exploits that an extension is
positive for one of $\sigma, \tau$, negative for the other -- apply
\emph{(Extension)} for the positive, and $\sim$-receptivity for the
other. The details are in Appendix \ref{app:pb_exist}.
\end{proof}

\subsubsection{Equivalences between strategies.} So, should we base our
composition of strategies with symmetry on pullbacks or
pseudo-pullbacks? To make up our mind another aspect must weight in:
the universal property used to compose strategies impacts the
equivalence up to which strategies may be considered. Indeed in $\CG$
strategies are naturally considered up to isomorphism (Definition
\ref{def:2cell_strat}), and that is of course preserved by pullbacks
(pullbacks along isomorphic maps being isomorphic).
In contrast, as pointed out in Section \ref{subsubsec:weakiso} in the
presence of replication it is crucial to consider strategies \emph{up
to symmetry}, and there is no reason for an equivalence such as that of
Definition \ref{def:weak_mor} to be preserved by pullback. However,
\emph{pseudo-pullbacks} do preserve notions of \emph{equivalence}:

\begin{defi}\label{def:weak_mor_sim}
Two maps of ess $\sigma : \S \to \A$ and $\sigma' : \S' \to \A$ are
\textbf{weak-equivalent} iff there are maps of ess $f : \S \to \S'$, $g : \S'
\to \S$ such that $g \circ f \sim \id_\S, f \circ g \sim \id_\T$, and
the two triangles below commute \emph{up to symmetry}:
\[
\xymatrix@R=10pt@C=10pt{
\S	\ar[rr]^{f}
	\ar[dr]_\sigma&~\ar@{}[d]|\sim&
\S'	\ar[dl]^{\sigma'}\\
&\A
}
\qquad\qquad
\xymatrix@R=10pt@C=10pt{
\S'	\ar[rr]^{g}
	\ar[dr]_{\sigma'}&~\ar@{}[d]|\sim&
\S	\ar[dl]^{\sigma}\\
&\A
}
\]

They are \textbf{strong-equivalent} iff the triangles further commute
on the nose.
\end{defi}

It follows from their definition that pseudo-pullbacks
preserve strong equivalence, and that bi-pullbacks
preserve weak equivalence. Being a particular case of bi-pullbacks,
pseudo-pullbacks preserve both. Weak equivalence looks like the notion
of Definition \ref{def:weak_mor} and indeed can serve to compare
strategies up to their choice of copy indices (we will come back later to the
subtle differences between weak equivalence and Definition \ref{def:weak_mor}),
while the requirement that the triangles commute
on the nose makes strong equivalence look less obviously relevant for that
purpose. 
In any case, the take-away message for this discussion seems to be:
\emph{just use pseudo-pullbacks for interaction!} 

Mathematically, this seems an elegant approach to
concurrent games with symmetry: as $\ESS$ is enriched over equivalence
relations, in building concurrent games with symmetry it is natural to
mimic the constructions of $\CG$ from $\ES$ using the corresponding
operations on ess that take account of the enrichment.
Indeed, this natural solution was the basis for our first account of
symmetry in concurrent games \cite{lics14}. It works 
as a general framework
(indeed \cite{lics14} contains a construction
of HO and AJM-style games), but suffers from significant drawbacks
concerning its applicability for further semantic purposes.

\subsubsection{Saturated uniformity} \label{subsubsec:sat_unif}
Let us investigate some
consequences of computing interactions via pseudo-pullbacks. Consider
maps of essp $\sigma : \S \to \A^\perp \parallel \B$ and $\tau : \T \to
\B^\perp \parallel \C$ (regarded as strategies with symmetry -- though
we leave for later the precise definition of those), and say that
mimicing the construction in Section \ref{subsubsec:rep_interaction},
we compute the pseudo-pullback:
\[
\xymatrix@R=15pt@C=0pt{
&\T\circledast \S
        \ar[dl]_{\Pi_1}
        \ar[dr]^{\Pi_2}
        \pb{270}\\
\S\parallel \C
        \ar[dr]_{\sigma \parallel \C}
	\ar@{}[rr]|\sim&&
\A\parallel \T
        \ar[dl]^{A\parallel \tau}\\
&\A\parallel \B \parallel \C
}
\]

Unlike pullbacks, pseudo-pullbacks commute \emph{up to symmetry}.
Consequently, interaction via pseudo-pullback allows one to synchronize
configurations that do not match on the nose, but \emph{up to
symmetry}. 
Accordingly, the pseudo-pullback analogue of Proposition
\ref{prop:rep_int} says that configurations in $\conf{\T \inter \S}$
correspond to the data of $x_S \parallel x_C \in \conf{S \parallel C},
x_A \parallel x_T \in \conf{A\parallel T}$ and $\theta \in
\tilde{A}\parallel \tilde{B}\parallel \tilde{C}$ such that:
\[
\xymatrix@C=5pt{
x_S \parallel x_C&
\stackrel{\sigma}{\mapsto}&
x_A \parallel x_B \parallel x_C
&\stackrel{\theta}{\isf_{\tilde{A}\parallel \tilde{B}\parallel
\tilde{C}}} &y_A \parallel y_B \parallel y_C&
\raisebox{5pt}{$\stackrel{\tau}{\rotatebox{180}{$\mapsto$}}$}&
y_A \parallel y_T
}
\]

It carries a symmetry $\theta \in \tilde{A}
\parallel \tilde{B} \parallel \tilde{C}$ which mediates between
configurations of $S$ and $T$ not quite matching on the game, but can
also apply a symmetry on the visible output in the game. Everything
becomes ``up to symmetry'', including the visible actions of the interaction.
Concretely, because of that, composition based on pseudo-pullbacks has a ``saturation''
effect: for $x\in \conf{S}$ in a saturated $\sigma :
S \to A$, for all $\theta : \sigma x \isf_{\tilde{A}} y'$ there
must be $\varphi : x \iso_\S y$ such that $\sigma y = y'$. In other
words, if a saturated strategy is prepared to play a move, by necessity
it is also prepared to play non-deterministically all symmetric moves.
To behave well with pseudo-pullback-based composition, all strategies need to
be saturated, even copycat \cite{lics14}.
\begin{figure}
\[
\xymatrix@R=10pt@C=10pt{
&&&\oc \llbracket\tbool
        \ar@{}[r]|\to&
\tbool\rrbracket\\
&&&&\q^{-,i}
        \ar@{-|>}[dllll]
	\ar@{-|>}[dlll]
	\ar@{-|>}[dll]
	\ar@{-|>}[dl]\\
\q^{+, 0}
        \ar@{--}@/^/[urrrr]
	\ar@{~}[r]&
\q^{+, 1}
	\ar@{~}[r]
	\ar@{--}@/^/[urrr]&
\q^{+, 2}
	\ar@{}[r]|\dots
	\ar@{--}@/^/[urr]&
\q^{+,j}\ar@{}[r]|\dots
	\ar@{--}@/^/[ur]&
}
\]
\caption{Saturated interpretation of the strategies of Example
\ref{ex:wiso}}
\label{fig:ex_saturated}
\end{figure}
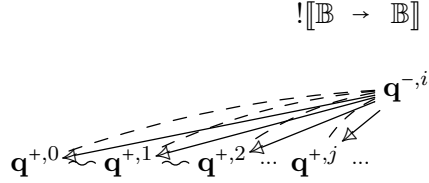
We display in Figure \ref{fig:ex_saturated} the saturated strategy
corresponding to the two diagrams at the end of Example \ref{ex:wiso}
-- to alleviate the diagram we only represent a few binary conflicts,
but in reality all positive events are pairwise incompatible and
symmetric: the strategy chooses non-deterministically one copy index
and plays it. 

The reader with a background in game semantics will recognize here a
phenomenon familiar from AJM games: in \cite{believe} was introduced a
variant of AJM games where strategies were similarly required to be
saturated under the action of the equivalence relation. Independently
of the specific purposes of \cite{believe} (which required it for
technical reasons), saturation provides a different approach to
uniformity in AJM games than the traditional one \cite{ajm}. While it
has remained rather marginal in game semantics, some recent works based
on AJM games are actually based on saturation
\cite{DBLP:journals/entcs/AbramskyJ09,DBLP:journals/iandc/VakarJA18}, as
it provides a slightly simpler mathematical foundation for uniformity.

While mathematically natural, saturation in concurrent games has some
significant drawbacks. First of all, it makes strategies more
``abstract'': even the underlying event structure becomes impossible to
represent faithfully. As copy indices are chosen non-deterministically, all
(non-trivial) strategies fail determinism in the usual sense of
concurrent games \cite{DBLP:journals/fac/Winskel12} and it is
tricky to recover a working notion of determinism \emph{up to
symmetry}. More generally, notions in $\CG$ that rely on
a concrete analysis of the shape of conflict (such as
\emph{single-threadedness} -- see Definition \ref{def:single-threaded}
-- or all our further work on \emph{innocence} \cite{lics15}) are
hard to accommodate with saturated symmetry as the non-deterministic
choice of symmetric events pollutes the genuine dynamic behaviour of the
strategy. Last but not least, the saturated framework is not
conservative over $\CG$: one cannot ``forget symmetry'' functorially
as composition mechanisms inherently involve it.
This means that developments in $\CG$ are less likely to
adapt transparently to saturated symmetry.

For these reasons, 
we have found it necessary 
to develop an
alternative approach to uniformity (first presented in \cite{lics15}), called
\emph{thin concurrent games} as opposed to \emph{saturated}, which is
conservative over $\CG$: in particular, the core components of the
categorical structure (composition, copycat) match those in $\CG$.
In that respect, it is analogous to the traditional AJM way
of handling uniformity \cite{ajm}. This has
been a subtle endeavour, but the effort pays off: conservativity over
$\CG$ means that many developments on $\CG$ extend transparently with
thin symmetry, undisturbed by the uniformity requirements. Such
successes include of course concurrent innocence \cite{lics15}, but
also the more recent probabilistic extension \cite{lics18}, and other
works as of now unpublished. So the
following construction of $\TCG$ is hard work, but experience tells
that the subtleties involved can be mostly abstracted away when relying
on $\TCG$, and one can work like in $\CG$ simply with the additional
proof obligations that further structure should be invariant under
symmetry.

\subsection{Thin concurrent games.}
\label{subsec:thin} Rather than changing the compositional machinery to
preserve weak equivalence, we instead seek further restrictions on strategies
for their standard interaction pullback to preserve it.
Interestingly the answer turns out to be a minimality condition 
mirroring saturation, suggesting a sort of duality that is still under
investigation.

\subsubsection{Thinness.}
Our starting point is the fact that, unfortunately,
weak equivalence of $\sim$-receptive maps is
\emph{not} preserved under composition or pullback -- a counter-example
appears in Appendix \ref{app:pentagram}. Our interpretation of that failure is 
that the \emph{(Extension)}
axiom is too permissive for strategies. Indeed, consider $\sigma : \S
\to \A$ a $\sim$-receptive map of essps, and $\theta : x
\isf_{\tilde{S}} y$, suppose further that $x$ extends by some positive
event $s^+ \in S$ (strategies will be strong-receptive, as such
extensions by negative events are inconsequential and entirely governed
by the game). By \emph{(Extension)}, we then have some $s'$ with:
\[
\xymatrix@R=10pt@C=20pt{
x \cup \{s\}
	\ar@{}[d]|{\rotatebox{90}{$\cov$}}
	\ar@{}[r]|{\stackrel{\theta'}{\isf_{\tilde{S}}}}&
y \cup \{s'\}
	\ar@{}[d]|{\rotatebox{90}{$\cov$}}\\
x	\ar@{}[r]|{\stackrel{\theta}{\isf_{\tilde{S}}}}&
y
}
\]

The crux of the issue is in how this $s'$ is chosen. For
saturated strategies the set of available choices for $s'$ is
\emph{canonical}: there are as many $s'$ as there are symmetric
events in the game. In contrast, in the counter-example of Appendix
\ref{app:pentagram}, there is no canonical choice
(more specifically, the
counter-example exploits a situation where no valid assignment from the
choices for $s$ to the choices for $s'$ can be described globally as a
map of event structures).

This need of a canonical choice for positive extensions motivates the
definition:

\begin{defi}\label{def:thin}
An essp $\A$ is \textbf{thin}\index{Thin!essp} if for $\theta : x
\isf_{\tilde{A}} y$ and $x \longcov{a_1^+}$, there is a
\emph{unique} $y \longcov{a_2^+}$ s.t.
\[
x \cup \{a_1\} \stackrel{\theta \cup \{(a_1, a_2)\}}{\isf_{\tilde{A}}}
y \cup \{a_2\}
\]
%
\end{defi}

The canonicity of the choice of positive extensions is ensured by
uniqueness. \emph{Thinness} is the main ingredient of
our forthcoming definition of \emph{$\sim$-strategies};
the next objective in our construction being to show that indeed (under mild
further conditions), the interaction between thin maps preserves weak
equivalence (a phenomenon we will from now on refer to as
\emph{congruence}, for brevity). For that we first mention a slightly
simpler equivalent formulation of thinness -- the equivalence proof is
an interesting exercise. 

\begin{lem}\label{lem:alt_thin}
An essp $\A$ is thin iff for all $x\in \conf{A}$, for all
$\id_x \subseteq^+ \theta \in \tilde{A}$, then $\theta = \id_y$ for
some $x \subseteq y \in
\conf{A}$.
\end{lem}

The \emph{$\sim$-strategies} will be certain $\sim$-receptive
$\sigma : \S \to \A$, with $\S$ thin (we will also say 
that $\sigma$ is thin). An intuitive reading of thinness
prompted by Lemma \ref{lem:alt_thin} is that $\sigma$ will not take the
initiative of declaring two positive events symmetric. As long as
negative extensions of the symmetry remain identity bijections,
positive extensions will too. As a consequence:

  \begin{lem}\label{lemma:pb_thin}
  Let $ \sigma  : \S  \rightarrow  \A$ and $ \tau  : \T  \rightarrow
\A^\perp$ be thin $\sim$-receptive maps of essp. 

Then, $\tilde S \wedge \tilde T$ is trivial (reduced to identities).
\end{lem}
\begin{proof}
  We prove by induction that all bijections in
  $\tilde S \wedge \tilde T$ are identities. Let
  $z \in \mathscr{C} (S \wedge T)$ and assume $\id_z$ extends by
  $(e, e')$ to $ \theta \in \tilde S \wedge \tilde T$. Assume for
  instance $ \Pi_2\,e$ is positive in $T$ (the other case is
  similar). By construction
  $ \id_{ \Pi_2\,z}$ extends to
  $ \theta_T = \Pi_2\,\theta \in \tilde{T}$ by positive events
$(\Pi_2\,e,\Pi_2\,e')$,
  hence $ \Pi_2\,e = \Pi_2\,e'$ and
  $ \theta _T$ is the identity because $\tilde T$ is thin. By local
injectivity of
  $\Pi_2$ it follows that $e$ and $e'$ must be equal, or incompatible
extensions of $z$.
  But if they are incompatible,
  by Lemma \ref{lem:pb_mconf} (and Proposition \ref{prop:rep_int}) it
means that
  $ \Pi _1\,e$ and $ \Pi _1\,e'$ are incompatible extensions of
$\Pi_1\,z$
  mapping to the same event in the game, contradicting the
$\sim$-receptivity
  of $ \sigma $. Hence $e = e'$ and $ \theta $ is the identity.
\end{proof}

In other words, a closed interaction between thin $\sim$-receptive maps
has a trivial symmetry. Of course that will not be the case for an
\emph{open} interaction between $\sigma : \S \to \A^\perp \parallel \B$
and $\tau : \T \to \B^\perp \parallel \C$, where the external Opponent
may contribute new symmetric pairs of negative events on $\A^\perp
\parallel \C$. Indeed, computing this interaction involves the
pullback
\[
\xymatrix@R=15pt@C=0pt{
&\T\circledast \S
        \ar[dl]_{\Pi_1}
        \ar[dr]^{\Pi_2}
        \pb{270}\\
\S\parallel \C
        \ar[dr]_{\sigma \parallel \C}&&
\A\parallel \T
        \ar[dl]^{A\parallel \tau}\\
&\A\parallel \B \parallel \C
}
\]
on which Lemma \ref{lemma:pb_thin} does not apply, because 
$\id_{\C^\perp} : \C^\perp \to \C^\perp$ and $\id_\A : \A \to \A$ are
\emph{not} thin -- indeed, $\A$ could be \emph{e.g.} the essp of
Proposition \ref{prop:bang_isos}, whose isomorphism family includes all
reindexing isomorphisms. In our proof of congruence, it will be of
great importance that we are able to consider a sub-structure of the open
interaction above where the external Opponent \emph{does not contribute new
symmetric pairs}, \emph{i.e.} restrict symmetries so as arrive in the realm of Lemma
\ref{lemma:pb_thin}. Concretely, we need a thin replacement $\A_+ \to
\A$ for $\A \to \A$, and likewise for $\C$. But no such replacement exist 
without further conditions on games\footnote{In fact
the counter-example of Appendix \ref{app:pentagram} can be adapted to show
that without further hypothesis on games, congruence fails even for
thin strategies.}.

\subsubsection{Thin concurrent games and $\sim$-strategies.} The
requirement above suggests that besides being essps, games with
symmetry $\A$ should feature a sub-symmetry $\tilde{A}_-$
such that $\id_\A : \A_- \to \A$ satisfies the conditions of Lemma
\ref{lemma:pb_thin}, \emph{i.e.} is thin and $\sim$-receptive. By
duality, we also need $\tilde{A}_+$ such that $\A_+^\perp \to
\A^\perp$ is thin $\sim$-receptive. These polarised sub-symmetries,
along with their desired properties, are nicely captured by the
following definition\footnote{This formulation was a later improvement
on the original definition of \cite{lics15}.}.

\begin{defi}\label{def:tcg}%
\nomenclature[bm]{$\tilde{A}_-, \tilde{A}_+$}{Negative/positive
isomorphism
families of a tcg $\A$}%
\nomenclature[bn]{$x \subseteq^p y, \theta \subseteq^p
\theta'$}{Extension with events of polarity
$p \in \{-, +\}$}%
  A \textbf{thin concurrent game (tcg)}\index{Thin!concurrent games
(tcg)} is an essp $\A$ with two additional isomorphism families $\tilde
A_-$ and $\tilde A_+$ on $A$ such that:
  \begin{enumerate}
  \item[(a)] The families $\tilde A_+$ and $\tilde A_-$ are subsets of
$\tilde A$,
  \item[(b)] If $ \theta   \in  \tilde A_+ \cap \tilde A_-$ then $
\theta $ is an identity bijection,
  \item[(c)] If $\theta \in \tilde A_-$ and $ \theta  \subseteq^-
\theta '  \in  \tilde A$ then
$ \theta '  \in  \tilde A_-$,
  \item[(d)] If $\theta \in \tilde A_+$ and $ \theta  \subseteq^+
\theta '  \in  \tilde A$ then
$ \theta '  \in  \tilde A_+$.
  \end{enumerate}
  where $\theta \subseteq^- \theta'$ (resp.
  $\theta \subseteq^+ \theta'$) means that $\theta \subseteq \theta'$
  such that $\theta' \setminus \theta$ only contains pairs of negative
(resp. positive) events.
  The triple
  $(\A, \tilde A_-, \tilde A_+)$ will be often written simply $\A$.
\end{defi}

Thin concurrent games support the basic operations on essps: the
\textbf{dual}\index{Dual!of a tcg} is 
$(\A, \tilde A_-, \tilde A_+)^\perp = (\A^\perp, \tilde{A}_+,
\tilde{A}_-)$, and the \textbf{simple parallel composition}
\index{Simple parallel composition!of thin concurrent games (tcg)}
is performed componentwise:
\[
(\A, \tilde A_-, \tilde A_+) \parallel (\B, \tilde B_-, \tilde B_+) = 
(\A \parallel \B, \tilde A_- \parallel \tilde B_-, \tilde
A_+ \parallel \tilde B_+).
\]

Of course, thin concurrent games include our guiding example:

\begin{prop}
Let $A$ be an arena. Then, $(\oc A, \tilde{\oc A}, \tilde{\oc A}_-,
\tilde{\oc A}_+)$ is a thin concurrent game.
\end{prop}

Through this example one may get a more concrete understanding of the
sub-symmetries: $\tilde{A}_+$ comprises the symmetries where only
Player has performed non-trivial exchanges between moves (so ``copy
indices'' of negative events are preserved), and dually for
$\tilde{A}_-$.

We mention right away two important properties of tcgs. Firstly, it
follows as required that $\A_-$ is thin (and likewise for
$\A_+^\perp$). Indeed, using the characterisation of Lemma
\ref{lem:alt_thin}, consider $x \in \conf{A}$, and $\id_x \subseteq^+
\theta \in \tilde{\A}_-$. But $\id_x$ is in all three isomorphism
families and in particular in $\tilde{A}_+$, so $\theta \in
\tilde{\A}_+$ as well by \emph{(d)}. Hence, $\theta = \id_y$ for some
$y\in \conf{A}$ by \emph{(b)}.

Secondly, it follows from the axiom that any $\theta \in \A$ can be
factored uniquely:

\begin{lem}[Decomposition lemma]\label{lemma:decomp}
  For $\A$ a tcg, we have an order-isomorphism:
$$\begin{aligned}
&\tilde A_-  \times _A \tilde A_+  &&\xrightarrow{} &&\tilde A \\
&( \theta^-,  \theta^+) &&\mapsto  &\theta^- &\circ  \theta^+
\end{aligned}$$
where
$\tilde A_-  \times_A  \tilde A_+ = \{ ( \theta^-,  \theta^+)  \in
\tilde A_-  \times  \tilde A_+\mid
\codom  \theta^+ = \dom  \theta^- \}$
is ordered by pairwise inclusion and $\tilde A$ is ordered by
inclusion.
\end{lem}
\begin{proof}
  The map is clearly well defined because $\tilde A_-$ and
  $\tilde A_+$ are included in $\tilde A$.

\emph{Injectivity.}
   Assume we have
    $ \theta  =  \theta _1^- \circ  \theta _1^+ =  \theta _2^- \circ
\theta _2^+ : x  \isf _{\tilde A} y$. 
%
  By using groupoid laws we get that
  $ \theta _1^+ \circ ( \theta _2^+)^{-1} =  (\theta _1^-)^{-1} \circ
\theta _2^- : z_2  \isf  z_1  \in 
  \tilde A_- \cap \tilde A_+$
  hence both are equal to the identity.

\emph{Surjectivity.} By induction on $ \theta   \in  \tilde A$ we build
a preimage. Assume we have the decomposition of $\theta = \theta^- \circ
\theta^+$, and that $\theta$ extends to $ \theta ' : x'  \isf  y'$ by a
pair of fixed polarity, say positive.
We use the extension axiom on $ \theta ^-$ to get
$ \theta ^- \subseteq  \theta '^- : z'  \isf_{\tilde A_-}  y'$. It
follows that
  $ \theta ^+ \subseteq ( \theta '^-)^{-1} \circ  \theta ' : x'  \isf
_{\tilde A} z'$ is a
  positive extension of $ \theta ^+$ so it must belong to $\tilde A_+$
by \emph{(d)}. Hence
  $ \theta ' =  \theta '^- \circ \left(( \theta '^-)^{-1} \circ  \theta
'\right)$
  provides the required decomposition.

\emph{Monotonicity.} Monotonicity of the decomposition follows from
uniqueness.
\end{proof}

So the full $\tilde{A}$ is actually redundant,
and can be uniquely recovered from $\tilde{A}_-$ and
$\tilde{A}_+$. This structure of a game with two isomorphism families
is strongly reminiscent of Melli\`es' earlier approach to uniformity by
bi-invariance under \emph{two} group actions \cite{ag1}. This suggests
that there is something intrinsic in this decomposition of symmetry as
compositions of Opponent's reindexings and Player's reindexing. This
structure, that alternation and sequentiality allows to keep hidden in
AJM's equivalence relations, seems to become inevitable in a thin treatment of
symmetry in games where sequentiality is not hard-wired.

Before going on to the proof of congruence, as we are finally 
in a position to formalize what we mean by \emph{strategy with
symmetry}, we start by doing that.

\begin{defi}
Let $\A$ be a tcg. A \textbf{pre-$\sim$-strategy} on $\A$ is a thin,
$\sim$-receptive map of essp $\sigma : \S \to \A$.

It is a \textbf{$\sim$-strategy} if it is also
receptive and courteous (\emph{i.e.} $\sigma : S \to A$ is a strategy).
\end{defi}

We are mainly interested in the interaction and composition of
$\sim$-strategies, but we perform some of the developments with the
slightly more permissive conditions of pre-$\sim$-strategies as it is
occasionally useful (in particular when we model state in Section
\ref{subsec:state}) to be able to compose ``strategies'' that are not
courteous or receptive.

\subsubsection{Congruence.} We start by stating the property
ensuring congruence: the \emph{bi-pullback property}.
Our next objective is to prove it for all interactions between
pre-$\sim$-strategies.

\begin{defi}[Bi-pullback property]\label{def:bip}
  Let $ \sigma : \S \rightarrow \A^\perp  \parallel  \B$ and
  $ \tau : \T \rightarrow \B^\perp  \parallel  \C$ be $\sim$-receptive
  maps of essps. Their interaction has the \textbf{bi-pullback property} iff 
  the pullback
$$\xymatrix@R=10pt@C=0pt{
& \T  \circledast  \S 
        \ar[dl]^{\Pi_1}
        \ar[dr]_{\Pi_2} 
        \pb{270}\\
\S  \parallel  \C
        \ar[dr]_{\sigma\parallel \C}&&
\A  \parallel  \T 
        \ar[dl]^{\,\,\,\A \parallel \tau}  \\
& \A \parallel \B \parallel \C
}$$
is a bi-pullback.
 \end{defi}


In essence, the definition of interaction of strategies via pullback
expresses that given $x_S \in \conf{S}$ and $x_T \in \conf{T}$ that
match on $B$ (\emph{i.e.} $\sigma x_S = x_A \parallel x_B$ and $\tau
x_T = x_B \parallel x_C$) and are causally compatible (see Proposition
\ref{prop:rep_int}), we can form their synchronization uniquely as some
$z \in \conf{T\inter S}$ such that $\Pi_1 z = x_S \parallel x_C$ and
$\Pi_2 z = x_A \parallel x_T$. The bipullback property extends this method
to construct synchronized states: it asserts in essence that if such
$x_S \in \conf{S}$ and $x_T \in \conf{T}$ only match up to symmetry
(\emph{e.g.} they fail to agree because of copy index mismatches), one
can always find a ``common ground'', both strategies changing their
choice of symmetric events, yielding some $x_S \isf_{\tilde{S}} x'_S$
and $x_T \isf_{\tilde{T}} x'_T$ where $x'_S$ and $x'_T$ now match on
the nose and can be synchronized directly.

This process of tweaking ``copy indices'' to find a common ground can be
done interactively, and -- beyond the bi-pullback property -- is at the heart of the
congruence problem. To illustrate that and showcase the inductive process at
play, we include the following example of transporting a weak equivalence through
composition. 

\begin{exa}
  Consider the two following strategies on $\oc \llbracket \com \rrbracket$

  $$\xymatrix@R=5pt{
     \sigma _1:\oc  \llbracket \com \rrbracket & &  \sigma _2:\oc
\llbracket  \com  \rrbracket \\
    \ar@{-|>}[d]\run^{-, i} & & \run^{-, i}\ar@{-|>}[d] \\
    \done^{+, 0}\ar@{--}@/^/[u] & & \done^{+, 1}\ar@{--}@/^/[u]
  }$$
  only differing by the choice of copy index for $\done$. There is an
  obvious weak equivalence between them,
  call $\varphi : S_1 \to S_2$ the obvious invertible map.
  Consider now the following strategy $ \tau $ (which, in IPA,
represents $x:\com \vdash x; x : \com$):
$$\xymatrix@R=5pt{
  \oc \llbracket  \com \rrbracket \ar@{}[rr]|\to|| &   & \oc \llbracket
\com  \rrbracket \\
  & & \run^{-, i} \ar@{-|>}[dll]\ar@/^2pc/@{--}[ddddd]\\
  \run^{+,  \langle i, 0 \rangle } \ar@{-|>}[d] \\
  \done^{-, j} \ar@{-|>}[d]\ar@{--}@/^/[u] \\
  \run^{+,  \langle i, j+1 \rangle } \ar@{-|>}[d] \\
  \done^{-, k} \ar@{-|>}[drr]\ar@{--}@/^/[u] \\
  & & \done^{+, \langle j, k \rangle } }$$

In order to build a weak equivalence between the resulting compositions
$\tau \odot \sigma_1$ and
$\tau \odot \sigma_2$, a reasonable first step is to build a weak
equivalence between the
interactions $\tau \circledast \sigma_1$ and $\tau \circledast
\sigma_2$. In particular, given a
configuration of $T \circledast S_1$, we should be able to build a
corresponding configuration of $T
\circledast S_2$. Consider \emph{e.g.} the following configuration of
$T \circledast S_1$.

$$\xymatrix@R=5pt{
  \oc \llbracket  \com \rrbracket \ar@{}[rr]|\to|| &   & \oc \llbracket
\com  \rrbracket \\
  & & \run^{-, i} \ar@{-|>}[dll]\ar@/^/@{--}[ddddd]\\
  \run^{\langle i, 0 \rangle } \ar@{-|>}[d] \\
  \done^{0} \ar@{-|>}[d]\ar@{--}@/^/[u] \\
  \run^{\langle i, 1 \rangle } \ar@{-|>}[d] \\
  \done^{0} \ar@{-|>}[drr]\ar@{--}@/^/[u] \\
  & & \done^{+, \langle 0, 0 \rangle } }$$
where events on the left hand side are drawn without polarity, as they
are synchronised between $\sigma_1$
and $\tau$. By projections, we get configurations $x\in \conf{S_1 \parallel
\oc \intr{\com}}$ and $y \in \conf{T}$ such that
$(\sigma_1 \parallel \oc \intr{\com})\,x = \tau\,y$
and such that the induced bijection is secured.

In order to construct a configuration in $T \circledast S_2$, it is
natural to try and replace $x$
with $\varphi(x)$ -- and that would work out if $\varphi$ was a
\emph{strong} equivalence. But as it
is only a weak equivalence, we do not have $(\sigma_2 \parallel \oc
\intr{\com})\,(\varphi\,x) =
\tau\,y$, only
\[
(\sigma_2 \parallel \oc \intr{\com})\,(\varphi\,x) \isf_{\tilde{\oc
\intr{\com}\parallel \oc
\intr{\com}}} \tau\,y
\]

Here we observe the phenomenon hinted at above: the need to extract
from $\varphi\,x\in \conf{S_2}$ and $y\in \conf{T}$, only matching up
to symmetry, a valid configuration of $T \circledast S_2$. For our
example, the only possibility is:

$$
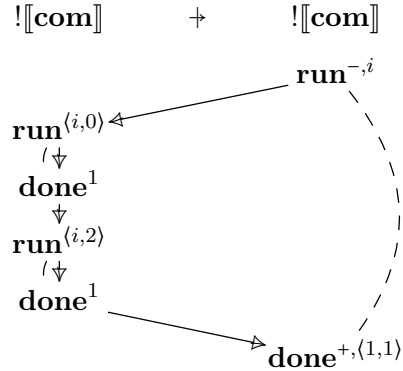
\xymatrix@R=5pt{
  \oc \llbracket  \com \rrbracket \ar@{}[rr]|\to|| &   & \oc \llbracket
\com  \rrbracket \\
  & & \run^{-, i} \ar@{-|>}[dll]\ar@/^2pc/@{--}[ddddd]\\
  \run^{\langle i, 0 \rangle } \ar@{-|>}[d] \\
  \done^{1} \ar@{-|>}[d]\ar@{--}@/^/[u] \\
  \run^{\langle i, 2 \rangle } \ar@{-|>}[d] \\
  \done^{1} \ar@{-|>}[drr]\ar@{--}@/^/[u] \\
  & & \done^{+, \langle 1, 1 \rangle } }$$

It appears that \emph{both} $\varphi\,x$ and $y$ had to change, in
order to find an agreement as to
the choice of copy indices. To compute it,
we replay the interaction up to the first disagreement between $\sigma_1$
and $\sigma_2$. By hypothesis, this disagreement yields symmetric
configurations of the game. Hence by $\sim$-receptivity, $\tilde{T}$
comprises a bijection:
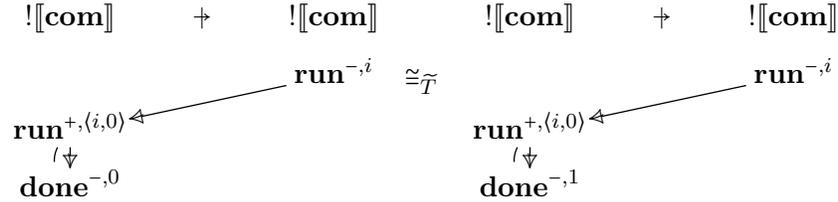
\[
\raisebox{25pt}{\xymatrix@R=5pt{
  \oc \llbracket  \com \rrbracket \ar@{}[rr]|\to|| &   & \oc \llbracket
\com  \rrbracket \\
  & & \run^{-, i} \ar@{-|>}[dll]\\
  \run^{+,  \langle i, 0 \rangle } \ar@{-|>}[d] \\
  \done^{-, 0} \ar@{--}@/^/[u] 
}}
~~~~~\isf_{\tilde{T}}~~~~~
\raisebox{25pt}{
\xymatrix@R=5pt{
  \oc \llbracket  \com \rrbracket \ar@{}[rr]|\to|| &   & \oc \llbracket
\com  \rrbracket \\
  & & \run^{-, i} \ar@{-|>}[dll]\\
  \run^{+,  \langle i, 0 \rangle } \ar@{-|>}[d] \\
  \done^{-, 1} \ar@{--}@/^/[u] 
}}
\]

By (Extension) in $\tilde{T}$, we know that this bijection can be
extended to some:
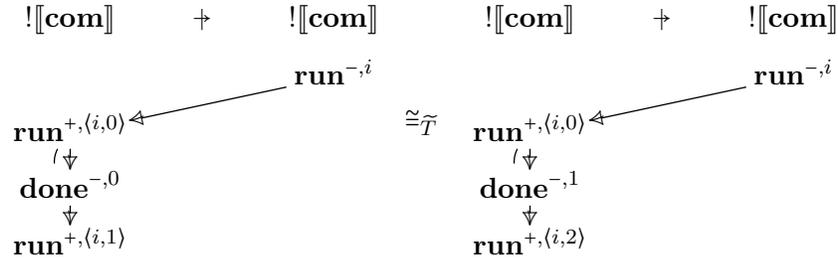
\[
\raisebox{40pt}{\xymatrix@R=5pt{
  \oc \llbracket  \com \rrbracket \ar@{}[rr]|\to|| &   & \oc \llbracket
\com  \rrbracket \\
  & & \run^{-, i} \ar@{-|>}[dll]\\
  \run^{+,  \langle i, 0 \rangle } \ar@{-|>}[d] \\
  \done^{-, 0} \ar@{--}@/^/[u] \ar@{-|>}[d]\\
  \run^{+, \tuple{i,1}}
}}
~~~~~\isf_{\tilde{T}}~~~~~
\raisebox{40pt}{
\xymatrix@R=5pt{
  \oc \llbracket  \com \rrbracket \ar@{}[rr]|\to|| &   & \oc \llbracket
\com  \rrbracket \\
  & & \run^{-, i} \ar@{-|>}[dll]\\
  \run^{+,  \langle i, 0 \rangle } \ar@{-|>}[d] \\
  \done^{-, 1} \ar@{--}@/^/[u] \ar@{-|>}[d]\\
  \run^{+,\tuple{i,2}}
}}
\]

Likewise, by $\sim$-receptivity of $\sigma_2 \parallel \oc \intr{\com}$
this extension is lifted to
$\tilde{S_2} \parallel \tilde{\oc \intr{\com}}$, and we then apply
(Extension) on $\tilde{S_2}$. And
the process goes on, interactively between $\sigma_2$ and $\tau$, until
we get
$x' \isf_{\tilde{S}\parallel \tilde{\oc \intr{\com}}}
\varphi\,x$ and $y' \isf_{\tilde{T}} y$ such that $(\sigma_2 \parallel
\oc \intr{\com})\,x' =
\tau\,y'$ (which in our example, is the configuration of the
interaction represented above).
\end{exa}

Formalizing this interactive process of using $\sim$-receptivity on one strategy
and extension on the other yields the following lemma:

 \begin{lem}[Weak bipullback property]\label{lemma:pb2}
   Let $ \sigma  : \S  \rightarrow  \A$ and $ \tau  : \T  \rightarrow
\A^\perp$ be pre-$\sim$-strategies.
   Let $x \in \mathscr{C} (S)$ and $y \in \mathscr{C} (T)$ and
   $ \theta : \sigma x \isf _{\tilde A} \tau y$, such that the
composite bijection
   \[
   x \stackrel{\sigma}{\sbij} \sigma\,x
\stackrel{\theta}{\isf_{\tilde{A}}} \tau\,y
\stackrel{\tau}{\sbij} y
   \]
   is secured. Then, there exists
   $z \in \mathscr{C} (S \wedge T)$ along
   with $ \theta _S : x \isf _{\tilde{S}} \Pi _1 z$ and
   $ \theta _T: \Pi _2 z \isf _{\tilde T} y$, such that
   $ \tau \theta _T \circ \sigma \theta _S = \theta $. Moreover, $z$ is
unique up to symmetry.
 \end{lem}
 \begin{proof}
   \emph{Uniqueness.} Assume we have such $(z, \theta_S, \theta_T)$ and
   $(z', \theta_S', \theta_T')$. Then it is easy to see that
   $ \theta'_S \circ \theta_S^{-1}:\Pi _1 z \isf _{\tilde S} \Pi _1 z'$
   and similarly
   $ \theta'_T \circ \theta_T^{-1} : \Pi _2 z \isf _{\tilde T} \Pi _2
z'$.
   Those match on the game $\A$, so they
   induce a $z \isf z'$ in $\tilde S \wedge \tilde T$ as desired.

   \emph{Existence.} We proceed by induction on $ \theta $; the
   base case is trivial. Assume $ \theta $ extends by $( \sigma s,
\tau t)$ to
   $ \theta ' :  \sigma x'  \isf   \tau y'$. For instance, $s$ is
positive. We have
   $ \theta _S : x  \isf   \Pi _1 z$ and $x$ can be extended to $x'$ by
$s$, so by the
   extension property of the symmetry $ \theta _S$ extends to
   $ \theta _S' : x'  \isf  z_S'$. This means that $ \tau\,\theta _T$
can be extended by
   symmetric negative (for $T$) events so by $\sim$-receptivity, $
\theta _T$
   can extend to $ \theta _T': z_T'  \isf _{\tilde T} y'$, with $
\sigma z_S' =  \tau z_T'$ by
   construction. Since the bijection $z_S'  \isf  z_T'$ is obviously
   secured, we get $z'  \in   \mathscr{C} (S  \wedge  T)$ that
satisfies
   our property.
 \end{proof}

 Note that we did not need that $ \sigma $ and $ \tau $ are thin --
 only $\sim$-receptivity. This statement is a step in
 the right direction, however the non-uniqueness of $z$ (only up to
 symmetry) is problematic: it cannot be used to build maps, in
 particular it cannot be used to lift a weak equivalence $ \sigma
\rightarrow   \sigma '$ to $ \tau   \circledast   \sigma   \rightarrow   \tau   \circledast
\sigma '$. However, we will see now that if the interacting strategies
are \emph{thin}, we can use Lemma \ref{lemma:pb_thin} to ``tighten the
screws'' and show that then the choice of $z$ is unique; from this
observation the required map will follow.

We now prove the main technical result of this section, the \emph{bi-pullback
lemma}.

\begin{lem}\label{lem:pib}
Let $\sigma : \S \to \A^\perp \parallel \B$ and $\tau : \T \to \B^\perp
\parallel \C$ playing on tcgs. Then, their interaction has the
bi-pullback property.
\end{lem}
\begin{proof}
Recall that for $f : \X \rightarrow \S  \parallel  \C$,
$g : \X \rightarrow \A  \parallel  \T$ such that
  $ \tau \circ g \sim_{\tilde A  \parallel  \tilde B  \parallel  \tilde
C} \sigma \circ f$, we need $ h : \X \rightarrow \T
\circledast  \S$, unique \emph{up to symmetry}, such that $ \Pi_1 \circ
h \sim_{\tilde{S}\parallel \tilde{C}} f$ and
  $ \Pi_2 \circ h \sim_{\tilde{A}\parallel \tilde{T}} g$.

\emph{Uniquess} up to symmetry follows from $ \Pi_1 \circ
h \sim_{\tilde{S}\parallel \tilde{C}} f$ and
  $ \Pi_2 \circ h \sim_{\tilde{A}\parallel \tilde{T}} g$ and
definition of the symmetry on the pullback. The main difficulty is
\emph{existence}. As hinted above, the trick is to apply Lemma
\ref{lemma:pb2}, not on the raw interaction pullback, but on that
between:
\[
\begin{aligned}
 (\sigma   \parallel  (\C^\perp)_-) &: \S  \parallel  (\C^\perp)_-
\rightarrow  \A^\perp  \parallel  \B  \parallel  \C^\perp \\
(\A_-   \parallel  \tau ) &:\A_-  \parallel  \T  \rightarrow  \A
\parallel  \B^\perp  \parallel  \C
   \end{aligned}
\]

Write $\T \ointer \S$ for this pullback, with projections $\Pi_1 : \T
\ointer \S \to \S \parallel (\C^\perp)_-$ and $\Pi_2 : \T \ointer \S
\to \A_- \parallel \T$. The underlying event structure of $\T \ointer
\S$ is the same as for $\T \inter \S$, but the symmetry is tighter:
intuitively, it is that where the external Opponent \emph{does not
change their copy indices}. In fact, the two maps above are
\emph{thin}, so by Lemma \ref{lemma:pb_thin}, the symmetry of $\T
\ointer \S$ is very tight indeed: it is restricted to identities.
Nevertheless, for $x\in \conf{X}$ we can apply Lemma \ref{lemma:pb2} 
to $f\,x  \in   \mathscr{C} (S  \parallel  C)$ and $g\,x  \in
\mathscr{C} (A  \parallel  T)$, and get $z\in   \mathscr{C} (T
\circledast  S)$ -- but its uniqueness up to symmetry now holds in
$\T\ointer \S$ with trivial symmetry, so $z$ is unique.
By uniqueness, this association induces a function
   $ \psi : \mathscr{C} (X) \rightarrow \mathscr{C} (T \circledast S)$
   such that $ \Pi _1\,(\psi\,x) \isf_{\tilde{S}\parallel \tilde{C}}
f\,x$ and $ \Pi _2\,(\psi\,x) \isf_{\tilde{A}\parallel \tilde{T}}
g\,x$. It is then routine to verify that this function is monotonic,
preserves cardinality and unions (hence it is generated by a map of
event structures); and that it preserves symmetry -- details are
omitted.
\end{proof}

This concludes our proof of congruence for interactions, \emph{i.e.} the
following corollary, simply proved by applying the bi-pullback property.

\begin{cor}
Weak equivalence is preserved by interactions of pre-$\sim$-strategies
on thin concurrent games.
\end{cor}

It will of course follow immediately, once composition is defined in
Section \ref{subsubsec:composition}, that it preserves weak equivalence as well.

\subsubsection{Weak isomorphism.}\label{subsubsec:weak_iso}
 Weak equivalence is a natural lax
version of the isomorphism of strategies in $\CG$ in the presence of
symmetry: all equalities in the definition of isomorphism are replaced
with symmetry. However, as the reader may have noticed, there is a
mismatch between weak equivalence and the \emph{weak isomorphisms} of
Definition \ref{def:weak_mor}, used in Section \ref{sec:replication} to
compare strategies up to copy indices.
Let us start this discussion by recasting weak isomorphism in the
context of thin concurrent games.

\begin{defi}
Let $\sigma : \S \to \A$ and $\sigma' : \S' \to \A$ be maps of ess,
where $\A$ is a tcg. A \textbf{positive morphism} from $\sigma$ to
$\sigma'$ is a map of ess $f : \S \to \S'$ such that $\sigma' \circ f
\sim_{\A_+} \sigma$.

We say that $f$ is a \textbf{weak isomorphism} if it is furthermore
invertible \emph{on the nose}, \emph{i.e.} there is $g : \S' \to \S$
such that $g \circ f = \id_\S$ and $f \circ g = \id_{\S'}$ -- we write
$\sigma \wiso \sigma'$ for the corresponding equivalence relation.
\end{defi}

The thin concurrent games presented in \cite{lics15} relied solely
on weak equivalence to compare strategies up to symmetry; its
refinement with \emph{weak isomorphism} presented here came later.
At first sight it looks like in switching from weak equivalence to weak
isomorphisms we are trading an arguably mathematically canonical notion for one
that is more concrete, but also possibly more ad-hoc. Of course, if weak
isomorphism is a congruence, then the change is convenient.
Indeed, having mediating maps be inverses on the nose makes the
equivalence more conservative over plain event
structures: if $\sigma \wiso \sigma'$ as above, then $S$ and $S'$ are
isomorphic as plain event structures (though of course the projection to the game
does not commute on the nose). The tighter the equivalence is, the easier it is
to transport properties and structure across.
But in fact, we will see in this section that it is not a compromise at
all: if $\sigma : \S \to \A$ and $\sigma' : \S' \to \A$ are
$\sim$-strategies, then they are weakly isomorphic if and only if they
are weakly equivalent\footnote{For that, thinness plays a crucial role:
if $\sigma : \S \to \A^\perp \parallel \B$ and $\tau : \T \to \B^\perp
\parallel \C$ are pre-$\sim$-strategies, their composition via pullback
and pseudo-pullback (as in Section \ref{subsubsec:sat_unif}) are weakly
equivalent as both are given by a bi-pullback, but they are certainly not weakly
isomorphic.}!

In order to prove that, the first step is to observe that the mediating maps
being inverses on the nose comes for free, provided one insists on using
\emph{positive} weak equivalences.

\begin{lem}\label{lem:pos_iso}
Let $\sigma : \S \to \A$ and $\sigma' : \S' \to \A$ be
pre-$\sim$-strategies, and $f : \S \to \S', g : \S' \to \S$ forming a
\emph{positive} weak equivalence, \emph{i.e.} $\sigma' \circ f
\sim_{\A_+} \sigma$ and $\sigma \circ g \sim_{\A^+} \sigma'$.

Then, $f$ and $g$ are actually inverse on the nose (and so form a weak
isomorphism).
\end{lem}

The proof of that is obvious in the light of the following lemma, which
shows plainly the phenomenon at play.

\begin{lem}\label{lemma:pos_symm}
Let $\sigma : \S \to \A$ be a pre-$\sim$-strategy on a tcg $\A$, and
let $\theta : x \isf_{\S} y$ such that $\sigma \theta \in \tilde{A}_+$.
Then, $x = y$ and $\theta = \id_x$.
\end{lem}
\begin{proof}
By induction on $ \theta $. If $ \theta $ is empty, it is clear. For $\id_x
\longcov {(s,s')}  \theta   \in  \tilde S$. If $s$ and $s'$ are positive, then by
thin $s = s'$. If negative, then $\sigma \theta \in \tilde{A}_+ \cap
\tilde{A}_-$, hence is an identity. So $\theta$ is a negative extension of
$\id_x$, whose image in $A$ is an identity; hence it is an identity by
$\sim$-receptivity.
\end{proof}

It is of course not the case that every weak equivalence is positive.
However, every map between pre-$\sim$-strategies is symmetric to one
which does preserve the projection on the game up to positive symmetry:
intuitively, if $f$ sends negative $s^- \in S$ to $f s$ with a different ``copy
index'', we set $f' s$ to the unique matching move positively symmetric to $f s$.

\begin{restatable}{lemma}{positivisation}\label{lem:positivisation}
Let $\sigma : \S \to \A$ and $\sigma' : \S' \to \A$ be pre-$\sim$-strategies, and
$f : \S \to \S'$ such that $\sigma' \circ f \sim_{\tilde{A}} \sigma$. Then, there
exists a unique $f' : \S \to \S'$ such that $f \sim_{\tilde{S'}} f'$,
and $\sigma' \circ f' \sim_{\tilde{A}_+} f$.
\end{restatable}
\begin{proof}[Sketch]
For $x\in \conf{S}$, the hypotheses give us $\theta_x : \sigma' (f x)
\isf_{\tilde{A}} \sigma x$, which we need to make positive. For that,
we first use Lemma \ref{lemma:decomp} to decompose $\theta_x$ as
\[
\sigma' (f x) \stackrel{\theta_x^-}{\isf_{\tilde{A}_-}} y
\stackrel{\theta_x^+}{\isf_{\tilde{A}_+}} \sigma x
\]
such that $\theta_x = \theta_x^+ \circ \theta_x^-$.
The key idea is then to \emph{transport} $f x$ over
this negative symmetry $\theta_x^-$, yielding $x' \isf_{\tilde{S'}} f
x$ such that $\sigma' x' \isf_{\tilde{A}_+} y$ (which can be done by
induction on $f x$ and $\theta_x^-$), so
that $\sigma x \isf_{\tilde{A}_+} \sigma' x'$ as well. We then set
$f' x$ to be $x'$, and extract from this a map $f' : \S
\to \S'$. 

The details appear in Appendix \ref{app:positivisation}. 
\end{proof}

This is a rather powerful result: it entails that every symmetry class of
such weak maps from $\sigma$ to $\sigma'$ has a canonical
representative, namely the unique equivalent map for which the
projection to the game commutes up to positive symmetry.

As an immediate corollary, we have:

\begin{cor}
Let $\sigma : \S \to \A$ and $\sigma' : \S' \to \A$ be pre-$\sim$-strategies.

Then, they are weakly isomorphic if and only if they are weakly equivalent.
\end{cor}
\begin{proof}
If $\sigma$ and $\sigma'$ are weakly equivalent, from Lemma
\ref{lem:positivisation} one easily gets a \emph{positive} weak equivalence. By
Lemma \ref{lem:pos_iso}, it is a weak isomorphism. The other direction
is trivial.
\end{proof}

As mentioned earlier, from Lemma \ref{lem:pib} it will be obvious once
composition is defined that it preserves weak equivalence. But from the
above, each weak equivalence is canonically
represented by a weak isomorphism; the induced equivalence relation on
$\sim$-strategies is the same. In particular, it will follow just as
directly that weak isomorphism is a congruence.

\subsection{Categorical structure.} 
\label{subsec:compact} Since the beginning of Section
\ref{sec:gameswithsymmetry} we have focused on the crucial problem of
\emph{congruence}, which imposes the most constraints on the design of games with
symmetry. Now that this is solved, we unfold the rest of the work required to
build a core setting for game semantics, \emph{i.e.} a \emph{compact
closed category} of tcgs and $\sim$-strategies.

\subsubsection{Composition.}\label{subsubsec:composition}
 As the previous section focused on interaction,
it makes sense to start the construction by completing it
to get \emph{composition}. Composition of pre-$\sim$-strategies will
be defined by simply enriching the composition of Section
\ref{subsubsec:hiding} with symmetry.

Let $\sigma : \S \to \A^\perp \parallel \B$ and $\tau : \T \to \B^\perp \parallel
\C$ be pre-$\sim$-strategies.
Ignoring symmetry, recall from Section \ref{sec:replication} that
$\tau \odot \sigma : T\odot S \to A^\perp \parallel C$ is obtained
using \emph{projection} (Definition \ref{def:proj}):
given $V = \{ p \in T \circledast S \mid (\tau \circledast \sigma)\,p \not\in
B\}$
we set $T\odot S = T\circledast S \proj V$, and $\tau \odot \sigma$ to be the
corresponding
restriction of $\tau \circledast \sigma$. We now extend this in the
presence of symmetry.

\begin{lem}\label{lem:proj_symm}
Let $\E$ be an ess and $V \subseteq E$ \emph{closed under symmetry}, in
the sense that for all $\theta : x \isf_{\tilde{E}} y$, for all $e\in V\cap x$,
we have
$\theta\,e\in V$ as well.
Then, defining
\[
\tilde{E}\proj V = \{\theta : x \bij y \mid x, y\in \conf{E\proj V},~\exists
\theta\subseteq
\theta' \in \tilde{E},~\theta' : [x]_E \isf_{\tilde{E}} [y]_E\}
\]
we have that $\tilde{E}\proj V$ is an isomorphism family, making $\E \proj V =
(E\proj V,
\tilde{E}\proj V)$ into an event structure with symmetry.
\end{lem}
\begin{proof}
As usual the axiom (Groupoid) is clear. In this proof we abbreviate $[x]_E$ to
$[x]$ for $x \in
\conf{E\proj V}$ for clarity reasons.

\emph{(Restriction)} Let $\theta : x \bij y \in \tilde{E}\proj V$, and $x_0 \in
\conf{E\proj V}$
such that $x_0 \subseteq x$. By definition there is $\theta \subseteq \theta' :
[x] \isf_{\tilde{E}} [y]$.
We have $[x_0] \subseteq [x]$. Therefore, by (Restriction) on
$\tilde{E}$ we have $\theta'_0 \subseteq \theta'$ with $\theta'_0 : [x_0]
{\isf_{\tilde{E}}} y'_0$.
Since $V$ is closed under symmetry, $\theta'_0 \cap V^2 : x_0 \bij y'_0 \cap V$
is still a
bijection, which by definition is in $\tilde{E}\proj V$. It is clear by
construction that
$\theta'_0 \cap V^2 \subseteq \theta$.

\emph{(Extension)} Let $\theta : x \bij y \in \tilde{E}\proj V$, and $x \subseteq
x_0 \in
\conf{E\proj V}$. By definition there is $\theta \subseteq \theta' : [x]
\isf_{\tilde{E}} [y]$. We
have $[x] \subseteq [x_0] \in \conf{E \proj V}$, therefore by (Extension) for
$\tilde{E}$ there is
$\theta'_0 : [x_0] \isf_{\tilde{E}} y'_0$. Again since $V$ is closed under
symmetry,
$\theta'_0 \cap V^2 : x_0 \bij y'_0 \cap V$ is still a bijection. By definition
it is in
$\tilde{E}\proj V$, and by construction it contains $\theta$.
\end{proof}

Given $\sim$-receptive $\sigma : \S \to \A^\perp \parallel \B$ and $\tau : \T \to
\B^\perp \parallel \C$ (where $\A, \B$ and $\C$ are tcgs), 
$V = \{ p \in S \circledast T \mid (\tau \circledast \sigma)\,p \not\in B\}$
is closed under symmetry; we can thus apply Lemma \ref{lem:proj_symm}.
Accordingly we set $\tilde{T} \odot \tilde S$ as $\tilde{T} \circledast
\tilde{S} \proj V$,
\emph{i.e.} comprising bijections $ \theta : x \bij y$ such that there is
\[
\theta \subseteq \bar{ \theta } : [x]_{T \circledast S} \isf
_{\tilde{T} \circledast \tilde{S}} [y]_{T \circledast S}.
\]

This makes $\T \odot \S = (T \odot S, \tilde{T}\odot \tilde{S})$ an event
structure with symmetry.
In fact, we will show in Lemma \ref{lemma:uniq_witness} that if $\S$ and $\T$ are
\emph{thin}, the witnessing symmetry $\bar{\theta}$ is unique.

Summing up, we state:

\begin{lem}
  If $\sigma : \S \to \A^\perp \parallel \B$ and $\tau : \T \to \B^\perp
\parallel \C$ are
  pre-$\sim$-strategies, then
  \[
    \tau \odot \sigma : \T \odot \S \to \A^\perp \parallel \C
  \]
  is a map of ess.
\end{lem}
\begin{proof}
We prove that $\tau \odot \sigma$ preserves symmetry. Let
$\theta : x \isf_{\tilde{T}\odot \tilde{S}} y$.
By definition, there is $\theta \subseteq \bar{\theta} : [x]
\isf_{\tilde{T}\circledast \tilde{S}}
[y]$. Then, $(\tau \circledast \sigma)\,\bar{\theta}$ is some
\[
\theta_A \parallel \theta_B \parallel \theta_C : x_A \parallel x_B \parallel x_C
\isf_{\tilde{A}\parallel \tilde{B} \parallel \tilde{C}} y_A \parallel y_B
\parallel y_C
\]
since $(\tau \circledast \sigma)$ preserves symmetry. But then $(\tau \odot
\sigma)\,\theta$ is
\[
\theta_A \parallel \theta_C : x_A \parallel x_C \isf_{\tilde{A}\parallel
\tilde{C}} y_A \parallel
y_C
\]
which is a valid symmetry in $\tilde{A} \parallel \tilde{C}$ as required.
\end{proof}

In order to get a notion of composition for $\sim$-strategies, we need to show
that composition preserves thinness, and $\sim$-receptivity. We will treat them
in that order.

The preservation of thinness under composition boils down to one crucial
property: the fact that any symmetry between configurations of the composition
has a \emph{unique} witness in the interaction. Indeed, recall from Lemma
\ref{lemma:pb_thin} that the closed interaction between dual thin maps has a
trivial symmetry. Of course composition is obtained via an \emph{open}
interaction, which does \emph{not} have trivial symmetry as the external Opponent
can contribute new isomorphic pairs of negative events. Nevertheless, 
whenever the interaction stays within $B$ the phenomenon above applies, and the
symmetry is fixed: only the external Opponent
can put in relation two non-identical events first. As a result, a bijection in
the symmetry of the interaction is fully determined by its restriction
to visible events:

 \begin{lem}[Unique witness]\label{lemma:uniq_witness}
   Let $ \sigma  : \S  \rightarrow  \A^\perp  \parallel  \B$ and $ \tau  : \T
\rightarrow
\B^\perp \parallel  \C$ be pre-$\sim$-strategies. Recall the set of visible
events of the
interaction:
\[
V = \{e \in T\circledast S\mid (\tau \circledast \sigma)\,e \not \in B\}
\] 

   Let $ \theta  : x  \isf_{\tilde{T}\circledast \tilde{S}}  y$ and $ \theta ' :
x
\isf_{\tilde{T} \circledast \tilde{S}}  y'$ such that
   $ \theta  \cap V^2 =  \theta ' \cap V^2$. Then $ \theta  =  \theta '$.
 \end{lem}
 \begin{proof}

   By hypothesis, we have that $y \cap V = y' \cap V$.  Note that
   $ \theta \circ \theta '^{-1} : y' \isf y \in (\tilde{S} \parallel
   \tilde{C}) \wedge (\tilde{A} \parallel \tilde{T})$ and contains
   $\id_{y \cap V}$. So necessarily, the projection of $\theta \circ \theta
'^{-1}$ to $\A$ and
   $\C$ is an identity bijection. As a result, the symmetry $ \theta \circ \theta
'^{-1}$ actually
   belongs to $(\tilde{S} \parallel (\C_+)^\perp) \wedge (\A_- \parallel
   \tilde{T})$. This is a pullback of pre-$\sim$-strategies, so
   $ \theta \circ \theta '^{-1}$ is an identity by Lemma \ref{lemma:pb_thin}, so
   $\theta = \theta'$.
 \end{proof}

 Using this, we can prove that thinness is stable
 under composition.

\begin{lem}\label{lem:comp_thin}
  For $ \sigma  : \S  \rightarrow  \A^\perp  \parallel  \B$ and
  $ \tau  : \T  \rightarrow  \B^\perp  \parallel  \C$ pre-$\sim$-strategies,
  $\tau \odot \sigma $ is thin.
\end{lem}
\begin{proof}
  Let $z \in \mathscr{C} (T \odot S)$ such that $\id_z$ extends by
  positives $(e, e')$ to $ \theta : x \isf y \in \tilde {T} \odot \tilde{S}$
  with witness $\bar{ \theta } : [x] \isf _{T \circledast S} [y]$.
  Write $ \theta _0$ for
  $\bar{ \theta } \setminus\{(e, e')\} : x_0  \isf  y_0$. By hypothesis,
  $ \theta _0$ behaves like the identity on the visible part of $x_0$. Hence,
  by Lemma \ref{lemma:uniq_witness}, $ \theta _0$ is the identity on $x_0$.

  Since $\id_{x_0} = \theta _0$ can be extended by $(e, e')$ to
  $\bar{ \theta }$ which is positive in $T \odot S$ we can assume
  eg. $ \Pi _2\,e$ and $ \Pi _2\,e'$ are positive in $T$. Hence
  $ \Pi _2\,\theta _0$ (which is also an identity) extends by positive
  $( \Pi _2\,e, \Pi _2\,e')$. Since $ \tau $ is thin, we have
  $ \Pi _2\,e = \Pi _2\,e'$ from which $e = e'$ follows ($e$ and $e'$
  are positive), as desired.
\end{proof}

We now focus on $\sim$-receptivity. Unlike thinness, it turns out that
$\sim$-receptivity is not preserved by composition without further hypotheses,
so pre-$\sim$-strategies are not stable under composition. To ensure preservation
of $\sim$-receptivity one needs courtesy, however it is sometimes necessary to
consider ``strategies'' that are not quite $\sim$-strategies -- in particular
for the interpretation of state in Section \ref{subsec:state}.
So we introduce a more restricted form of courtesy sufficient to ensure
preservation of $\sim$-receptivity.

\begin{defi}\label{def:comp_court}
Let $\sigma : \S \to \A^\perp \parallel \B$ be a pre-$\sim$-strategy. We say that
$\sigma$ is $(A,
B)$-courteous iff for all $s_1 \imc s_2$ in $S$, if $\pol_S(s_2) = -$
(\emph{i.e.} $\pol_{A^\perp
\parallel B}(\sigma\,s_2) = -$), then $s_1$ and $s_2$ map to the same $A/B$
component.

We will also say that $\sigma : \S \to \A^\perp \parallel \B$ is
\textbf{componentwise
courteous}\index{Courteous!componentwise} to
mean that it is $(A, B)$-courteous, when $\A$ and $\B$ are clear from the
context.
\end{defi}

So $\sigma$ is not necessarily courteous, but is not allowed to influence
negative moves accross
components.
As announced, we have the following.

\begin{restatable}{lemma}{compcourt}
\label{lem:comp_court}
Let $\sigma : \S \to \A^\perp \parallel \B$ and $\tau : \T \to \B^\perp \parallel
\C$ be
pre-$\sim$-strategies, such that $\sigma$ is $(A, B)$-courteous and $\tau$ is
$(B, C)$-courteous.
Then, $\tau \odot \sigma$ is $\sim$-receptive and $(A, C)$-courteous.
\end{restatable}
\begin{proof}
The key ingredient of the proof is that thanks to componentwise courtesy of
$\sigma$ and $\tau$, the immediate dependency of a negative event has to be a
visible event in the same component (and not a neutral event); hence availability
of a negative extension is entirely determined by the visible part of the
interaction, and $\sim$-receptivity follows. Technical details are fairly
tedious, and relegated to Appendix \ref{app:comp_simrec_court}.
\end{proof}

So, componentwise courteous pre-$\sim$-strategy are
stable under composition. The $\sim$-strategies are precisely those that are
furthermore courteous and receptive as plain strategies, and we know from
\cite{cg1} that those are stable under composition; so $\sim$-strategies are
preserved under composition as well. We will see in Section
\ref{subsubsec:compact} that composition is associative (up to \emph{strong}
isomorphism) -- however, before then we now focus on a crucial element of the
compositional structure: its identity, the $\sim$-strategy copycat.

\subsubsection{Copycat.}
Recall that the copycat strategy on game $A$ is a labeled event structure:
\[
\cc_A : \CC_A \to A^\perp \parallel A
\]
where $\CC_A$ has the same events as $A^\perp \parallel A$, but additional
immediate causal links
from negative events on one side to matching positive events on the other side.
Consequently,
configurations $x\in \conf{\CC_A}$ decompose as $x = x_1 \parallel x_2 \in
\conf{A^\perp
\parallel A}$.

The following definition is forced by the requirement that the map $\cc_A$ should
be a map of ess,
and that each symmetry should be an order-iso.

\begin{defi}\label{def:symm_cc}%
\nomenclature[bo]{$\CC_{\tilde{A}}$}{Isomorphism family on copycat}%
Let $\A$ be a tcg.
Given $x = x_1 \parallel x_2 \in \conf{\CC_A}$, $y = y_1 \parallel y_2 \in
\conf{\CC_A}$, the set of
\emph{symmetries} between $x$ and $y$ (written $\CC_{\tilde{A}}$)
comprises any bijection $\theta = \theta_1 \parallel \theta_2$ such that
$\theta_1, \theta_2 \in \tilde{A}$, and which is an order-iso (for the order on
$x, y$
induced by $\leq_{\CC_A}$).
\end{defi}

This definition is forced by necessity. However, to reason on such
symmetries, it will
be convenient to rely on a more high-level characterisation that does not
explicitly require an
order-isomorphism. To introduce it, recall first from
\cite{cg1} that configurations $x\in \conf{\CC_A}$ are exactly those $x_1
\parallel x_2 \in \conf{A
\parallel A}$ such that (with polarity as in $A\parallel A$):
\[
x_2 \supseteq^- x_1 \cap x_2 \subseteq^+ x_1
\]

Furthermore, it is observed in \cite{DBLP:conf/fossacs/Winskel13,cg1} that this
relation between $x_2$ and $x_1$ is a%
\nomenclature[bp]{$x \sqsubseteq_A y$}{The Scott order $y \supseteq^- x \cap y
\subseteq^+ x$}%
partial order called the ``Scott order'', written $x_2 \sqsubseteq_A x_1$. This
order is of crucial
importance in the construction and study of the bicategory $\CG$.

\begin{restatable}{proposition}{symmcc}\label{prop:symm_cc}
The set $\CC_{\tilde{A}}$ is equivalently defined as comprising the bijections
\[
\theta _1 \parallel \theta _2 : x_1 \parallel x_2 \bij _{\tilde
  A^\perp \parallel \tilde A} y_1 \parallel y_2
\]
satisfying the further condition that for all $a \in x_1 \cap x_2$, we
have $ \theta _1(a) = \theta _2(a)$.
\end{restatable}
\begin{proof}
Fairly straightforward, details are in Appendix \ref{app:copycat}.
\end{proof}

In other words, $\CC_{\tilde{A}}$ comprises those $\theta_1 \parallel
\theta_2 \in
\tilde{A}^\perp \parallel \tilde{A}$ such that $\theta_2 \supseteq^- \theta_1
\cap \theta_2
\subseteq^+ \theta_1$, \emph{i.e.}
$\theta_2 \sqsubseteq_{\tilde{A}} \theta_1$.
This justifies the notation $\CC_{\tilde{A}}$, as this agrees with the
description of configurations of copycat
via the Scott order. 
Wrapping up this construction, we state:

\begin{restatable}{proposition}{ccsimstrat}\label{prop:cc_sim_strat}
Let $\A$ be a tcg. Then, writing $\CC_\A = (\CC_A, \CC_{\tilde{A}})$, the map
\[
\cc_A : \CC_\A \to \A^\perp \parallel \A
\]
is a $\sim$-strategy.
\end{restatable}
\begin{proof}[Sketch]
The details are fairly tedious and relegated to Appendix \ref{app:copycat}.
Interestingly it relies on $\A$ being a tcg: for arbitrary essp $\A$, the
set $\CC_{\tilde{\A}}$ fails \emph{(Extension)} (see Appendix
\ref{app:failextrace}).
%
\end{proof}

\subsubsection{Compact closed structure.}\label{subsubsec:compact}
We now describe the categorical structure of the constructions above.
As all our constructions are conservative extensions of those in $\CG$ (for which
categorical laws are proved in details in \cite{cg1}), the proofs of categorical
laws boil down to showing that all the isomorphisms involved preserve symmetry.
In fact all laws can be actually \emph{deduced} directly from those in $\CG$
established in \cite{cg1}, by
exploiting the representation of event structures with symmetry as spans (Section
\ref{subsubsec:essp}). As the details are at the same time unsurprising and
rather tedious, we chose to omit them.

\begin{prop}
We have a category $\TCG$ having tcgs as objects, and as morphisms from $\A$ to
$\B$ the $\sim$-strategies $\sigma : \S \to \A^\perp \parallel \B$, up to weak
isomorphism.

Furthermore, composition of componentwise courteous pre-$\sim$-strategies is
associative.
\end{prop}

We also write $\sigma : \A \tcgto \B$ if $\sigma : \S \to \A^\perp \parallel \B$
is a $\sim$-strategy from $\A$ to $\B$, keeping the $\S$ anonymous.
Note that the associativity and unity laws actually hold up to \emph{strong
isomorphism}, \emph{i.e.} the projection to the game is preserved on the nose.
The situation will be the same for all laws relative to the compact closed
structure. Of course, this implies that they hold up to weak isomorphism as well.

We now equip $\TCG$ with a monoidal structure. For that, observe first that the
category $\ESSP$ of essps and maps preserving symmetry and polarities already has
a monoidal structure, with tensor $\parallel$ extending to maps in the obvious
way. The tensor for $\TCG$ is defined on tcgs $\A$ and $\B$ to be $\A
\tensor \B = \A \parallel \B$ (below Definition \ref{def:tcg}). Likewise, the
tensor of $\sim$-strategies $\sigma_1 : \S_1 \to \A_1^\perp \parallel \B_1$ and
$\sigma_2 : \S_2 \to \A_2^\perp \parallel \B_2$ is the map
$\sigma_1 \tensor \sigma_2$, defined as the composite
\[
\xymatrix{
\S_1 \parallel \S_2 
	\ar[r]^{\!\!\!\!\!\!\!\!\!\!\!\!\sigma_1 \parallel
\sigma_2\,\,\,\,\,\,\,\,\,\,\,\,}&
(\A_1^\perp \parallel \B_1) \parallel (\A_2^\perp \parallel \B_2)
	\ar[r]^{\gamma}&
(\A_1 \parallel \A_2)^\perp \parallel (\B_1 \parallel \B_2)
}
\]
where $\gamma$ is the obvious relabeling. Again, as this is compatible with the
tensor of strategies in $\CG$, establishing that we have a
bifunctor $- \tensor - : \TCG \times \TCG \to \TCG$ just amounts to the fact that
the corresponding isomorphisms in $\CG$ additionally preserve symmetry.

In order to construct the compact closed structure of $\TCG$, we need to define
all the required structural isomorphisms, such as \emph{e.g.} the strategy
$(\A \parallel \B) \parallel \C \profto \A \parallel (\B \parallel \C)$ expressing
that the monoidal product is associative. For that we follow \cite{cg1} and
simply \emph{lift} them from the corresponding isomorphisms from the monoidal
structure of the category $\ESSP$ of essps and maps preserving both polarity and
symmetry between them.

\begin{defi}\label{def:liftingtcg}
Let $f : \A \to \B$ be a strong-receptive courteous polarities-preserving map
between tcgs. Then its
\textbf{lifting}\index{Lifting!of maps of $\ESSP$} is
the $\sim$-strategy
\[
\lift{f} = (\A^\perp \parallel f) \circ \cc_\A : \CC_\A \to \A^\perp \parallel \B
\]
which is a $\sim$-strategy from $\A$ to $\B$ (in particular, it is thin).
\end{defi}

We then have:

\begin{thm}
The category $\TCG$ is compact closed.
\end{thm}
\begin{proof}
For completeness, we list here all structural morphisms for the symmetric
monoidal structure of $\ESSP$.
\[
\begin{array}{rcrcl}
\rho_\A &:& \A\parallel 1 &\to & \A\\
\lambda_\A &:& 1\parallel \A &\to& \A\\
s_{\A, \B} &:& \A\parallel \B &\to& \B \parallel \A\\
\alpha_{\A, \B, C} &:& (\A \parallel \B) \parallel \C &\to& \A \parallel
(\B\parallel \C)
\end{array}
\]
These isomorphisms are then lifted to $\sim$-strategies.
\[
\begin{array}{rcrcl}
\lift{\rho_{\A}} &:& \A \tensor 1 &\tcgto & \A\\
\lift{\lambda_{\A}} &:& 1\tensor \A &\tcgto& \A\\
\lift{s_{\A, \B}}  &:& \A\tensor \B &\tcgto& \B \tensor\A\\
\lift{\alpha_{\A, \B, \C}} &:& (\A \tensor \B) \tensor \C &\tcgto& \A \tensor
(\B\tensor \C)
\end{array}
\]

As before, coherence and naturality laws are easily adapted from those in $\CG$.
Finally, there are (copycat) $\sim$-strategies
\[
\eta_\A : 1 \tcgto \A^\perp \tensor \A\qquad\qquad\epsilon_\A : \A\tensor
\A^\perp \tcgto 1
\]
satisfying the necessary equations up to isomorphism of $\sim$-strategies.
\end{proof}

Finally, we adapt from \cite{cg1} the \emph{lifting lemma}, which we will
use later. It characterises the effect of composition of a $\sim$-strategy via a
lifted map.

\begin{lem}\label{lem:lifting}
Let $f: \B \to \C$ be a strong-receptive courteous polarities-preserving map
between tcgs, and $\sigma : \S \to \A^\perp \parallel \B$ a $\sim$-strategy.
Then, those $\sim$-strategies are isomorphic:
\[
\begin{array}{rcrcl}
\lift{f}\odot \sigma &:& \CC_\B \odot \S &\to& \A^\perp \parallel \C\\
(\A^\perp \parallel f)\circ \sigma &:& \S &\to& \A^\perp \parallel \C
\end{array}
\]
\end{lem}

Note that for $f : \B^\perp \to \A^\perp$ strong-receptive and courteous,
we have the dual lifting $\lift{f}^\perp = (f \parallel \B)\circ \cc_{\B} :
\A\tcgto \B$; and, by duality, the symmetric lemma to the above holds: for
$\sigma : \B \tcgto \C$, $\sigma \odot \lift{f}^\perp \siso (f \parallel \C)
\circ \sigma$. Finally we note: 

\begin{lem}\label{lem:lift_colift}
Let $f : \A \to \B$ be an isomorphism of tcgs -- so both $f$ and $f^{-1}$ are
strong-receptive courteous. Then, $\lift{f} \siso \lift{f^{-1}}^\perp$.
\end{lem}

All this work to build just a compact closed category may feel a little bit
anticlimactic to the reader, and understandably so since by itself, a compact
closed category can only be used to interpret a simple logic such as
Multiplicative Linear Logic. For us however, it represents a crucial achievement
in the construction of our framework for game semantics: it covers all the basic
compositional properties, while remaining completely agnostic as to the language
under study, its features, evaluation strategy, \emph{etc}. 

The focus of the present paper is the construction of the category of Concurrent
Hyland-Ong games in the next section; but we consider $\TCG$ as a crucial
contribution of the paper in itself, and it can certainly be used in other ways
than just through Concurrent Hyland-Ong games. For instance, just like in
Hyland-Ong games, replication in $\CHO$ is ubiquitous: one cannot speak anymore
of linear resources. In contrast, building directly on $\TCG$ (as we did in
\cite{lics18}) with a AJM-style exponential allows for a more traditional
approach to replication (\emph{i.e.} a model of ILL), and allows mixing linear
and non-linear resources. 

\subsubsection{AJM-style exponentials.}\label{subsubsec:ajm}
For completeness, and to emphasize the different ways in which one can work with
$\TCG$, we conclude this section by building an AJM-style exponential -- we only
show that we have an exponential modality, and leave the construction of a full
model of Intuitionnistic Linear Logic as out of scope. The remaining sections of
this paper are independent of this construction.

Of course, to construct an AJM-style exponential in $\TCG$ the first step is to
extend the construction of $\ocajm \A$ in Definition \ref{def:bang_ajm} to tcgs. On
that front the first news is bad: $\ocajm \A$ cannot be made a tcg in general.
For instance, write $\ominus \oplus$ for the tcg with two events, one negative
and one positive, the rest of the structure being trivial. Then, $\tilde{\ocajm
(\ominus \oplus)}$ cannot be decomposed into $\tilde{\ocajm (\ominus \oplus)}_-$
and $\tilde{\ocajm (\ominus \oplus)}_+$ satisfying the axioms of tcgs.
Intuitively, that is because $\tilde{\ocajm (\ominus \oplus)}$ forces symmetry
constraints across polarities, \emph{e.g.} $\{(1, \ominus)\} \isf \{(1,
\ominus)\}$ and $\{(1, \oplus)\} \isf \{(2, \oplus)\}$ are incompatible, although
they have dual polarities and their projections are compatible. The issue
runs deeper than a lack of generality of tcgs: $\CC_{\tilde{\ocajm (\ominus
\oplus)}}$ fails the \emph{(Extension)} axiom of isomorphism families (see
Appendix \ref{app:failextrace}).

Fortunately, this phenomenon is circumscribed to \emph{non-polarized} games with
minimal events belonging to both players. Say that a game $\A$ is
\textbf{negative} (resp. \textbf{positive}) if all its minimal events have
negative (resp. positive) polarity. Then we have:

\begin{prop}
Let $\A$ be a negative tcg. We define $\tilde{\ocajm \A}_-$ to include all bijections
$\theta :\,\parallel_{i\in I} x_i \bij\,\parallel_{j\in J}
y_j$ such that there exists a permutation $\pi : I \bij J$, and for all
$i\in I$ a symmetry $\theta_i : x_i \isf_{\tilde{A}_-} y_{\pi(i)}$ such that
for all $(i, a) \in\,\parallel_{i\in I} x_i$, we have $\theta(i, a) =
(\pi(i), \theta_i(a))$. 
Likewise, $\tilde{\ocajm \A}_+$ is defined in the same way, with
$\tilde{A}_+$ replacing $\tilde{A}_-$ and $\pi$ restricted to be the identity.

Then, $(\ocajm \A, \tilde{\ocajm \A}_-, \tilde{\ocajm \A}_+)$ is a tcg.
\end{prop}
\begin{proof}
Straightforward.
\end{proof}

It \emph{is} possible to have an AJM-style exponential in a non-polarized
setting in the \emph{saturated} games of \cite{lics14} mentioned at the beginning
of the section. Hence the inability to cover uniform replication on non-polarized
games is a restriction of $\TCG$, however we
believe the gains from $\TCG$ far outweight this cost,
especially since there are very few situations in game semantics that require
uniform replication in non-polarized games (in fact, the only example we are
aware of is a games model of classical Linear Logic -- interestingly, this is
also the paper in which the saturated variant of AJM games was introduced
\cite{believe}).

We now prove that $\ocajm$ is an exponential modality in the
\emph{negative} subcategory of $\TCG$, that we define now.

\begin{prop}\label{prop:tcg_minus}
There is a symmetric monoidal subcategory $\TCG_-$ of $\TCG$ having as objects
the negative tcgs, and as morphisms the \textbf{negative} $\sim$-strategies
$\sigma : \S \to \A^\perp \parallel \B$, where the negativity of $\sigma$ means
that the minimal events of $S$ are negative.
\end{prop}
\begin{proof}
The only non-trivial thing to check is that negative (pre)strategies are stable
under composition, which will be proved as Lemma \ref{lem:comp_neg} in the next section.
\end{proof}

We finally conclude this section:

\begin{thm}
The operation $\ocajm$ extends to a \emph{linear exponential comonad}:
\[
\ocajm : \TCG_- \to \TCG_-
\]
\end{thm}
\begin{proof}
First we define the functorial action: for $\sigma : \S \to \A^\perp \parallel
\B$ between negative tcgs, we have $\ocajm \sigma : \ocajm \S \to (\ocajm
\A)^\perp \parallel \ocajm \B$ defined in the obvious way; functoriality is a
variant of that of $\tensor$. As for the monoidal structure, the
components are lifted from maps in $\ESSP$. More precisely, for any negative tcgs
$\A, \B$ there are maps defined in the standard way:
\[
\begin{array}{rclcrcl}
(\ocajm \A)^\perp \tensor (\ocajm \A)^\perp &\to& (\ocajm \A)^\perp &\qquad&
	\A^\perp &\to& (\ocajm \A)^\perp\\
(\ocajm \ocajm \A)^\perp &\to& (\ocajm \A)^\perp&\qquad&
	1&\to& (\ocajm \A)^\perp\\
(\ocajm (\A \tensor \B))^\perp &\to& (\ocajm \A)^\perp \tensor (\ocajm
\B)^\perp&\qquad&
	\ocajm 1 &\iso& 1
\end{array}
\]

All those maps are courteous and strong-receptive, so by the ``dual lifting''
operation below Lemma \ref{lem:lifting} they yield $\sim$-strategies in the other
direction, \emph{e.g.} $\ocajm \A \profto \ocajm \A \tensor \ocajm \A$ and
$\ocajm \A \profto \ocajm \ocajm \A$. It follows direction from Lemma
\ref{lem:lifting} that these $\sim$-strategies are natural and satisfy the
required coherence laws up to positive symmetry.
\end{proof}

\section{Concurrent Hyland-Ong games}
\label{sec:cho}
We have constructed a compact closed category $\TCG$, which is equipped to deal with the 
problem evoked at the end of Section \ref{sec:replication}. Using it, we can revisit (more
formally) the interpretation sketched in Subsection \ref{subsec:replication}. Exploiting the 
developments of the previous section, and in particular the fact that weak isomorphism is a 
congruence, it will follow that the two terms of Example \ref{ex:wiso} cannot be distinguished by
any strategy in the model. Indeed, we will get a cartesian closed category supporting \emph{e.g.}
the interpretation of IPA.

From now on, all event structures are assumed to have binary conflict. All the operations we will
consider on them (simple parallel composition, composition, interaction, \emph{etc}.) have been
established throughout the development to preserve that property.

\subsection{The cartesian category $\CHO$}

We now construct the category $\CHO$ proper, of \emph{Concurrent Hyland-Ong games}; and prove that
it is \emph{cartesian}.
The \textbf{objects} of $\CHO$ will
be negative arenas, as in Definition \ref{def:arena} -- with the further restriction that arenas
should have a \emph{countable} set of events, assumed from now on. The \textbf{morphisms} from 
arena $A$ to arena $B$ will be certain $\sim$-strategies from $\oc A$ to $\oc B$
(up to weak isomorphism):
\[
\sigma : \S \to (\oc A)^\perp \parallel (\oc B)
\]

Just as in standard HO games, we restrict 
strategies in order to satisfy the laws of a cartesian category.
We will now inspect the different requirements of a cartesian category, and introduce the additional
conditions on strategies as they are required.

\subsubsection{Terminal object and negativity} 
First of all, a cartesian category has a terminal
object. In our case, this will be the \textbf{empty arena}\index{Empty arena} $\emptyar$, defined as having
an empty set of events -- note that $\oc \emptyar$ also has an empty set of events.
However, as it is, $\emptyar$ is not a terminal object. For each 
negative arena $A$, it is easy to see that the unique labelling function
\[
\e_A : \emptyar \to (\oc A)^\perp \parallel (\oc \emptyar)
\]
is a $\sim$-strategy. Crucially, it is receptive since, by negativity of $A$, the minimal
events of $(\oc A)^\perp$ are all positive.
However, $\e_A$ might not be \emph{unique} from $\oc A$ to $1$, as
illustrated below. 

\begin{exa}
The following diagram represents a $\sim$-strategy from $\oc \com$ to $\oc \emptyar$.
\[
\xymatrix@R=0pt{
(\oc \com)^\perp	\ar@{}[r]|{\parallel}&
(\oc \emptyar)\\
\run^{+,0}
}
\]
\end{exa}

The answer to this issue is clear: we need to require morphisms in $\CHO$ to 
be negative, just as arenas. As in Proposition \ref{prop:tcg_minus},
$\sigma : \S \to (\oc A)^\perp \parallel (\oc B)$
is \textbf{negative} \index{Negative!(pre)strategy} whenever the underlying event
structure $S$ is negative, \emph{i.e.} its minimal events are negative -- note
that this definition makes sense in general without symmetry, for a prestrategy
$\sigma : S \to A$. 

We then easily have:

\begin{prop}
For any negative arena $A$, the empty $\sim$-strategy:
\[
\e_A : \emptyar \to (\oc A)^\perp \parallel (\oc \emptyar)
\]
is the unique negative $\sim$-strategy from $\oc A$ to $\oc \emptyar$.
In more generality, the only negative prestrategy $\sigma : S \to A^\perp \parallel \emptyar$ for a
negative game $A$ is the empty prestrategy.
\end{prop}
\begin{proof}
Immediate, as in a negative prestrategy $\sigma : S \to A^\perp \parallel \emptyar$, 
any hypothetical minimal events in $S$ have nowhere to map to.
\end{proof}

Thus, in order to get a category of $\sim$-strategies with a terminal object, we will require that
all $\sim$-strategies are negative. Clearly copycat is negative, along with all
$\sim$-strategies obtained by lifting. Moreover, negative $\sim$-strategies are
stable under composition. Since negativity makes sense without symmetry, we 
prove that in slightly greater generality. 

\begin{lem}\label{lem:comp_neg}
Let $\sigma : S \to A^\perp \parallel B$ and $\tau : T \to B^\perp \parallel C$ be
negative prestrategies (with $A, B, C$ negative). Then, $\tau \odot \sigma$ is still negative.
\end{lem}
\begin{proof}
First, maps of event structures preserve minimal events: for $f : A \rightarrow
B$ and $a$ minimal in $A$, it follows easily from the axioms that $f a$ is
minimal in $B$. 
Hence, minimal events of $T  \circledast  S$ are
projected to minimal events of $S  \parallel  C$ and $A  \parallel  T$. Take
$e  \in  T  \circledast  S$ a minimal event. If $( \tau   \circledast   \sigma )\,e$ is in $A$, then
$ \Pi _1 e$ is a minimal event of $S$ projected to a (necessarily positive) minimal event of $A$ -- absurd because
$ \sigma $ is negative. Likewise, if $( \tau   \circledast   \sigma )\,e$ is in $B$, this
contradicts the negativity of $ \tau $. So minimal events of $T  \circledast  S$ are
visible and are in $C$.

Now, take any minimal event $e \in T \odot S$. Since minimal events of
$T \circledast S$ are visible, $e$ is also minimal in
$T \circledast S$.  By the previous remark, $( \tau \odot \sigma )\,e$
is in $C$ and is minimal. It is also negative because $C$ is
negative.
\end{proof}

Therefore, the category having arenas as objects and as morphisms from $A$ to $B$,
\emph{negative} $\sim$-strategies $\sigma : \S \to (\oc A)^\perp \parallel (\oc B)$
up to weak isomorphism, has a terminal object $\emptyar$. We now investigate 
the existence of products.

\subsubsection{Binary products and single-threadedness} 

For two arenas $A$ and $B$, their
\textbf{product}\index{Product arena} $A \times B$ is defined as the parallel
composition $A\parallel B$,%
\nomenclature[da]{$A\times B$}{Product of arenas $A$ and $B$, defined as
$A\parallel B$}%
which is still a negative arena. 

\subsubsection*{Projections.}
Note that there is an injection map of event structures with symmetry:
\[
\begin{array}{rcccc}
\mathrm{i}_A &:& \oc A &\to& \oc (A\times B)\\
&& (\alpha : [a] \to \omega) &\mapsto& \left(\begin{array}{rcrcl}
						\alpha' &:& [(1, a)] &\to& \omega\\
							& &  (1, a') &\mapsto& \alpha(a')
					      \end{array}\right)
\end{array}
\]

Likewise, there is $\mathrm{i}_B : \oc B \to \oc (A\times B)$. Using those, we define the
\textbf{projections} \index{Projections (strategies)}
\[
\varpi_A : \CC_{\oc A} \to \oc (A\times B)^\perp \parallel \oc A\qquad
\varpi_B : \CC_{\oc B} \to \oc (A\times B)^\perp \parallel \oc B
\]
by lifting the injections, \emph{i.e.} $\varpi_A = \lift{\mathrm{i}_{A^\perp}}^\perp$ and $\varpi_B =
\lift{\mathrm{i}_{B^\perp}}^\perp$ (see Definition \ref{def:liftingtcg}).%
\nomenclature[db]{$\varpi_A : \oc (A \times B) \tcgto A$}{Left projection in
$\CHO$}%
\nomenclature[dc]{$\varpi_B : \oc (A \times B) \tcgto B$}{Right projection in
$\CHO$}%

\subsubsection*{Pairing.}
Now, for negative $\sim$-strategies $\sigma : \S \to \oc A^\perp \parallel \oc B$ and
$\tau : \T \to \oc A^\perp \parallel \oc C$, we wish to define their pairing
$\tuple{\sigma, \tau}$, a $\sim$-strategy from $\oc A$ to $\oc (B \times C)$. 
This $\sim$-strategy will simply be obtained by relabeling the parallel composition
of $\S$ and $\T$. In simple cases, it suffices to take the co-pairing:
\[
\begin{array}{rcl}
\ttuple{\sigma, \tau} 	&:& \S \parallel \T \to \oc A^\perp \parallel \oc (B \times C)\\
			&=& [(\oc A^\perp \parallel \mathrm{i}_{B})\circ \sigma, (\oc A^\perp \parallel
\mathrm{i}_C)\circ \tau]
\end{array}
\]

However, this is not always well-defined as a $\sim$-strategy. Indeed, it might fail local injectivity if some events in
$S$ and $T$ have the same image in $\oc A^\perp$. As a first step towards the general construction
of pairing, let us prove that this gives a well-defined $\sim$-strategy when the images of $\sigma$
and $\tau$ \emph{are} disjoint.

\begin{lem}\label{lem:pairing_disjoint}%
\nomenclature[dd]{$\ttuple{\sigma, \tau}$}{Pre-pairing of $\sim$-strategies $\sigma : \oc A \tcgto \oc B$ and $\tau :
\oc A \tcgto \oc C$ with disjoint codomains}%
If negative $\sim$-strategies $\sigma : \S \to \oc A^\perp \parallel \oc B$ and
$\tau : \T \to \oc A^\perp \parallel \oc C$ have \emph{disjoint codomain} on $\oc A^\perp$,
then $\ttuple{\sigma, \tau}$ as above is a negative $\sim$-strategy.
\end{lem}
\begin{proof}
\red{First, we prove that it is a $\sim$-strategy. That it is a map of essps along
with courtesy and thinness are direct verifications.}
%
Strong-receptivity needs further attention. Take $\theta : x_S \parallel x_T \isf y_S \parallel y_T
\in \sym{S} \parallel \sym{T}$, write $\theta = \theta_S \parallel \theta_T$. Its projection to
the game is 
\[
\ttuple{\sigma, \tau}\,\theta = ((\oc A^\perp \parallel \mathrm{i}_{B})\circ \sigma\,\theta_S) \uplus
((\oc A^\perp \parallel \mathrm{i}_C)\circ \tau\,\theta_T)
\]
which is a valid symmetry in $G = \oc A^\perp \parallel \oc (B \times C)$. Assume it
extends by a pair $(c^-_1, c^-_2)$. Since dependency in the game is forest-shaped, there are unique
$d_1 \imc_G c^-_1$ and $d_2 \imc c^-_2$, and since symmetries are order-preserving, 
we have $(d_1, d_2) \in \ttuple{\sigma, \tau}\,\theta$. But that means that 
it must be either in $((\oc A^\perp \parallel \mathrm{i}_{B})\circ \sigma)\,\theta_S$, or
in $((\oc A^\perp \parallel \mathrm{i}_C)\circ \tau)\,\theta_T$. We can then apply strong-receptivity 
of $\sigma$, $\tau$, and the injection maps, to produce the extension to $\theta = \theta_S
\parallel \theta_T$.
\end{proof}

Now, we prove that this simple pairing behaves well \emph{w.r.t.} projections.

\begin{prop}\label{prop:pairing_disjoint}
Assume negative $\sim$-strategies $\sigma : \S \to \oc A^\perp \parallel \oc B$ and
$\tau : \T \to \oc A^\perp \parallel \oc C$ as in the previous lemma. Then, we
have (strong) isomorphisms:
\[
\varpi_B \odot \ttuple{\sigma, \tau} \siso \sigma \qquad\qquad \varpi_C \odot \ttuple{\sigma, \tau} \siso
\tau
\]
\end{prop}
\begin{proof}
Let us prove the first. More precisely, we prove that the interactions $\varpi_B \circledast
\ttuple{\sigma, \tau}$ and $\cc_{\oc B} \circledast \sigma$ are isomorphic. This will entail by
restriction an isomorphism between the corresponding compositions, and the latter is isomorphic to
$\sigma$ as copycat is the identity.

We establish the isomorphism between $\varpi_B \circledast \ttuple{\sigma, \tau}$ 
and $\cc_{\oc B} \circledast \sigma$ first for the plain event structures -- by Lemma 2.12 of
\cite{cg1} it suffices to prove that they have an isomorphic domain of configurations.
Using Proposition \ref{prop:rep_int}, we know that configurations of the event structure for the
former interaction correspond to secured bijections
\[
(x_S \parallel x_T) \parallel x_B \,\,\,\, \sbij \,\,\,\, y_A \parallel (y_B^1 \parallel y_B^2)
\]
where $x_S \in \conf{S}, x_T \in \conf{T}, \ttuple{\sigma, \tau}\,(x_S \parallel x_T) = y_A
\parallel (\mathrm{i}_B\,y_B^1)$, and $y_B^1 \parallel y_B^2 \in \conf{\CC_{\oc B}}$, and where the
bijection is the unique such that image of events through the labelings $\ttuple{\sigma, \tau}
\parallel \oc B$ and $\oc A \parallel \varpi_B$ match. In particular, $\ttuple{\sigma,
\tau}\,(x_S \parallel x_T)$ does not reach $\oc C$. But any minimal events of $x_T$ are negative by
negativity of $\tau$, and hence must be in $\oc C$ (since $A$ is negative). Therefore, $x_T$ is empty.
Getting rid of $x_T$ yields a secured bijection corresponding to a configuration of the event
structure of $\cc_{\oc B} \circledast \sigma$. This association is bijective, and yields the
required isomorphism between domains of configurations.
By construction, it is clear that this isomorphism preserves symmetry.
\end{proof}

So, we know how to construct a pairing behaving well with projections, 
when the paired strategies happen to
have a disjoint codomain. However, for arbitrary $\sigma : \S \to \oc A^\perp \parallel \oc B$ and
$\tau : \T \to \oc A^\perp \parallel \oc C$, there might in general be collisions:
events $s \in S$ and $t\in T$ such that $\sigma\,s = \tau\,t$. In such a case, the
co-pairing as above fails local injectivity, and therefore does not correspond to 
a strategy. Fortunately, we can \emph{relabel} moves of $\S$ and $\T$, not changing
their weak isomorphism class, to ensure that there are no such collisions. For that,
we note that there are maps of event structures with symmetry
\[
\iota_e: \oc A^\perp \to \oc A^\perp \qquad\qquad \iota_o: \oc A^\perp \to \oc A^\perp
\]
such that $\iota_e \sim_+ \iota_o \sim_+ \id_{\oc A^\perp}$, but such that $\iota_e$ and
$\iota_o$ have disjoint codomain. For definiteness, say that
$\iota_e$ sends (necessarily positive) minimal events with copy index $i$
to the same events with copy index $2i$, and preserves the index of other
events. Likewise, $\iota_o$ follows the injection $i \mapsto 2i + 1$. These
maps preserve the index of negative events, so that $\iota_e \sim_+ \iota_o \sim_+ \id_{\oc A^\perp}$.

Given arbitrary $\sigma : \S \to \oc A^\perp \parallel \oc B$ and
$\tau : \T \to \oc A^\perp \parallel \oc C$, we define:
\[
\sigma_e = (\iota_e \parallel \oc B)\circ \sigma \qquad \qquad \tau_o = (\iota_o \parallel \oc C) \circ
\tau
\]

From $\iota_e \sim_+ \iota_o \sim_+ \id_{\oc A^\perp}$ it is obvious that
$\sigma \wiso \sigma_e$ and $\tau \wiso \tau_o$, but $\sigma_e$
and $\tau_o$ now have disjoint codomains: $\sigma_e$ (resp. $\tau_o$) only reaches indexing functions
in $\oc A$ whose index for minimal events is even (resp. odd). Therefore, using Proposition
\ref{prop:pairing_disjoint}, we define:
\[
\tuple{\sigma, \tau} = \ttuple{\sigma_e, \tau_o}
\]

\nomenclature[de]{$\tuple{\sigma, \tau}$}{Pairing of $\sim$-strategies $\sigma : \oc A \tcgto \oc B$
and $\tau : \oc A \tcgto \oc C$}%
The \textbf{pairing}\index{Pairing} of arbitrary negative $\sim$-strategies $\sigma : \S \to \oc A^\perp \parallel
\oc B$ and $\tau : \T \to \oc A^\perp \parallel \oc C$ is defined as $\tuple{\sigma, \tau}$.
We have, as required, $\varpi_B \odot \tuple{\sigma,
\tau} =
\varpi_B \odot \ttuple{\sigma_e, \tau_o} \siso \sigma_e \wiso \sigma$, and
for the same reason $\varpi_C \odot \tuple{\sigma, \tau} \wiso \tau$. 
It is an immediate verification that $\tuple{-,-}$ preserves weak isomorphism,
so it will still make sense as an operation on the quotient category.

\begin{exa}\label{ex:contraction}
Consider the copycat strategy $\cc_{\oc \intr{\com}}$ on $\oc \intr{\com}$.
\[
\xymatrix@R=5pt{
\oc \intr{\com}
	\ar@{}[r]|{\to}||&
\oc \intr{\com}\\
&\run^{-,i}
	\ar@{-|>}[dl]\\
\run^{+,i}
	\ar@{-|>}[d]\\
\done^{-,j}
	\ar@{--}@/^/[u]
	\ar@{-|>}[dr]\\
&\done^{+,j}
	\ar@{--}@/_/[uuu]
}
\]
Following the definition above, $\tuple{\cc_{\oc \intr{\com}}, \cc_{\oc
\intr{\com}}}$ is the $\sim$-strategy illustrated below.
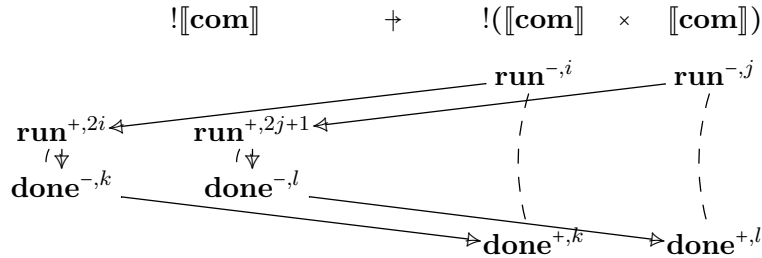
\[
\xymatrix@R=5pt{
&\!\!\!\!\!\!\!\!\!\!\!\!\!\!\!\oc \intr{\com}~~~~~~~~~~~~~~~
	\ar@{}[rr]|\to||&~~~~~~~~~~&
\oc (\intr{\com}
	\ar@{}[r]|\times&
\intr{\com})\\
&&&\run^{-,i}
	\ar@{-|>}[dlll]&
\run^{-,j}
	\ar@{-|>}[dlll]\\
\run^{+,2i}
	\ar@{-|>}[d]&
\run^{+,2j+1}
	\ar@{-|>}[d]\\
\done^{-,k}
	\ar@{-|>}[drrr]
	\ar@{--}@/^/[u]&
\done^{-,l}
	\ar@{-|>}[drrr]
	\ar@{--}@/^/[u]\\
&&&\done^{+,k}
	\ar@{--}@/^/[uuu]&
\done^{+,l}
	\ar@{--}@/^/[uuu]
}
\]
As prescribed by the construction, the positive moves on the left hand side had to be relabeled
to avoid the collision in the case where $i=j$. 
\end{exa}

Note that the above only displays the event part of the $\sim$-strategy $\tuple{\cc_{\oc \intr{\com}},
\cc_{\oc \intr{\com}}}$, but its construction also equips it with a symmetry ensuring its
uniformity.

\subsubsection*{Surjective pairing.} In order to obtain a product, we also need to prove surjective
pairing, that is, that for all $\sigma : \S \to \oc A^\perp \parallel \oc (B \times C)$, we have:
\[
\sigma \wiso \tuple{\varpi_B \odot \sigma, \varpi_C \odot \sigma}
\]
However, as it stands, this is in general not the case. 

\begin{exa}
In Figure \ref{fig:fail_sp}, we display on the left hand side two $\sim$-strategies $\sigma_1,
\sigma_2 : \oc \emptyar \tcgto \oc
(\intr{\com} \times \intr{\com})$, and on the right hand side the corresponding distinct $\sim$-strategies obtained
by projection and pairing.
\begin{figure}
\begin{center}
\[
\begin{array}{ccc}
\xymatrix@C=5pt@R=5pt{
\sigma_1 &:& \oc(\intr{\com}
	\ar@{}|\times[r]&
\intr{\com})\\
&&\run^{-,i}
	\ar@{-|>}[d]&
\run^{-,j}
	\ar@{-|>}[dl]\\
&&\done^{+,j}
	\ar@{--}@/^/[u]
}&&
\xymatrix@C=5pt@R=5pt{
\tuple{\varpi_l \odot \sigma_1, \varpi_r \odot \sigma_1} &:& \oc(\intr{\com}
        \ar@{}|\times[r]&
\intr{\com})\\
&&\run^{-,i}&
\run^{-,j}\\
&&~~
}\\\\
\xymatrix@C=5pt@R=5pt{
\sigma_2 &:& \oc(\intr{\com}
        \ar@{}|\times[r]&
\intr{\com})\\
&&\run^{-,i}
        \ar@{-|>}[d]&
\run^{-,j}
        \ar@{-|>}[d]\\
&&\done^{+,0}
        \ar@{--}@/^/[u]
	\ar@{~}[r]&
\done^{+,0}
	\ar@{--}@/^/[u]
}&&
\xymatrix@C=5pt@R=5pt{
\tuple{\varpi_l \odot \sigma_2, \varpi_r \odot \sigma_2} &:& \oc(\intr{\com}
        \ar@{}|\times[r]&
\intr{\com})\\
&&\run^{-,i}
	\ar@{-|>}[d]&
\run^{-,j}
	\ar@{-|>}[d]\\
&&\done^{+,0}
	\ar@{--}@/^/[u]&
\done^{+,0}
	\ar@{--}@/^/[u]
}
\end{array}
\]
\end{center}
\caption{Failures to surjective pairing}
\label{fig:fail_sp}
\end{figure}

We observe that surjective pairing fails for these strategies, as behaviours that span both
components get erased through composition with the projections.
\end{exa}

This is analogous as in standard Hyland-Ong games
\cite{russphd}, where \emph{single-threadedness} ensures that strategies treat
independently events hereditarily caused by distinct minimal events. The
definition is independent from symmetry, so we state it first in more
generality.

\begin{defi}\label{def:single-threaded}%
\nomenclature[df]{$\init(s)$}{Unique minimal event of $s\in S$ for a single-threaded prestrategy
$\sigma : S \to A$}%
Let $\sigma : S \to A$ be a prestrategy. We say that $\sigma$ is
\textbf{single-threaded}\index{Single-threaded} if
it satisfies the following two conditions.
\begin{enumerate}
\item For any $s\in S$, $[s]$ has exactly one minimal event written $\init(s)$.
\item Whenever $s_1 \conflict s_2$ in $S$, $\init(s_1) = \init(s_2)$.
\end{enumerate}
\end{defi}

Single-threaded $\sim$-strategies always satisfy surjective pairing.

\begin{prop}
Let $\sigma : \S \to \oc A^\perp \parallel \oc (B \times C)$ be a single-threaded
$\sim$-strategy. Then:
\[
\sigma \wiso \tuple{\varpi_B \odot \sigma, \varpi_C \odot \sigma}
\]
\end{prop}
\begin{proof}
First of all, we define two subsets of $S$ as follows:
\[
\begin{array}{rcl}
S_B &=& \{s\in S \mid \sigma\,(\init(s)) \in B\}\\
S_C &=& \{s\in S \mid \sigma\,(\init(s)) \in C\}
\end{array}
\]
(we abuse notations slightly with $\in B, \in C$).

By single-threadedness, $S_B$ and $S_C$ are disjoint and down-closed, with no immediate conflict spanning both
components -- in other words, $S = S_B \uplus S_C$. They are obviously still event structures.
The restrictions of $\sigma$ (along with a simple relabeling to $\oc B /
\oc C$)
\[
\sigma_B : S_B \to \oc A^\perp \parallel \oc B \qquad \qquad \sigma_C : S_C \to \oc A^\perp \parallel \oc C
\]
are receptive and courteous, \emph{i.e.} are strategies.

This decomposition also works at the level of symmetries.
Any $\theta \in \sym{S}$ preserves $S_B$ and $S_C$. Indeed if $(s_B, s_C) \in \theta$, then $(\init(s_B),
\init(s_C))\in \theta$ as well: absurd, since one maps to $\oc B$ and the other to $\oc C$. 
It follows that $\theta = \theta_B \uplus \theta_C$ where
$\theta_B$ and $\theta_C$ are bijections between configurations of $S_B$ and $S_C$ respectively.
The set of restrictions to $S_B$ (resp. $S_C$) of symmetries in $\sym{S}$ yields a set of 
bijections between configurations of $S_B$ (resp. $S_C$), which is easily checked to satisfy the
conditions for an isomorphism family $\sym{S_B}$ (resp. $\sym{S_C}$). The labeling functions
$\sigma_B$ and $\sigma_C$ preserve symmetry. Strong-receptivity and thinness follow directly from
those for $\sigma$, so $\sigma_B$ and $\sigma_C$ are $\sim$-strategies.

By construction, $\sigma_B$ and $\sigma_C$ have disjoint codomain; so we can form their pairing
$\ttuple{\sigma_B, \sigma_C}$ without relabeling.
Then, the obvious bijection $S = S_B \uplus S_C \iso S_B \parallel S_C$ is an isomorphism of event structures, preserves
symmetry, and preserves labeling so as to yield an isomorphism of $\sim$-strategies:
\[
\sigma \siso \ttuple{\sigma_B, \sigma_C}
\]
By Proposition \ref{prop:pairing_disjoint}, it follows that $\varpi_B \odot \sigma \siso \sigma_B$
and $\varpi_C \odot \sigma \siso \sigma_C$. But clearly, $\ttuple{\sigma_B, \sigma_C} \wiso
\tuple{\sigma_B, \sigma_C}$, and $\tuple{-,-}$ preserves weak isomorphism, so we have 
surjective pairing.
\end{proof}

So, single-threadedness ensures surjective pairing. It is clear that copycat $\sim$-strategies -- and lifted
$\sim$-strategies in general -- on (expanded) arenas are single-threaded, since $\CC_{\oc A}$ has the
shape of a conflict-free forest. In order to get a cartesian category, the last thing to check is
that single-threaded strategies are stable under composition.

Single-threadedness and its stability under composition is independent from symmetry, so we state it
and prove it below in greater generality.

\begin{restatable}{proposition}{compst}\label{prop:comp_st}
Let $\sigma : S \to A^\perp \parallel B$ and $\tau : \T \to B^\perp \parallel C$ be
negative single-threaded prestrategies. Then, $\tau \odot \sigma$ is single-threaded.
\end{restatable}
\begin{proof}
\red{The details are rather tedious; 
we postpone them to Appendix
\ref{app:single_threaded}.}
\end{proof}

We have finished constructing our basic category of Concurrent Hyland-Ong games. Let
us call $\CHO$ the category having: as objects, negative arenas; and as
morphisms from $A$ to $B$, negative single-threaded $\sim$-strategies $\sigma : \S \to \oc A^\perp \parallel
\oc B$, up to weak isomorphism. 

As for $\TCG$ we will also sometimes write $\sigma : A \choto
B$ to mean that%
\nomenclature[dfa]{$\sigma : A \choto B$}{Morphism from $A$ to $B$ in $\CHO$,
\emph{i.e.}
single-threaded negative $\sigma : \oc A \tcgto \oc B$}%
$\sigma$ is a morphism from $A$ to $B$ in $\CHO$ (or just
$\sigma : A \profto B$ when $\CHO$ is clear from the context). We get:

\begin{prop}
The category $\CHO$ has finite products.
\end{prop}

In particular, it follows as usual that $\times$ is a bifunctor $\CHO^2 \to \CHO$, by setting
$\sigma_1 \times \sigma_2 = \tuple{\sigma_1 \odot \varpi_{A_1}, \sigma_2 \odot \varpi_{A_2}}$,
for $\sigma_1 : A_1 \profto B_1$ and $\sigma_2 : A_2 \profto B_2$. 

When constructing the cartesian closed structure, we will leverage the compact closed structure of
$\TCG$. Therefore, it is useful to connect the cartesian structure of $\CHO$
with the monoidal structure of $\TCG$. For that, we note that there is an isomorphism of
essps:
\[
m_{A, B} : \oc (A\times B) \iso \oc A \tensor \oc B
\]

Using Definition \ref{def:liftingtcg}, it lifts to an iso $\lift{m_{A,B}}$
in $\TCG$ between them. Consequently:

\begin{lem}\label{lem:prodmon}
Let $\sigma_1 : A_1 \choto B_1$ and $\sigma_2 : A_2 \choto B_2$. Then,
\[
\sigma_1 \times \sigma_2 \wiso \lift{m_{B_1, B_2}}^{-1} \odot (\sigma_1 \tensor \sigma_2) \odot
\lift{m_{A_1, A_2}}
\]
\end{lem}
\begin{proof}
Write $\sigma_1 : \S_1 \to \oc A_1^\perp \parallel \oc B_1$ and $\sigma_2 : \S_2 \to \oc A_2^\perp
\parallel \oc B_2$. By definition, 
\[
\sigma_1 \times \sigma_2 = \tuple{\sigma_1 \odot \varpi_{A_1}, \sigma_2 \odot \varpi_{A_2}}
\]

By Lemma \ref{lem:lifting}, $\sigma_1 \odot \varpi_{A_1} \siso (\mathrm{i}_{A_1^\perp} \parallel \oc
B_1) \circ \sigma_1$ and $\sigma_2 \odot \varpi_{A_2} \siso (\mathrm{i}_{A_2^\perp} \parallel \oc
B_2) \circ \sigma_2$. These 
maps have disjoint codomain, so up to weak isomorphism their pairing is
\[
\ttuple{(\mathrm{i}_{A_1^\perp} \parallel \oc
B_1) \circ \sigma_1, (\mathrm{i}_{A_2^\perp} \parallel \oc
B_2) \circ \sigma_2} : \S_1 \parallel \S_2 \to \oc (A_1 \times A_2)^\perp \parallel \oc (B_1 \times
B_2)
\]

Likewise by Lemma \ref{lem:lifting}, $ \lift{m_{B_1, B_2}}^{-1} \odot
(\sigma_1 \tensor \sigma_2) \odot \lift{m_{A_1, A_2}}$ has (up to isomorphism) essp $\S_1 \parallel \S_2$ and labeling
function the obvious relabeling of $\sigma_1 \tensor \sigma_2$ by $m_{A_1, A_2}$ and $m_{B_1, B_2}$.
It is a simple verification that these two coincide.
\end{proof}

\subsection{Cartesian closure}

We finish the construction of our cartesian closed category by describing the cartesian closure.
We have constructed $\CHO$ as a subcategory of $\TCG$ -- which, as a compact closed category, is
symmetric monoidal closed. We wish to leverage this closed structure of $\TCG$ in order to transfer
it to $\CHO$. 

\subsubsection{Arrow arena.}
For two thin concurrent games $\A$ and $\B$ in $\TCG$, the corresponding exponential object
(following the compact closed structure) is obtained as $\A^\perp \parallel \B$. In $\CHO$, where
objects are arenas, this hints at defining the exponential object of $A, B$ as $A^\perp \parallel
B$. Indeed, it is easy to check that $\oc (A^\perp \parallel B) \iso \oc A^\perp \parallel \oc B$, so
this matches the closed structure of $\TCG$. However, objects in $\CHO$ are required to be
\emph{negative} arenas, and $A^\perp \parallel B$ is no longer negative.
Therefore, we are brought to introduce a negative variant of $A^\perp \parallel B$, that would be an
object of $\CHO$.
The natural choice, familiar from Hyland-Ong games, is to make events in $A$ depend on
minimal events of $B$. It would be incorrect to make events of $A$ depend on \emph{all} minimal
events of $B$, so we will instead create as many copies of $A$ as they are minimal events in $B$.
Writing $\min(B)$ for the set of minimal events of $B$, we define:

\begin{defi}%
\nomenclature[dg]{$A\tto B$}{Arrow arena}%
Let $A, B$ be two negative arenas. Their \textbf{arrow}\index{Arrow arena} is $A\tto B$, with the following components.
\begin{itemize}
\item \emph{Events, and polarity.} Those of 
$(\parallel_{b\in \min(B)} A^\perp) \parallel B$.
\item \emph{Causality.} As follows:
\[
\leq_{(\parallel_{b\in \min(B)} A^\perp) \parallel B} \uplus \{((2, b), (1, (b, a)))\mid b\in
\min(B)~\&~a\in A\}
\]
\end{itemize}
\end{defi}

\begin{exa}
The reader can check that $\intr{\com} \tto \intr{\com}$ is the arena
presented as $\intr{\com \to \com}$ in Example \ref{ex:arenas}. As $\intr{\com}$ has only one
minimal event, there is no duplication of the left hand side. However, the arena 
$\intr{\com} \tto (\intr{\com} \times \intr{\com})$ is displayed below.
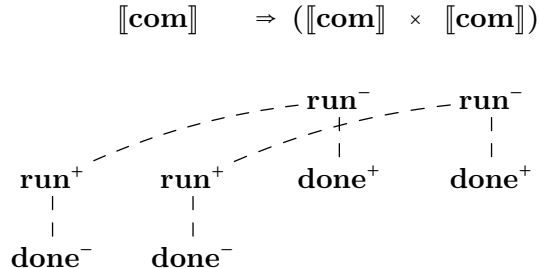
\[
\xymatrix@R=15pt@C=15pt{
&\!\!\!\!\!\!\!\!\!\!\!\!\!\!\intr{\com}~~~~~~~~~~~~~~
	\ar@{}[r]|\tto&
(\intr{\com}
	\ar@{}[r]|\times&
\intr{\com})\\
&& \run^{-}
	\ar@{--}@/_/[dll]
	\ar@{--}[d]&
\run^{-}
	\ar@{--}@/_/[dll]
	\ar@{--}[d]\\
\run^{+}
	\ar@{--}[d]&
\run^{+}
	\ar@{--}[d]&\done^{+}&\done^{+}\\
\done^{-}&\done^{-}
}
\]
\end{exa}

This is exactly the arena construction of \cite{ho}, where arenas are forests.

\subsubsection{Cartesian closed structure.} Our proof of cartesian closure will leverage the compact
closed structure of $\TCG$. More precisely, we will show that there is a bijection (up to weak
isomorphism) between negative single-threaded $\sim$-strategies playing respectively on $\oc A^\perp
\parallel \oc (B\tto C)$ and $\oc A^\perp \parallel (\oc B^\perp \parallel \oc C)$.
This bijection will leave the internal event structure of strategies unchanged, and will only
operate through relabeling.

First, we describe the action of the bijection from $\oc A^\perp \parallel \oc (B\tto C)$ to 
$\oc A^\perp \parallel (\oc B^\perp \parallel \oc C)$. Let us first explain it on an example.
Consider a $\sim$-strategy represented as below -- which is, in essence, a curried version of the
contraction on $\intr{\com}$ of Example \ref{ex:contraction}.

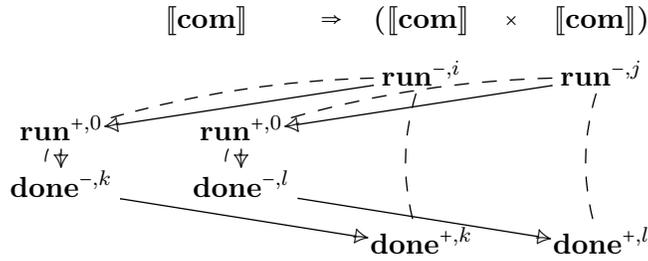
\[
\xymatrix@R=5pt{
&\!\!\!\!\!\!\!\!\!\!\!\!\!\!\intr{\com}~~~~~~~~~~~~~~
        \ar@{}[r]|\tto&
(\intr{\com}
        \ar@{}[r]|\times&
\intr{\com})\\
&&\run^{-,i}
	\ar@{-|>}[dll]&
\run^{-,j}
	\ar@{-|>}[dll]\\
\run^{+,0}
	\ar@{--}@/^/[urr]
	\ar@{-|>}[d]&
\run^{+,0}
	\ar@{--}@/^/[urr]
	\ar@{-|>}[d]\\
\done^{-,k}
	\ar@{--}@/^/[u]
	\ar@{-|>}[drr]&
\done^{-,l}
	\ar@{--}@/^/[u]
	\ar@{-|>}[drr]\\
&&\done^{+,k}
	\ar@{--}@/^/[uuu]&
\done^{+,l}
	\ar@{--}@/^/[uuu]
}
\]

Note that the positive moves on the left hand side have copy index $0$, whereas in Example
\ref{ex:contraction} they were carefully chosen so as to avoid collisions. This
makes sense because the current arena has more causal links: the two positive
moves on the left are already made distinct by their justification pointers, so
there is no need to distinguish them further via their copy indices. As this
example illustrates, we cannot simply relabel this $\sim$-strategy to
$\intr{\com}^\perp \parallel (\intr{\com} \times \intr{\com})$ without changing
copy indices, as that would result in a collision, \emph{i.e.} a failure of local
injectivity of the labeling function. 

Therefore, we use countability of the arena in order to do a collision-free relabeling.

\begin{lem}
There is a strong-receptive, courteous map of essps:
\[
\chi_{A, B} : \oc (A\tto B) \to \oc A^\perp \parallel \oc B
\]
which, additionally, preserves the copy index of negative events.
\end{lem}
\begin{proof}
For events $b\in B$ we use $\sharp b$ for the natural number associated to%
\nomenclature[dh]{$\sharp \_$}{Any injective encoding of the moves of an arena $B$ as natural
numbers}%
$b$ by the countability of $B$. As in Section \ref{subsec:replication}, we use $\tuple{-} : \omega^3 \to \omega$
for any injective function; the collision with the pairing operation should not
generate any confusion.

We set:

\[
\begin{array}{rcccl}
\chi_{A, B} &:& \oc (A\tto B) &\to& \oc A^\perp \parallel \oc B\\
&&(\alpha : [(1, (b, a))] \to \omega) &\mapsto& (1, \alpha')\\
&&(\beta : [(2, b)] \to \omega) &\mapsto& (2,\beta')
\end{array}
\]

where:
\[
\begin{array}{rccccl}
\alpha' &:& [a] &\to&\omega\\
&& a' &\mapsto& \tuple{\sharp b, \alpha((2, b)), \alpha((1, (b, a')))} &\text{(if $a'\in
\min(A)$)}\\
&& a' &\mapsto& \alpha((1, (b, a'))) &\text{(otherwise)}
\end{array}
\]

and:
\[
\begin{array}{rcccl}
\beta' &:& [b] &\to& \omega\\
&& b' &\mapsto& \beta((2, b'))
\end{array}
\]

This $\chi_{A, B}$ preserves symmetry, is strong-receptive (since minimal events
of $A^\perp$ are positive) and courteous (it only breaks immediate causal links
from minimal events of $B$ to minimal events of $A^\perp$, so from negative to
positive).
\end{proof}

This allows us, from $\sigma : \S \to \oc C^\perp \parallel \oc (A\tto B)$,
to define its relabeling:
\[
\begin{array}{rcl}
\Phi(\sigma) &:& \S \to \oc C^\perp \parallel (\oc A^\perp \parallel \oc B)\\
&=& (\oc C^\perp \parallel \chi_{A, B})\circ \sigma
\end{array}
\]
For well-chosen hashing function $\sharp$ and injection $\tuple{-}$, this relabeling applied to the
curried contraction above yields exactly the $\sim$-strategy of Example \ref{ex:contraction}.

Before going on to the other direction, we note a further property of this relabeling.

\begin{lem}\label{lem:st_tech}
Let $\sigma : \S \to \oc C^\perp \parallel \oc (A\tto B)$ be a negative single-threaded $\sim$-strategy.
Take $s_1, s_2 \in S$ such that $\sigma\,s_1$ has the form $(2, \beta)$ with $\lbl\,\beta = (2, b)$
($b \in \min(B)$), and $\sigma\,s_2 = (2, \alpha)$ with $\lbl\,\alpha = (1, (b', a))$. Then, $b=b'$ iff $s_1 = \init(s_2)$.
\end{lem}
\begin{proof}
Straightforward consequence of single-threadedness.
\end{proof}

\begin{lem}\label{lem:Phi_sym}
Let $\sigma_1, \sigma_2 : \S \to \oc C^\perp \parallel \oc (A \tto B)$ be two negative single-threaded 
$\sim$-strategies sharing the same internal ess. Then, $\sigma_1 \sim_+ \sigma_2$ iff
$\Phi(\sigma_1) \sim_+ \Phi(\sigma_2)$.
\end{lem}
\begin{proof}
\emph{if}. Assume $\Phi(\sigma_1) \sim_+ \Phi(\sigma_2)$. Take $x\in \conf{S}$, and form
$\theta = \{(\sigma_1\,s, \sigma_2\,s)\mid s \in x \}$.
We wish to prove that $\theta$ is a valid symmetry on $\oc C^\perp \parallel \oc (A\tto B)$. Firstly, we remark
that the following diagram of bijections commutes.
\[
\xymatrix{
&x	\ar[dl]_{\sigma_1}
	\ar[dr]^{\sigma_2}\\
\sigma_1\,x
	\ar[d]_{\oc C^\perp \parallel \chi_{A, B}}
	\ar[rr]^{\theta}&&
\sigma_2\,x
	\ar[d]^{\oc C^\perp \parallel \chi_{A, B}}\\
\Phi(\sigma_1)\,x
	\ar[rr]^{\theta'_{\oc C^\perp} \parallel (\theta'_{\oc A^\perp} \parallel \theta'_{\oc B})}&&
\Phi(\sigma_2)\,x
}
\]

It follows that $\theta$ decomposes as $\theta_{\oc C^\perp} \parallel \theta_{\oc (A\tto B)}$ with
$\theta_{\oc C^\perp} \in \sym{\oc C^\perp}$, and we are left to prove that $\theta_{\oc (A\tto B)}
\in \sym{\oc (A\tto B)}$. By construction it is a bijection, so we need to prove that it preserves
and reflects causality, that it preserves labels, and that it preserves indices of negative events
-- which is clear, as they are preserved throughout this diagram.

We prove that it preserves immediate causality. The only nontrivial case concerns
immediate causal links not preserved by $\chi_{A, B}$, \emph{i.e.} those of the form:
\[
\sigma_1\,s_1 = (2, \{(2, b)\mapsto n\}) \imc (2, \{(2, b)\mapsto n, (1, (b, a))\mapsto p\}) =
\sigma_1\,s_2
\]
But then, by Lemma \ref{lem:st_tech}, we have $s_1 = \init(s_2)$. Since labels are preserved by
$\theta'_{\oc A^\perp}$ and $\theta'_{\oc B}$, and using Lemma \ref{lem:st_tech} again, we still
have $\theta\,(\sigma_2\,s_1) \imc \theta\,(\sigma_2\,s_2)$. The argument also applies to the
$\theta^{-1}$, which therefore is an order-isomorphism.

Preservation of labels also follows directly from Lemma \ref{lem:st_tech}. Finally, $\theta$ is a
positive symmetry as all bijections involved preserve the copy index of negative events.

\emph{only if}. By preservation of symmetry for $\chi_{A, B}$, and the fact that it preserves the
copy index of negative events.
\end{proof}

Relabeling from $\oc C^\perp \parallel (\oc A^\perp \parallel \oc B)$ to $\oc
C^\perp \parallel \oc (A\tto B)$ is more subtle: we go from a
game having one copy of $A$ to one having as many as minimal moves in
$B$. Thus, choosing the label for events formerly mapping to $A$ requires
identifying a copy of $A$ corresponding to some minimal event in $B$. Here condition
\emph{(1)} of single-threadedness is crucial: each $s$ mapped to $A$ has a
unique minimal dependency $\init(s)$ mapped to a minimal event of
$B$, hence specifying the copy of $A$ that $s$ should be sent to. More
formally, we prove:

\begin{lem}
\label{lem:chi_factor}
For any single-threaded $\sim$-strategy $\sigma : \S \to \oc C \parallel (\oc A \parallel \oc B)$,
there is $\sigma' : \S \to \oc C \parallel \oc (A\tto B)$, unique up to positive
symmetry, such that 
$\sigma \sim_+ (\oc C \parallel \chi_{A, B}) \circ \sigma'$.
\end{lem}
\begin{proof}
We define $\sigma' : \S \to \oc C \parallel \oc (A\tto B)$.
For $s\in S$, if $\sigma(s) = (1, \gamma)$ we set $\sigma'(s) = (1, \gamma)$ still.
If $\sigma(s) = (2, (2, \beta))$ with $\beta : [b] \to \omega$, then we set $\sigma'(s) = (2,
\beta')$ with
\[
\begin{array}{rcccl}
\beta' &:& [(2, b)] &\to& \omega\\
&& (2, b') &\mapsto& \beta(b')
\end{array}
\]

If $\sigma(s) = (2, (1, \alpha))$ with $\alpha : [a] \to \omega$, then by condition \emph{(1)} of
single-threadedness it has a unique minimal dependency $\init(s) \leq s$. By hypothesis,
$\sigma\,(\init(s))$ has the form $(2, (2,
\beta))$
with $\beta = \{b\mapsto n\}$. Therefore we set:
\[
\begin{array}{rcccl}
\alpha' &:& [(1, (b, a))] &\to& \omega\\
&& (1, (b, a')) &\mapsto& \alpha(a')\\
&& (2, b) &\mapsto& n
\end{array}
\]
and we define $\sigma'(s) = (2, \alpha')$.

It is routine to check that this map is strong-receptive and courteous, and that its
composition with $\oc C \parallel \chi_{A, B}$ is positively symmetric to $\sigma$. It follows from
Lemma \ref{lem:Phi_sym} that it preserves symmetry, and that it is unique up to positive symmetry.
\end{proof}

From that, we deduce the following.

\begin{prop}\label{prop:phi}
There is a bijection $\Phi$ up to weak isomorphism, preserving and reflecting weak isomorphism,
between:
\begin{itemize}
\item Negative, single-threaded $\sim$-strategies $\sigma : \S \to \oc C^\perp \parallel \oc (A\tto
B)$,
\item Negative, single-threaded $\sim$-strategies $\sigma' : \S \to \oc C^\perp \parallel (\oc A^\perp
\parallel \oc B)$.
\end{itemize}
Moreover this bijection is compatible with pre-composition: for all $\tau : \T \to \oc D^\perp
\parallel \oc C$,
\[
\Phi(\sigma)\odot \tau \simeq \Phi(\sigma \odot \tau)
\]
\end{prop}
\begin{proof}
On the one hand $\Phi(\sigma)$ is obtained as $(\oc C^\perp \parallel \chi_{A, B})\circ \sigma$,
while
$\Phi^{-1}(\sigma')$ is obtained by the unique factorisation of Lemma \ref{lem:chi_factor}. The
bijection
up to weak isomorphism follows from Lemma \ref{lem:chi_factor} as well.

We now prove stability under composition. By definition, we have $\Phi(\sigma) = (\oc C^\perp
\parallel \chi_{A, B})\circ \sigma$. But by Lemma \ref{lem:lifting}
this is the same (up to isomorphism) as
$\lift{\chi_{A, B}} \odot \sigma$, so the action of $\Phi$ can be obtained by post-composition
via a lifted map. Stability under composition follows immediately by associativity of composition.
\end{proof}

And finally, we deduce:

\begin{thm}
The category $\CHO$ is cartesian closed.
\end{thm}
\begin{proof}
We already know that it is cartesian.
Throughout this proof, in the construction of the components of the cartesian closed structure,
we ignore the associativity and unity isomorphisms from
the compact closed structure of $\TCG$ -- those can be easily and uniquely recovered from the
context.

For any two arenas $A, B$, we first define the \emph{evaluation} $\sim$-strategy:
\[
\begin{array}{rcl}
\ev_{A, B} &:& A \times (A \tto B) \choto B\\
&=& (\epsilon_{\oc A} \tensor \oc B) \odot (\oc A \tensor \Phi(\cc_{\oc (A\tto B)})) \odot
\lift{m_{A, A\tto B}}
\end{array}
\]

Likewise, for any $\sigma : A\times C \choto B$, we define its \emph{curryfication} as:
\[
\begin{array}{rcl}
\Lambda(\sigma) &:& C \choto (A\tto B)\\
&=& \Phi^{-1}(\oc A^\perp \tensor (\sigma \odot \lift{m_{A,C}}^{-1})\odot (\eta_{\oc A} \tensor \oc
C))
\end{array}
\]

It is then a straightforward equational reasoning to prove the two equations \cite{lambekscott}, 
for $\sigma : A\times C \choto B$ and $\tau : C \choto (A\tto B)$,

\[
\begin{array}{rrcl}
(\beta) & \ev_{A, B} \odot (A\times \Lambda(\sigma)) &\wiso& \sigma\\
(\eta) & \Lambda(\ev_{A, B} \odot (A\times \sigma)) &\wiso& \sigma
\end{array}
\]
using mainly Proposition \ref{prop:phi} and the compact closed structure of $\TCG$, in
combination with Lemma \ref{lem:prodmon} to relate the cartesian structure of $\CHO$ and the
monoidal structure of $\TCG$ -- all the structural isomorphisms involved in the definition cancel
each other.
\end{proof}

\subsection{Recursion}
\label{subsec:recursion}

As the final technical part of this paper, we prove that $\CHO$ supports the interpretation of a
fixpoint combinator. 

Usually in game semantics, the interpretation of the fixpoint combinator $Y$ is obtained by showing
that the category of games and strategies is enriched over a category of sufficiently complete
partial orders. Here however it will not be the case: indeed, just as in AJM games \cite{ajm}, our
cartesian closed category is a quotient (its morphisms being weak isomorphism classes). It is not
clear that the natural ordering on weak isomorphism classes is complete. 
However, this is not a 
real  issue: although weak isomorphism classes of $\sim$-strategies might not form a
complete partial order, concrete $\sim$-strategies do. Therefore, when solving recursive strategy
equations, we will make sure to work with concrete $\sim$-strategies rather than weak isomorphism
classes.

Our first step will be to order (concrete) $\sim$-strategies.

\begin{defi}
Let $\sigma : \S \to \A$, $\tau : \T \to \A$ be two $\sim$-strategies on a tcg $\A$. We write
$\sigma \inc \tau$ iff $S\subseteq T$, the inclusion map $\S \hookrightarrow \T$ is a map of essps,
with all data in $\S$ coinciding with the restriction of that in $\T$,
and such that for all $s\in S$, $\sigma\,s = \tau\,s$
\end{defi}

The $\sim$-strategies on $\A$ ordered by $\inc$ form a directed complete partial order (dcpo). It is
not \emph{pointed} though -- it does not have a least element. Indeed, although a
$\inc$-minimal $\sim$-strategy only comprises (by receptivity) events matching
minimal negative events of $\A$, their \emph{name} in $\S$ is arbitrary, so there
is one $\inc$-minimal $\sim$-strategy on $\A$ for each renaming of the minimal
negative events of $\A$. For each $\A$ we distinguish one $\inc$-minimal
$\sim$-strategy
\[
\bot_\A : \mathrm{min}^-(\A) \to \A
\]
that has as events the negative minimal events of $\A$ with induced symmetry, and as labeling
function the identity.
Not every $\sim$-strategy is above $\bot_\A$. However, for every $\sim$-strategy $\sigma : \S \to
\A$, we pick one $\sigma \siso \sigma^\dagger$ such that $\bot_\A \inc \sigma^\dagger$ obtained
by renaming the minimal negative events of $S$.
We write $\D_\A$ for the pointed dcpo of $\sim$-strategies above $\bot_\A$.

\begin{lem}
For any tcg $\A$, $\D_\A$ is a pointed dcpo with $\bot_\A$ as minimal element.
\end{lem}
\begin{proof}
If $\Gamma = \{\gamma : \S_\gamma \to \A \} \subseteq \D_\A$ is a directed subset of $\D_\A$, we
form
\[
\vee \Gamma = \cup \gamma : \bigcup_{\gamma \in \Gamma} S_\gamma \to \A
\]
with all components defined as componentwise union. 

It is direct that this defines a $\sim$-strategy, which is the least upper bound of
$\Gamma$. 
\end{proof}

If all $\sim$-strategies in a directed set $\Gamma$ are negative or
single-threaded, so is $\vee \Gamma$.
We now note that all the operations we defined on $\sim$-strategies in this section are continuous
for $\inc$.

\begin{lem}\label{lem:comb_continuous}
Composition, tensor, pairing, curryfication and the $(-)^\dagger$ operation defined above are
continuous for $\inc$.
\end{lem}
\begin{proof}
Straightforward.
\end{proof}

From the above, we deduce the following.

\begin{cor}
For any arena $A$ there is a fixpoint combinator $\Y_A : (A\tto A) \choto A$, \emph{i.e.} a
single-threaded $\sim$-strategy such that:
\[
\Y_A \wiso ev_{A, A} \odot \tuple{\Y_A, \cc_{\oc (A\tto A)}}
\]
\end{cor}
\begin{proof}
First, we define the following operation, using the combinators on $\CHO$.
\[
\begin{array}{rcrcl}
F &:& \mathcal{D}_{\oc (A\to A)^\perp \parallel \oc A} &\to& \mathcal{D}_{\oc (A\to A)^\perp
\parallel \oc A}\\
&& \sigma &\mapsto& (ev_{A, A} \odot \tuple{\sigma, \cc_{\oc (A\tto A)}})^\dagger
\end{array}
\]
By Lemma \ref{lem:comb_continuous} it is continuous, and from the outermost dagger it has
indeed value in $\D_{\oc (A\tto A)^\perp \parallel A}$. Thus, we can take its least fixpoint
$\Y_A \in \D_{\oc (A\tto A)^\perp \parallel A}$. The weak isomorphism in the statement actually
follows as an \emph{equality}.
\end{proof}


\section{Interpretation of IPA}
\label{sec:interpretation}
In this section, we illustrate our model by defining the interpretation
of IPA, displaying the interpretation of programs of interest, and proving a few
properties along the way.

We 
emphasise here that our purpose is \emph{not} to prove full abstraction, nor to prove deep
properties of the interpretation. We feel indeed that given the length of the paper, the specifics
of such an endeavour are best left for later. Furthermore, it is our impression that it serves the
purpose of this paper better (introducing and developing Concurrent Hyland-Ong games) to give the
reader an understanding of what the model computes, what it can and cannot do, rather than delve
into additional technical developments.

Throughout this section, by \emph{strategy} we mean \emph{$\sim$-strategy} 
(symmetries will be implicit).

\subsection{Sequential innocent part}

In this subsection, we focus on the interpretation of the (sequential) innocent
part of IPA, \emph{i.e.} essentially PCF, plus the combinators for commands. In
other words, it lacks state and parallel composition of commands.

\subsubsection*{Interpretation of types.} The arenas for the types $\com$ and $\tbool$ were given in Example
\ref{ex:arenas}. The interpretation for $\tnat$ is a countably infinite variant of the
interpretation of $\tbool$:
\[
\xymatrix@C=5pt@R=10pt{
&\!\!\!\intr{\tnat}\!\!\!\\
&\!\!\!\q^{-}\!\!\!       \ar@{--}@/_/[dl]
        \ar@{--}[d]
	\ar@{--}@/^/[drr]\\
\!\!\!0^{+}\!\!\!&1^{+}& \dots & n^{+} & \dots
}
\]

The interpretation extends to all types (not containing $\var$), with 
$\intr{A\to B} = \intr{A}\tto \intr{B}$. 

\subsubsection*{Interpretation of terms.} The interpretation follows the standard lines of the
interpretation of the $\lambda$-calculus in a cartesian closed category. A \emph{context}
$\Gamma = x_1 : A_1, \dots, x_n : A_n$ is interpreted as the product
$\Pi_{1\leq i \leq n} \intr{A_i}$
(which is just the parallel composition of the $A_i$s). A \emph{typing sequent} $\Gamma \vdash M :
A$ is interpreted as a $\CHO$-morphism:
\[
\intr{\Gamma \vdash M : A} : \intr{\Gamma} \choto \intr{A}
\]

For the $\lambda$-calculus combinators -- variables, application, abstraction --, the 
interpretation is standard (and we do not detail it). For the fixpoint combinators, we use the
combinator $\Y$ of Section \ref{subsec:recursion}. The interpretation of constants is displayed in
Figure \ref{fig:constants}. Note that we only display representations, treating multiple copies of
Opponent moves symbolically. The reader should be able to expand them unambiguously to the full
event structures, and to detail their isomorphism families. Note also that we give these
interpretations over the empty context -- they can easily be relabeled to any context $\Gamma$.

\begin{figure}
\begin{mathpar}
\intr{\ttrue} = 
\xymatrix@R=10pt@C=0pt{
\!\!\!\!\!\!\oc \intr{\tbool}\\
\q^{-,i}
	\ar@{-|>}[d]\\
\ttrue^{+,0}
	\ar@{--}@/^/[u]
}
\and
\!\!\!\!\!\!\intr{\tfalse} =
\xymatrix@R=10pt@C=0pt{
\oc \intr{\tbool}\\
\q^{-,i}
        \ar@{-|>}[d]\\
\tfalse^{+,0}
        \ar@{--}@/^/[u]
}
\and
\!\!\!\!\!\!\intr{n} =
\xymatrix@R=10pt@C=0pt{
\oc \intr{\tnat}\\
\q^{-,i}
        \ar@{-|>}[d]\\
n^{+,0}
        \ar@{--}@/^/[u]
}
\and
\!\!\!\!\!\!\intr{\tskip} =
\xymatrix@R=10pt@C=0pt{
\oc \intr{\com}\\
\run^{-,i}
        \ar@{-|>}[d]\\
\done^{+,0}
        \ar@{--}@/^/[u]
}
\end{mathpar}
\caption{Interpretation of constants of IPA}
\label{fig:constants}
\end{figure}

Likewise, the interpretation of function symbols is given in Figure \ref{fig:functions}. We have
only one figure for a unary function $\mathrm{op} : \tx \to \tx$, which covers (up to obvious
relabeling) the cases of $\tsucc : \tnat \to \tnat, \pred : \tnat \to \tnat$ and $\iszero : \tnat
\to \tbool$. The interpretation of sequents involving those follows as usual, with \emph{e.g.} the
following composition in $\CHO$:
\[
\intr{\Gamma \vdash \ifpcf\,M\,N_1\,N_2: \tx} = \intr{\ifpcf} \odot \tuple{\intr{M}, \intr{N_1},
\intr{N_2}} : \intr{\Gamma} \choto \intr{\tx}
\]

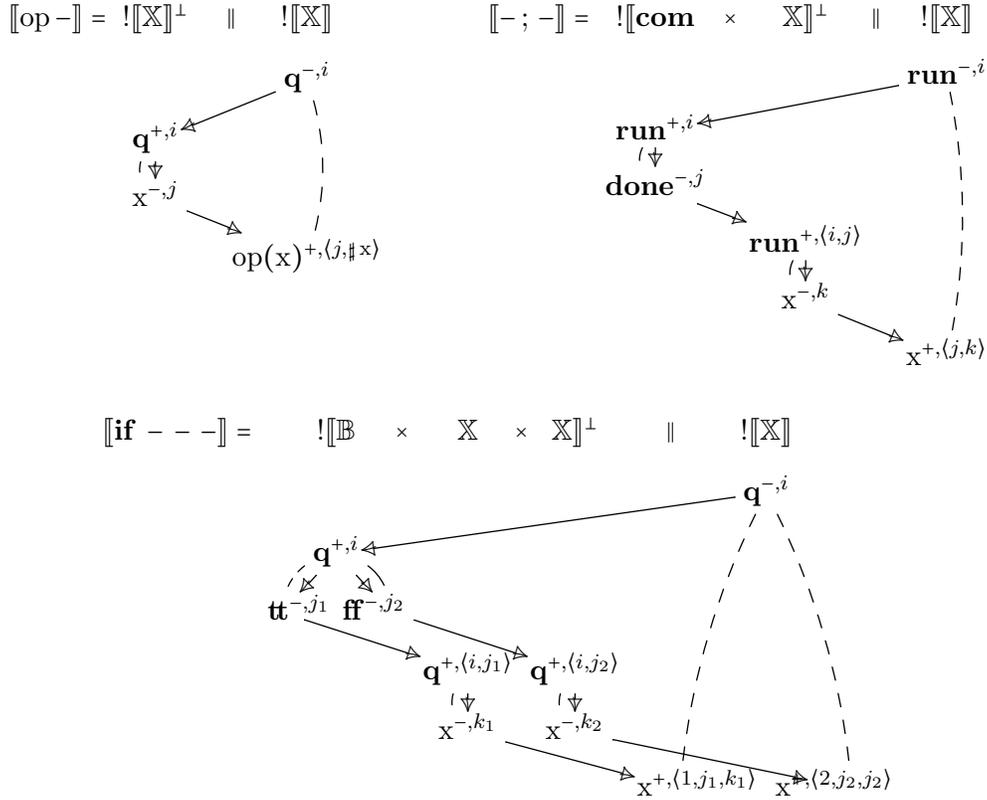
\begin{figure}
\begin{mathpar}
\intr{\mathrm{op}\,-} = 
\xymatrix@R=5pt@C=10pt{
\oc \intr{\tx}^\perp
	\ar@{}[r]|\parallel&
\oc \intr{\tx}\\
&\q^{-, i}
	\ar@{-|>}[dl]\\
\q^{+,i}\ar@{-|>}[d]\\
\mathrm{x}^{-,j}
	\ar@{--}@/^/[u]
	\ar@{-|>}[dr]\\
&\mathrm{op}(\mathrm{x})^{+,\tuple{j,\sharp \mathrm{x}}}
	\ar@{--}@/_/[uuu]
}
\and
\intr{-\,;\,-}=
\xymatrix@R=5pt@C=10pt{
\oc \llbracket\com
	\ar@{}[r]|\times&
\tx\rrbracket^\perp
	\ar@{}[r]|\parallel&
\oc \intr{\tx}\\
&&\run^{-,i}
	\ar@{-|>}[dll]\\
\run^{+,i}\ar@{-|>}[d]\\
\done^{-,j}
	\ar@{--}@/^/[u]
	\ar@{-|>}[dr]\\
&\run^{+,\tuple{i,j}}
	\ar@{-|>}[d]\\
&\mathrm{x}^{-,k}
	\ar@{--}@/^/[u]
	\ar@{-|>}[dr]\\
&&\mathrm{x}^{+,\tuple{j,k}}
	\ar@{--}@/_/[uuuuu]
}
\and
\intr{\ifpcf\,-\,-\,\,-} =
\xymatrix@R=5pt@C=0pt{
&\oc \llbracket \tbool
	\ar@{}[rr]|\times&&
\tx	\ar@{}[r]|\times&
\tx \rrbracket^\perp
	\ar@{}[rr]|\parallel&&
\oc \intr{\tx}\\
&&&&&&\q^{-,i}
	\ar@{-|>}[dlllll]\\
&\q^{+,i}
	\ar@{-|>}[dl]
	\ar@{-|>}[dr]\\
\ttrue^{-,j_1}\!\!\!\!\!\!\!\!
	\ar@{--}@/^/[ur]
	\ar@{-|>}[drrr]&&
\!\!\!\!\!\!\!\!\tfalse^{-,j_2}
	\ar@{--}@/_/[ul]
	\ar@{-|>}[drr]\\
&&&\q^{+, \tuple{i,j_1}}
	\ar@{-|>}[d]&
\q^{+,\tuple{i,j_2}}
	\ar@{-|>}[d]\\
&&&\mathrm{x}^{-,k_1}
	\ar@{--}@/^/[u]
	\ar@{-|>}[drr]&
\mathrm{x}^{-,k_2}\ar@{--}@/^/[u]
	\ar@{-|>}[drrr]\\
&&&&&\mathrm{x}^{+, \tuple{1,j_1, k_1}}\!\!\!\!\!\!\!\!
	\ar@{--}@/^/[uuuuur]&&
\!\!\!\!\!\!\!\!\mathrm{x}^{+,\tuple{2,j_2, j_2}}
	\ar@{--}@/_/[uuuuul]
}
\end{mathpar}
\caption{Interpretation of sequential function symbols of IPA}
\label{fig:functions}
\end{figure}

At this point, we have defined the interpretation of 
the sequential innocent fragment of IPA. Using the cartesian closed
structure and the definition of the fixpoint combinator, it would be
straightforward to prove soundness and adequacy of the
interpretation, \emph{e.g.} using logical relations. We refrain from detailing
this -- rather standard -- proof.
%

The paper already contains some examples of the interpretation of terms of the
fragment of IPA currently under study, most notably in Section
\ref{sec:replication} -- where for some, copy indices need to be adequately
adjoined. The interpretation of such terms yields rather simple event structures,
whose causal order is forest-shaped and without conflict. Modulo copy
indices, and as it was noted in Section \ref{sec:replication}, these forests
exactly coincide with the \emph{view functions} of standard Hyland-Ong games:
their branches are exactly the $P$-views. Hence, our interpretation computes the
composition of innocent strategies while staying within a causal representation
corresponding to view functions, never resorting to expanded plays. 

\subsubsection*{Non-determinism.} Although the fragment of the language currently under study is
deterministic, we find it interesting to study some examples given by its extension with a
non-deterministic primitive. Therefore, we add to the language a new constant $\coin : \tbool$ which
returns a random boolean. Its interpretation is (an obvious extension with copy indices of) the
strategy on the left hand side of Figure \ref{fig:nd_dialogues}. For $\Gamma \vdash M, N: A$, we
define as syntactic sugar a non-deterministic sum $\Gamma \vdash M + N : A$ as 
$\Gamma \vdash \ifpcf\,\coin\,M\,N : A$. 

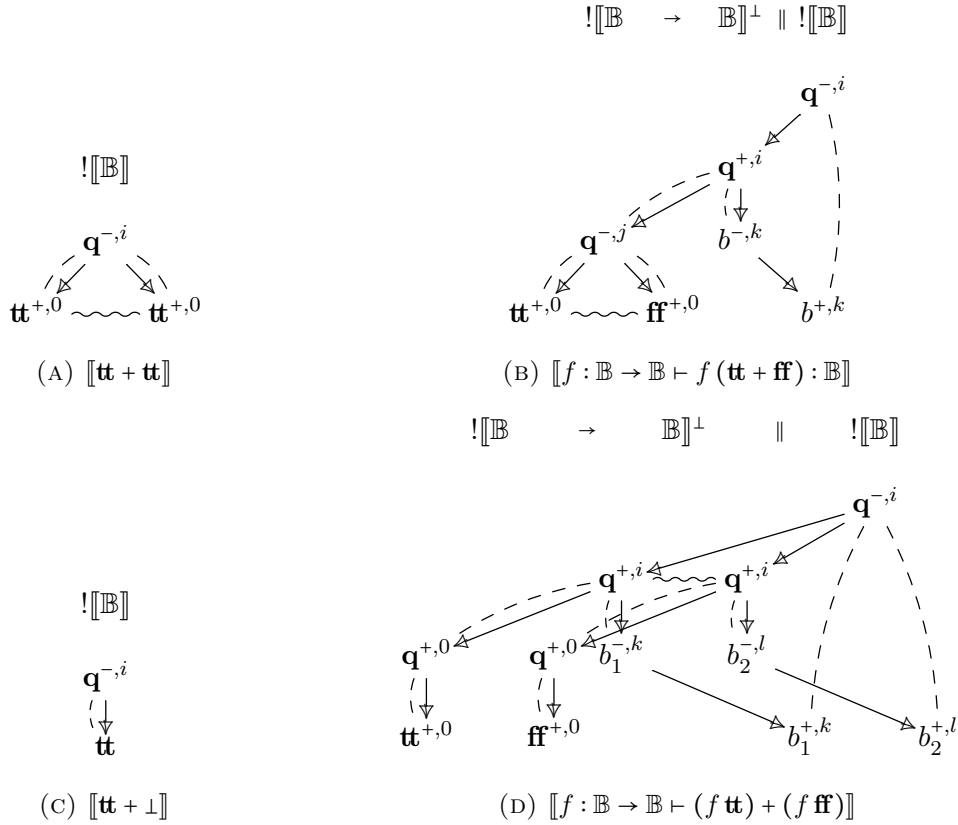
\begin{figure}[t!]
    \centering
    \begin{subfigure}[b]{0.3\textwidth}
        \centering
	\[
	\xymatrix@R=10pt@C=0pt{
	&\oc \intr{\tbool}\\
	&\q^{-,i}
	        \ar@{-|>}[dl]
	        \ar@{-|>}[dr]\\
	\ttrue^{+,0}
	        \ar@{--}@/^/[ur]
	        \ar@{~}[rr]&&
	\ttrue^{+,0}
	        \ar@{--}@/_/[ul]
	}
	\]
	\caption{$\intr{\ttrue + \ttrue}$}
	\label{subfig:non_idem}
    \end{subfigure}%
    \begin{subfigure}[b]{0.7\textwidth}
        \centering
	\[
	\xymatrix@R=10pt@C=0pt{
	&\oc \llbracket \tbool
	        \ar@{}[rr]|\to&&
	\tbool \rrbracket^\perp
	        \ar@{}[rr]|\parallel&&
	\oc \intr{\tbool}\\
	&&&&&\q^{-,i}
	        \ar@{-|>}[dll]\\
	&&&\q^{+,i}
	        \ar@{-|>}[dll]
	        \ar@{-|>}[d]\\
	&\q^{-,j}
	        \ar@{-|>}[dl]
	        \ar@{-|>}[dr]
	        \ar@{--}@/^/[urr]&&
	b^{-,k} \ar@{--}@/^/[u]
	        \ar@{-|>}[drr]\\
	\ttrue^{+,0}
	        \ar@{--}@/^/[ur]
	        \ar@{~}[rr]&&
	\tfalse^{+,0}
	        \ar@{--}@/_/[ul]&&&
	b^{+,k}        \ar@{--}@/_/[uuu]
	}
	\]
        \caption{$\intr{f:\tbool \to \tbool \vdash f\,(\ttrue + \tfalse) : \tbool}$}
	\label{subfig:branching1}
    \end{subfigure}\\
    \begin{subfigure}[b]{0.3\textwidth}
        \centering
        \[
        \xymatrix@R=10pt@C=0pt{
	\oc \intr{\tbool}\\
	\q^{-,i}\ar@{-|>}[d]\\
	\ttrue  \ar@{--}@/^/[u]
	}
        \]
        \caption{$\intr{\ttrue + \bot}$}
	\label{subfig:may}
    \end{subfigure}%
    \begin{subfigure}[b]{0.7\textwidth}
        \centering
        \[
	\xymatrix@R=10pt@C=0pt{
	&\oc \llbracket \tbool
	        \ar@{}[rrr]|\to&&&
	\tbool \rrbracket^\perp
	        \ar@{}[rrr]|\parallel&&&
	\oc \intr{\tbool}\\
	&&&&&&&\q^{-,i}
	        \ar@{-|>}[dllll]
		\ar@{-|>}[dll]\\
	&&&\q^{+,i}
		\ar@{~}[rr]
		\ar@{-|>}[d]
		\ar@{-|>}[dlll]&&
	\q^{+,i}\ar@{-|>}[d]
		\ar@{-|>}[dlll]\\
	\q^{+,0}\ar@{--}@/^/[urrr]
		\ar@{-|>}[d]&&
	\q^{+,0}\ar@{--}@/^/[urrr]
		\ar@{-|>}[d]&
	b_1^{-,k}	\ar@{--}@/^/[u]
		\ar@{-|>}[drrr]
	&&b_2^{-,l}
		\ar@{--}@/^/[u]
		\ar@{-|>}[drrr]\\
	\ttrue^{+,0}
		\ar@{--}@/^/[u]&&
	\tfalse^{+,0}
		\ar@{--}@/^/[u]&&&&
	b_1^{+,k}
		\ar@{--}@/^/[uuur]&&
	b_2^{+,l}
		\ar@{--}@/_/[uuul]
	}
        \]
        \caption{$\intr{f:{\tbool \to \tbool}\vdash (f\,\ttrue) + (f\,\tfalse)}$}
	\label{subfig:branching2}
    \end{subfigure}
    \caption{Intepretation of some non-deterministic terms}
    \label{fig:nondet_ex}
\end{figure}

We give in Figure \ref{fig:nondet_ex} representations of the interpretation of some well-chosen
terms. Copy indices are not exactly as given by the interpretation function (though they are up to
weak isomorphism): they have been relabeled for convenience of presentation. 

As Figure \ref{subfig:non_idem} illustrates, the model represents non-determinism in a
\emph{non-idempotent way}: redundant non-deterministic choices are kept separate by the
interpretation. In Figure \ref{subfig:may}, $\bot$ (which is syntactic sugar for $\Y\,(\lambda
x^\tbool.x)$) is interpreted as the empty strategy. The interpretation of $\ttrue + \bot$
illustrates that, despite displaying explicitely the point of non-deterministic branching, the
hiding step of the interpretation removes some diverging branches of the interaction. 
Figures \ref{subfig:branching1} and \ref{subfig:branching2} display two strategies which have the
same branches (P-views), but differ in their branching points. This gives an interpretation of
non-deterministic sequential programs that is similar to Tsukada and Ong's recent presheaf-based
model \cite{DBLP:conf/lics/TsukadaO15}, although our composition mechanism is very different.
It is fairly easy to capture exactly their category as a subcategory of $\CHO$,
whose morphisms are \emph{sequential innocent} \cite{lics14,lics15} but not deterministic.

\subsection{Concurrent innocent part}

Now, we go on to show how our model represents concurrent primitives. The only concurrent primitive
of IPA is parallel composition, whose interpretation relies on the following strategy
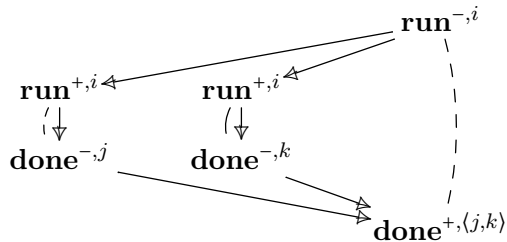
\[
\intr{\,\parallel\,} = 
\xymatrix@R=10pt{
\oc \llbracket \com
	\ar@{}[r]|\times&
\com \rrbracket^\perp
	\ar@{}[r]|\parallel&
\intr{\com}\\
&&\run^{-,i}
	\ar@{-|>}[dll]
	\ar@{-|>}[dl]\\
\run^{+,i}
	\ar@{-|>}[d]&
\run^{+,i}
	\ar@{-|>}[d]\\
\done^{-,j}
	\ar@{--}@/^/[u]
	\ar@{-|>}[drr]&
\done^{-,k}
	\ar@{--}@/^/[u]
	\ar@{-|>}[dr]\\
&&\done^{+,\tuple{j,k}}
	\ar@{--}@/_/[uuu]
}
\]

Using this strategy we can define $\intr{\Gamma \vdash M \parallel N : \com} = \intr{\,\parallel\,}
\odot \tuple{\intr{M}, \intr{N}}$.

This strategy is no longer a forest, but rather a directed acyclic graph. We
also note that this is a \emph{deterministic} strategy: there is no conflict in
its event structure. As we shall see later, without any non-deterministic
primitive, it is only in the presence of shared state that non-deterministic
strategies will arise. In fact, a major advantage of our approach to modeling
concurrent languages is that, not being based on interleavings, we represent the
execution of such non-interfering terms deterministically.

In \cite{lics15}, we exploit this property: we give a concurrent notion of innocence where
strategies are directed acyclic graphs rather than forests, and using this notion we give an
intensionally fully abstract model of a variant of PCF where independent computations are performed
in parallel. The detailed construction is out of the scope of this particular paper, but let us
illustrate it with two examples that are both concurrent innocent.
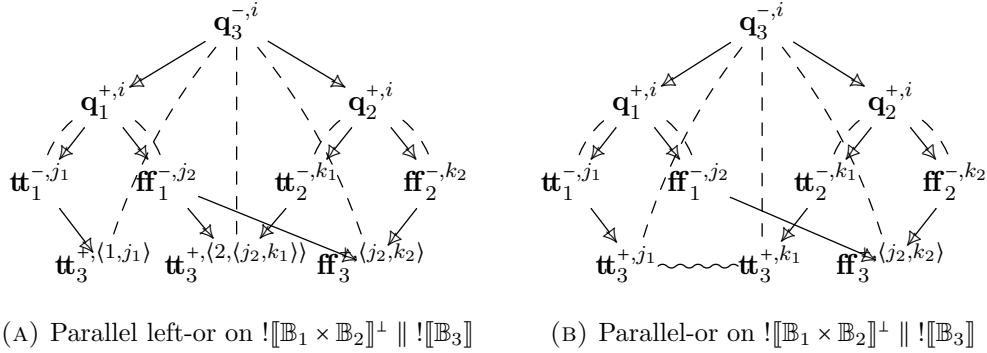
\begin{figure}
\begin{subfigure}[b]{0.45\textwidth}
\[
\xymatrix@R=10pt@C=0pt{
&&&\q^{-,i}_3
	\ar@{-|>}[dll]
	\ar@{-|>}[drr]\\
&\q^{+,i}_1
	\ar@{-|>}[dl]
	\ar@{-|>}[dr]&&&&
\q^{+,i}_2
	\ar@{-|>}[dl]
	\ar@{-|>}[dr]\\
\!\!\ttrue_1^{-,j_1}\!\!
	\ar@{-|>}[dr]
	\ar@{--}@/^/[ur]&&
\!\!\tfalse_1^{-,j_2}\!\!
	\ar@{-|>}[dr]
	\ar@{-|>}[drrr]
	\ar@{--}@/_/[ul]&&
\!\!\ttrue_2^{-,k_1}\!\!
	\ar@{-|>}[dl]
	\ar@{--}@/^/[ur]&&
\!\!\tfalse_2^{-,k_2}\!\!
	\ar@{-|>}[dl]
	\ar@{--}@/_/[ul]\\
&\!\!\!\!\!\!\!\!\!\!\ttrue^{+,\tuple{1, j_1}}_3\!\!\!\!\!\!\!\!\!\!
	\ar@{--}@/^/[uuurr]&&
\!\!\!\!\!\!\!\!\!\!\ttrue^{+,\tuple{2, \tuple{j_2, k_1}}}_3\!\!\!\!\!\!\!\!\!\!
	\ar@{--}[uuu]&&
\!\!\!\!\!\!\!\!\!\!\tfalse^{+, \tuple{j_2, k_2}}_3\!\!\!\!\!\!\!\!\!\!
	\ar@{--}@/_/[uuull]
}
\]
\caption{Parallel left-or on $\oc \intr{\tbool_1 \times \tbool_2}^\perp \parallel \oc \intr{\tbool_3}$}
\label{subfig:left-or}
\end{subfigure}
\begin{subfigure}[b]{0.45\textwidth}
\[
\xymatrix@R=10pt@C=0pt{
&&&\q^{-,i}_3
        \ar@{-|>}[dll]
        \ar@{-|>}[drr]\\
&\q^{+,i}_1
        \ar@{-|>}[dl]
        \ar@{-|>}[dr]&&&&
\q^{+,i}_2
        \ar@{-|>}[dl]
        \ar@{-|>}[dr]\\
\!\!\ttrue_1^{-,j_1}\!\!
        \ar@{-|>}[dr]
        \ar@{--}@/^/[ur]&&
\!\!\tfalse_1^{-,j_2}\!\!
        \ar@{-|>}[drrr]
        \ar@{--}@/_/[ul]&&
\!\!\ttrue_2^{-,k_1}\!\!
        \ar@{-|>}[dl]
        \ar@{--}@/^/[ur]&&
\!\!\tfalse_2^{-,k_2}\!\!
        \ar@{-|>}[dl]
        \ar@{--}@/_/[ul]\\
&\!\!\!\!\!\!\ttrue^{+,j_1}_3\!\!
	\ar@{~}[rr]
        \ar@{--}@/^/[uuurr]&&
\!\!\ttrue^{+, k_1}_3\!\!\!\!\!\!
        \ar@{--}[uuu]&&
\!\!\!\!\!\!\!\!\!\!\tfalse^{+, \tuple{j_2, k_2}}_3\!\!\!\!\!\!\!\!\!\!
        \ar@{--}@/_/[uuull]
}
\]
\caption{Parallel-or on $\oc \intr{\tbool_1 \times \tbool_2}^\perp \parallel \oc \intr{\tbool_3}$}
\label{subfig:parallel-or}
\end{subfigure}
\caption{Two concurrent innocent strategies}
\label{fig:conc_inn}
\end{figure}

Figure \ref{fig:conc_inn} displays two concurrent innocent strategies
(we associate moves to the corresponding sub-type using indices rather
than location). In Figure \ref{subfig:left-or}, we have a strategy for a parallel
implementation of the left or, that is strict in its left argument. Indeed,
although the strategy starts evaluating both its arguments in parallel, it can
only return at toplevel if its first argument has returned. However, this is not
true anymore for the strategy of Figure \ref{subfig:parallel-or}. There, it
suffices that \emph{one} argument returns $\ttrue$ for the overall computation to
return $\ttrue$ -- indeed, this strategy computes the well-known
\emph{parallel-or} function \cite{plotkin}.

\subsection{Stateful part}
\label{subsec:state}

Finally, we finish the interpretation of IPA and describe how to interpret the
primitives dealing with manipulations of state. For the simplicity of
presentation, references only store booleans; however the method
applies just as well to integers. 

A variable can be interacted with in two ways: via reading and writing. As usual in game semantics,
we follow this idea for the interpretation of variables, and take $\intr{\var}$ to be a product
arena comprising actions for reading the reference or writing on the reference. More precisely, we
define:
\[
\intr{\var} = 
\raisebox{15pt}{
\xymatrix@C=5pt@R=10pt{
&\tread^-
	\ar@{--}@/_/[dl]
	\ar@{--}@/^/[dr]&&
\twrite_\ttrue^-
	\ar@{--}[d]&
\twrite_\tfalse^-
	\ar@{--}[d]\\
\ttrue^+&&\tfalse^+&\tok^+&\tok^+
}}
\]

We now describe the interpretation of term constructors for the manipulation of state. As usual,
assignment and dereferenciation are simply interpreted as (sequential innocent) strategies that
interact with the memory cell. We give in Figure \ref{fig:comb_state} the strategies used in the
interpretation of those.
\begin{figure}
\begin{subfigure}[b]{0.45\textwidth}
\[
\xymatrix@R=5pt{
\oc \llbracket \var
	\ar@{}[r]|\times&
\tbool \rrbracket^\perp
	\ar@{}[r]|\parallel&
\oc \intr{\com}\\
&&\run^{-,i}
	\ar@{-|>}[dl]\\
&\q^{+,i}
	\ar@{-|>}[d]\\
&b^{-,j}\ar@{-|>}[dl]
	\ar@{--}@/^/[u]\\
\twrite_b^{\tuple{i,j}}
	\ar@{-|>}[d]\\
\tok^{-,k}
	\ar@{--}@/^/[u]
	\ar@{-|>}[drr]\\
&&\done^{+,\tuple{j,k,\sharp b}}
	\ar@{--}@/_/[uuuuu]
}
\]
\caption{$\intr{-\,:=\,-}$}
\label{subfig:assign}
\end{subfigure}
\begin{subfigure}[b]{0.45\textwidth}
\[
\xymatrix@R=10pt{
\oc \intr{\var}^\perp
	\ar@{}[r]|\parallel&
\oc \intr{\tbool}\\
&\q^{-,i}
	\ar@{-|>}[dl]\\
\tread^{+,i}
	\ar@{-|>}[d]\\
b^{-,j}	\ar@{--}@/^/[u]
	\ar@{-|>}[dr]\\
&b^{+,j}\ar@{--}@/_/[uuu]
}
\]
\caption{$\intr{!\,-}$}
\label{subfig:deref}
\end{subfigure}
\caption{Strategies for assignment and dereferenciation}
\label{fig:comb_state}
\end{figure}
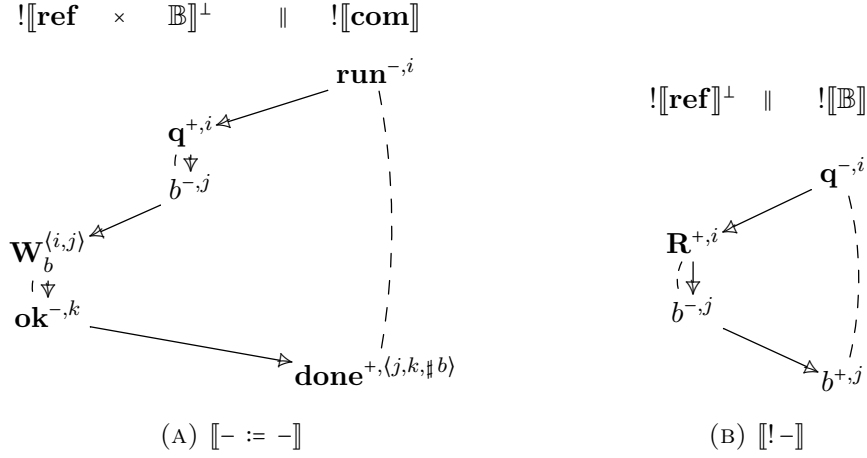
Using those, we can define:
\[
\begin{array}{rcl}
\intr{\Gamma \vdash M\,:=\,N: \com} &=& \intr{-\,:=\,-} \odot \tuple{\intr{M}, \intr{N}}\\
\intr{\Gamma \vdash \oc\,M : \tbool} &=& \intr{!\,-} \odot \intr{M}
\end{array}
\]

Before giving the interpretation of genuine references, we mention that the interpretation of
$\mkvar$ exploits as usual the isomorphism between $\intr{\var}$ and $\intr{\tbool} \times
\intr{com}^2$ \cite{am}.

\subsubsection*{New reference.}
As usual, the subtle part is the interpretation of $\newref$.
Indeed, whereas the strategies for assignment and dereferenciation only interact
with the interface of the variable in an innocent way, it is $\newref$ that
provides an implementation for the memory. 

If $\Gamma, x:\var \vdash M : A$ depends on a reference $x$, its interpretation plays on (up to
iso) $\oc \intr{\var}^\perp \parallel \oc \intr{\Gamma}^\perp \parallel \oc \intr{A}$.
Naively (we will see that this is a slight simplification), all we have to do is to build a strategy
$\cell : \oc \intr{\var}$, and compose $\intr{M}$ with it to obtain $\intr{\newref\,x\,\tin\,M}$. 

To define $\cell$, we keep in mind the operational behaviour of a memory cell. In our (sequentially
consistent) understanding of memory in a concurrent setting, although reads and writes are
called concurrently, they are performed in some sequential
order by the central memory. Thus the behaviour of a boolean memory cell is best described as the
prefix language of the infinite traces: 

\[
\begin{array}{rcl}
\Cell_{\ttrue} &::=& \tread^- \cdot \ttrue^+ \cdot \Cell_{\ttrue} \mid \twrite_{\ttrue}^- \cdot \tok^+ \cdot
\Cell_{\ttrue} \mid \twrite_{\tfalse}^-\cdot \tok^+ \cdot \Cell_{\tfalse}\\
\Cell_{\tfalse} &::=& \tread^- \cdot \tfalse^+ \cdot \Cell_{\tfalse} \mid \twrite_{\ttrue}^- \cdot \tok^+ \cdot
\Cell_{\ttrue} \mid \twrite_{\tfalse}^-\cdot \tok^+ \cdot \Cell_{\tfalse}
\end{array}
\]

This language is ordered by prefix, so that $\Cell_{\tfalse}$ is a forest. Setting all
incomparable words to conflict with each other, we get an event structure whose events are
words, and configurations are prefix-closed sets of prefixes of a word -- so in one-to-one
correspondence with words. This event structure, with the obvious labeling function, can be regarded as a
prestrategy on $\intr{\var}$ (not on $\oc \intr{\var}$). But in order to fit in our framework, we
need to equip it with copy indices (and symmetry). This calls for extra bookkeeping, as we need to
make sure that the same copy index is not used twice in the same branch. We define

\[
\begin{array}{rcll}
\Cell_{\ttrue}^{I_{\tread}, I_\ttrue, I_\tfalse} &::=& \tread^{-, i} \cdot \ttrue^{+, 0} \cdot
\Cell_{\ttrue}^{I_\tread \cup \{i\}, I_\ttrue, I_\tfalse}  & (i\not \in I_\tread) \\
&& \mid \twrite_\ttrue^{-,i} \cdot \tok^{+,0} \cdot \Cell_{\ttrue}^{I_{\tread}, I_\ttrue \cup \{i\},
I_\tfalse} & (i\not \in I_{\ttrue})\\
&& \mid \twrite_\tfalse^{-,i} \cdot \tok^{+,0} \cdot \Cell_{\tfalse}^{I_{\tread}, I_\ttrue,
I_\tfalse \cup \{i\}} & (i\not \in I_{\tfalse})
\end{array}
\]
and similarly for $\Cell_{\tfalse}^{I_{\tread}, I_\ttrue, I_\tfalse}$. Then we define the event
structure $\Cell$ via $\Cell_{\tfalse}^{\emptyset, \emptyset, \emptyset}$, as we did above. It has an
isomorphism family, that relates any two words differing only on their copy indices.
Moreover the names of the events denote the labeling function to $\oc \intr{\var}$ (with all positive moves pointing
-- that is, being immediately dependent in the game -- to the prevous move). Overall, we get a map
of essp:
\[
\cell : \Cell \to \oc \intr{\var}
\]

\begin{exa}
The following diagram represents a sub-event structure of $\Cell$.
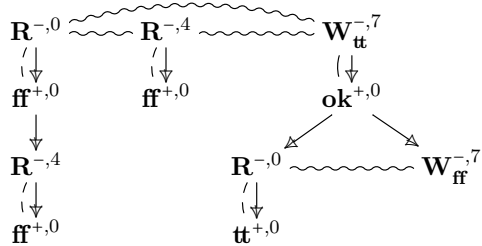
\[
\scalebox{.9}{$
\xymatrix@R=10pt@C=10pt{
\tread^{-,0}
	\ar@{~}[rr]
	\ar@{~}@/^1pc/[rrrr]
	\ar@{-|>}[d]&&
\tread^{-,4}
	\ar@{~}[rr]
	\ar@{-|>}[d]&&
\twrite_{\ttrue}^{-,7}
	\ar@{-|>}[d]\\
\tfalse^{+,0}
	\ar@{--}@/^/[u]
	\ar@{-|>}[d]&&
\tfalse^{+,0}
	\ar@{--}@/^/[u]&&
\tok^{+,0}
	\ar@{--}@/^/[u]
	\ar@{-|>}[dl]
	\ar@{-|>}[dr]\\
\tread^{-,4}
	\ar@{-|>}[d]&&&
\tread^{-,0}
	\ar@{~}[rr]
	\ar@{-|>}[d]&&
\twrite_{\tfalse}^{-,7}\\
\tfalse^{+,0}
	\ar@{--}@/^/[u]&&&
\ttrue^{+,0}
	\ar@{--}@/^/[u]
}$}
\]
\end{exa}

We have constructed $\cell : \Cell \to \oc \intr{\var}$ a map of essps.
Unfortunately, $\cell$ is \emph{not} a $\sim$-strategy: it is neither receptive
(after playing $\tread^{-,4}$ above one cannot play $\tread^{-,4}$, although it
is compatible in the game) nor courteous (we have $\tfalse^{+,0} \imc
\tread^{-,4}$ which does not hold in $\oc \intr{\var}$). However, $\cell$ is a
\emph{thin pre-$\sim$-strategy}, and as such can be composed with $\intr{M} : \oc
\intr{\var}^\perp \parallel \oc \intr{A}$ to obtain $\intr{M}\odot \cell$ -- and
it turns out that $\intr{M}\odot \cell$ is \emph{always} a valid
$\sim$-strategy.

Still, that is not quite what we want. The intended semantics for
$\newref\,x\,\tin\,M$ is that each of its evaluations spawns a new, independent
memory cell, whereas the operation above would have it spawned once and for all
and shared over all copies of $M$.  In other words, $\intr{M}\odot \cell$ is a
valid $\sim$-strategy indeed, but it might not be \emph{single-threaded}. So
finally, we build another pre-$\sim$-strategy displayed in Figure
\ref{fig:newcell}, where $\Cell$ means a copy of the pre-$\sim$-strategy above,
with minimal events pointing as indicated.

\begin{figure}
\[
\xymatrix@R=5pt@C=10pt{
\oc \llbracket \var
	\ar@{}[r]|\to&
\tx \rrbracket^\perp
	\ar@{}[r]|\parallel&
\oc \intr{\tx}\\
&&\q^{-,i}
	\ar@{-|>}[dl]\\
&\q^{+,i}
	\ar@{-|>}[dl]
	\ar@{-|>}[d]\\
\Cell	\ar@{--}@/^/[ur]&
\mathrm{x}^{-,j}
	\ar@{-}@/^/[u]
	\ar@{-|>}[dr]\\
&&\mathrm{x}^{+,j}
	\ar@{--}@/_/[uuu]
}
\]
\caption{The pre-$\sim$-strategy $\newcell$}
\label{fig:newcell}
\end{figure}
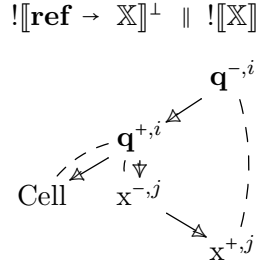

Finally, from $\Gamma, x: \var \vdash M : \tx$, we define:
\[
\intr{\newref\,r\,\tin\,M} = \newcell \odot \Lambda(\intr{M})~:~\oc \intr{\Gamma} \tcgto \oc \intr{\tx}
\]

Then, despite $\newcell$ being a pre-$\sim$-strategy rather than a $\sim$-strategy, we have:

\begin{prop}
For any $\Gamma, x: \var \vdash M : \tx$, the thin pre-$\sim$-strategy:
\[
\intr{\newref\,r\,\tin\,M} = \newcell \odot \Lambda(\intr{M})
\]
is a single-threaded $\sim$-strategy.
\end{prop}
\begin{proof}
The composition is well-defined (as a map of essps) since both compounds are $\sim$-receptive. Moreover, both compounds
are also \emph{componentwise courteous} (see Definition \ref{def:comp_court}), so by Lemma
\ref{lem:comp_court} the composition $\newcell \odot \Lambda(\intr{M})$ is a componentwise courteous
pre-$\sim$-strategy. It is also thin, negative and single-threaded as these properties are stable
under composition (respectively Lemmas \ref{lem:comp_thin}, \ref{lem:comp_neg} and Proposition
\ref{prop:comp_st}).

It remains to check that it is receptive and courteous. But that does not involve symmetry at all;
and by the results of \cite{lics11,cg1} it suffices to check that 
\[
\cc_{\oc \intr{\tx}} \odot (\newcell \odot \Lambda(\intr{M})) \odot \cc_{\oc \intr{\Gamma}} \siso
\newcell \odot \Lambda(\intr{M})
\]
but that follows from the composition of componentwise courteous
pre-$\sim$-strategies being associative, $\Lambda(\intr{M})$ being a strategy, and
the easy fact that $\cc_{\oc \intr{\tx}} \odot \newcell \siso \newcell$.
%
\end{proof}

This concludes the definition of the interpretation of IPA in $\CHO$. As said before, we do not aim
in this paper to prove properties of this interpretation, such as soundness or adequacy --
those could be either proved directly as in \cite{am}, or more easily by constructing a
functor to the interpretation of \cite{am} linearizing the partial orders. In any case, the proof
would take additional space without bringing much insight 
or taking advantage of the more refined
representation offered by our event structures strategies, so we chose not to include it.
However, we will now illustrate this interpretation by providing some examples.

\subsection{Some examples}

\subsubsection*{First example: strictness test.}
As a first example, we detail the interpretation of the term of Example \ref{ex:state1}. Recall that
it was:
\[
\newref~b~\tin~\lambda f^{\com\to \com}.~f~(b:=\ttrue);~!b : (\com \to \com)
\to \tbool
\]

As the constructor $\mathtt{new}$ is only defined on terms of ground type, this is just syntactic
sugar for 
$
\lambda f^{\com\to \com}.~\newref~b~\in~f~(b:=\ttrue);~!b : (\com \to \com)
\to \tbool
$.
In order to define its interpretation, the first step is to define:
\[
\intr{f : \com \to \com,~b:\var \vdash f~(b:=\ttrue);~!b : \tbool} : \intr{\com \to \com}
\times \intr{\var} \choto \intr{\tbool}
\]

This is covered by the definitions above, using the cartesian closed structure and the strategies of
Figure \ref{fig:comb_state} for assignment and dereferenciation. Computing this yields the strategy
represented below (again, the copy indices given by the actual interpretation function differ, but
this is irrelevant up to weak isomorphism).
Again, this deterministic event structure is forest-shaped and its branches are
versions with explicit copy indices of the P-views of the corresponding innocent
strategy in Hyland-Ong games. 

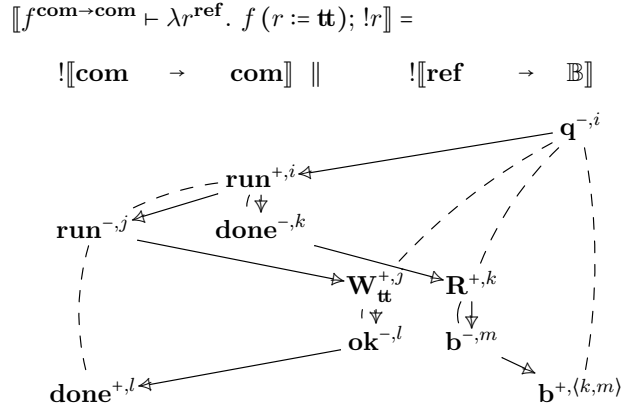
\[
\scalebox{.9}{$
\xymatrix@R=5pt@C=10pt{
\intr{f^{\com \to \com}\vdash \lambda r^{\var}.~f\,(r:=\ttrue);\,!r} = \hspace{-100pt}\\
\oc \llbracket\com   \ar@{}[r]|\to&
\com\rrbracket   \ar@{}[r]|~~~~\parallel\!\!\!\!\!\!\!\!&&
\!\!\!\!\!\!\!\!\!\!\!\!\!\!\!\!\oc \llbracket\var~~~~~~~~ \ar@{}[r]|\to&
\tbool\rrbracket\\
&&&&\q^{-,i}
	\ar@{-|>}[dlll]\\
&\run^{+,i}
        \ar@{-|>}[dl]
        \ar@{-|>}[d]
\\
\run^{-,j}
        \ar@{-|>}[drr]
        \ar@{--}@/^/[ur]
&\done^{-,k}
        \ar@{--}@/^/[u]
        \ar@{-|>}[drr]\\
&&\twrite_{\ttrue}^{+,j}
        \ar@{-|>}[d]
	\ar@{--}@/^/[uuurr]
&\tread^{+,k}
	\ar@{--}@/^/[uuur]
        \ar@{-|>}[d]	
\\
&&\tok^{-,l}
        \ar@{--}@/^/[u]
        \ar@{-|>}[dll]
&\mathbf{b}^{-,m}
        \ar@{--}@/^/[u]
        \ar@{-|>}[dr]\\
\done^{+,l}
        \ar@{--}@/^/[uuu]
&&&&\mathbf{b}^{+,\tuple{k,m}}
        \ar@{--}@/_/[uuuuu]
}$}
\]

Now, we compose it with $\newcell$. We represent below the event structure
resulting from their interaction. Events of the hidden/synchronised part of the
interaction no longer have a well-defined polarity, hence we set it to $0$.
After hiding, the minimal conflict between the first two events in $\cell$ is inherited by the final
positive events. The reader can check that hiding yields (up to the copy indices) the event
structure of Example \ref{ex:state1}. 

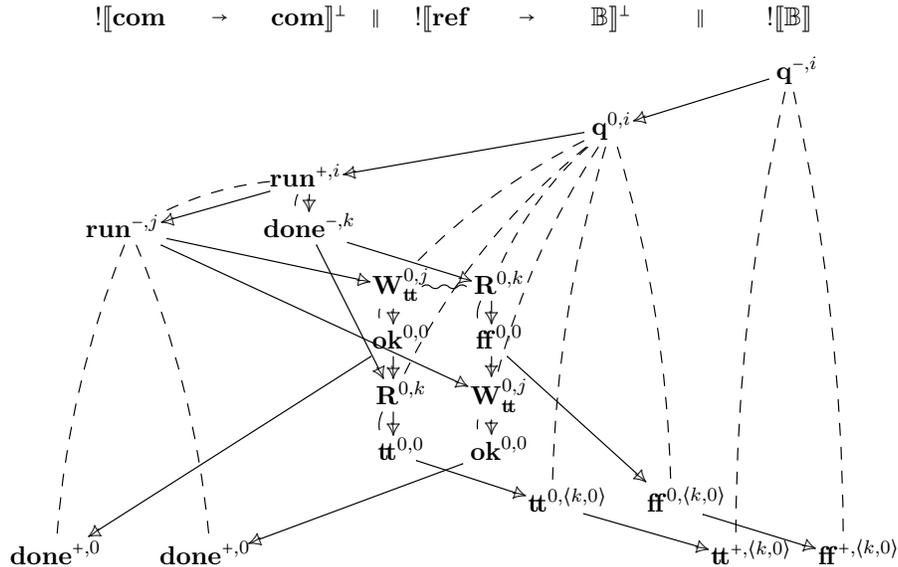
\[
\scalebox{.9}{$
\xymatrix@R=5pt@C=0pt{
&\!\!\!\!\oc \llbracket \com\!\!\!\!
	\ar@{}[rrr]|\to&&&
\!\!\!\!\com\rrbracket^\perp\!\!\!\!
	\ar@{}[rrr]|\parallel&&&
\!\!\!\!\oc \llbracket \var\!\!\!\!
	\ar@{}[rrr]|\to&&&
\!\!\!\!\tbool \rrbracket^\perp\!\!\!\!
	\ar@{}[rrr]|\parallel&&&
\!\!\!\!\oc \intr{\tbool}\!\!\!\!\\
&&&&&&&&&&&&&\q^{-,i}\!\!\!\!
	\ar@{-|>}[dlll]\\
&&&&&&&&&&\q^{0,i}
	\ar@{-|>}[dllllll]\\
&&&&\!\!\run^{+,i}
	\ar@{-|>}[d]
	\ar@{-|>}[dlll]\\
&\!\!\!\!\run^{-,j}
	\ar@{--}@/^/[urrr]
	\ar@{-|>}[drrrrr]
	\ar@{-|>}[dddrrrrrrr]&&&
\!\!\!\!\done^{-,k}\!\!\!\!
	\ar@{--}@/^/[u]
	\ar@{-|>}[drrrr]
	\ar@{-|>}[dddrr]\\
&&&&&&\twrite_{\ttrue}^{0,j}\!\!\!\!
	\ar@{--}@/^/[uuurrrr]
	\ar@{~}[rr]
	\ar@{-|>}[d]&&
\tread^{0,k}\!\!\!\!
	\ar@{--}@/^/[uuurr]
	\ar@{-|>}[d]\\
&&&&&&\tok^{0,0}\!\!\!\!
	\ar@{--}@/^/[u]
	\ar@{-|>}[d]
	\ar@{-|>}[ddddllllll]&&
\tfalse^{0,0}\!\!\!\!
	\ar@{--}@/^/[u]
	\ar@{-|>}[d]
	\ar@{-|>}[dddrrr]\\
&&&&&&\tread^{0,k}\!\!\!\!
	\ar@{--}@/^.8pc/[uuuuurrrr]
	\ar@{-|>}[d]&&
\twrite_{\ttrue}^{0,j}\!\!\!\!
	\ar@{--}@/^/[uuuuurr]
	\ar@{-|>}[d]\\
&&&&&&\ttrue^{0,0}\!\!\!\!
	\ar@{--}@/^/[u]
	\ar@{-|>}[drrr]&&
\tok^{0,0}\!\!\!\!
	\ar@{--}@/^/[u]
	\ar@{-|>}[ddllllll]\\
&&&&&&&&&\ttrue^{0,\tuple{k,0}}\!\!\!\!\!\!\!\!
	\ar@{--}@/^/[uuuuuuur]
	\ar@{-|>}[drrr]&&
\tfalse^{0,\tuple{k,0}}\!\!\!\!\!\!\!\!
	\ar@{--}@/_/[uuuuuuul]
	\ar@{-|>}[drrr]\\
\!\!\!\!\done^{+,0}\!\!\!\!
	\ar@{--}@/^/[uuuuuur]&&
\!\!\!\!\done^{+,0}\!
	\ar@{--}@/_/[uuuuuul]
&&&&&&&&&&\ttrue^{+,\tuple{k,0}}\!\!\!\!\!\!\!\!
	\ar@{--}@/^/[uuuuuuuuur]&&
\tfalse^{+,\tuple{k,0}}\!\!\!\!\!\!\!\!
	\ar@{--}@/_/[uuuuuuuuul]
}$}
\]

The reader familiar with Abramsky and McCusker's model for IA will see that taking the plays --
\emph{i.e.} alternating well-bracketed linear orderings of configurations, without copy indices -- 
yields the expected sequential strategy. But our model says more,
\emph{e.g.} it specifies the behaviour of the strategy if Opponent \emph{both}
asks its argument \emph{and} returns in parallel.

\subsubsection*{Second example: synchronization through state.} We interpret the
following term of IPA.

\[
\begin{array}{l}
x : \com,~y:\com \vdash \newref\,r\,\tin\\
~~~~~~~~\ifpcf~(!r)~\bot~(x;r:=\ttrue) \parallel\\
~~~~~~~~\ifpcf~(!r)~y~\bot\\
: \com
\end{array}
\]

\red{This term simulates sequential composition 
through parallel composition and
state up to may-equivalence: the only execution that survives divergence is the
one where the first thread is executed before the second, so that $y$ is 
run after $x$ has terminated.}

As before, we first compute the interpretation of the variant of this term where the variable has
been abstracted away, obtaining the following strategy.

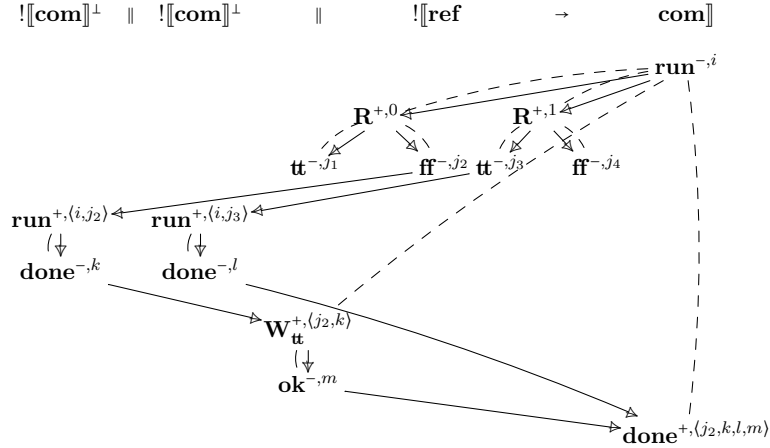
\[
\scalebox{.8}{$
\xymatrix@C=0pt@R=8pt{
&\oc \intr{\com}^\perp
	\ar@{}[rrr]|\parallel&&&
\oc \intr{\com}^\perp
	\ar@{}[rrr]|\parallel&&&
\oc \llbracket \var
	\ar@{}[rrrr]|\to&&&&
\com \rrbracket\\
&&&&&&&&&&&\run^{-,i}
	\ar@{-|>}[dlllll]
	\ar@{-|>}[dll]\\
&&&&&&\!\!\!\!\tread^{+,0}
	\ar@{--}@/^/[urrrrr]
	\ar@{-|>}[dl]
	\ar@{-|>}[dr]
&&&\!\!\!\!\tread^{+,1}
	\ar@{--}@/^/[urr]
	\ar@{-|>}[dl]
	\ar@{-|>}[dr]\\
&&&&&\ttrue^{-,j_1}\!\!\!\!
	\ar@{--}@/^/[ur]
&&\tfalse^{-,j_2}\!\!\!\!
	\ar@{--}@/_/[ul]
	\ar@{-|>}[dllllll]&
\ttrue^{-,j_3}\!\!\!\!
	\ar@{--}@/^/[ur]
	\ar@{-|>}[dllll]
&&\tfalse^{-,j_4}\!\!\!\!
	\ar@{--}@/_/[ul]\\
&\run^{+,\tuple{i,j_2}}
	\ar@{-|>}[d]&&&
\run^{+,\tuple{i,j_3}}
	\ar@{-|>}[d]\\
&\done^{-,k}	
	\ar@{--}@/^/[u]
	\ar@{-|>}[drrrr]&&&
\done^{-,l}
	\ar@{--}@/^/[u]
	\ar@/^.5pc/@{-|>}[dddrrrrrrr]\\
&&&&&\twrite_{\ttrue}^{+,\tuple{j_2,k}}
	\ar@{--}@/^/[uuuuurrrrrr]
	\ar@{-|>}[d]\\
&&&&&\tok^{-,m}
	\ar@{--}@/^/[u]
	\ar@{-|>}[drrrrrr]\\
&&&&&&&&&&&\done^{+,\tuple{j_2,k,l,m}}\!\!\!\!\!\!
	\ar@{--}@/_/[uuuuuuu]
}$}
\]

We now compute a part of the interaction with $\newcell$, pictured below.

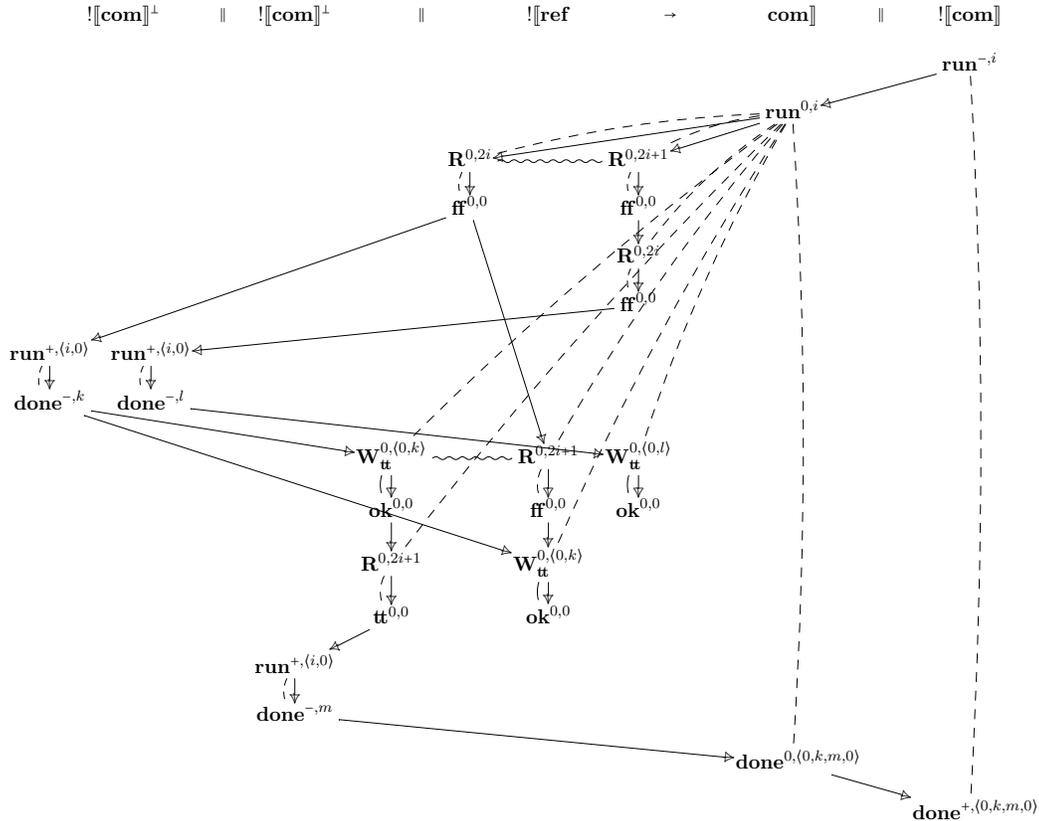
\[
\scalebox{.7}{
\xymatrix@C=5pt@R=10pt{
&\!\!\!\!\!\!\!\!\!\!\!\!\!\!\!\!\oc \intr{\com}^\perp~~~~~~~~
        \ar@{}[rrr]|\parallel&&&
\oc \intr{\com}^\perp
        \ar@{}[rrr]|\parallel&&&
\oc \llbracket \var
        \ar@{}[rrrr]|\to&&&&
\com \rrbracket
	\ar@{}[rrr]|\parallel&&&
\oc \intr{\com}\\
&&&&&&&&&&&&&&\run^{-,i}
	\ar@{-|>}[dlll]\\
&&&&&&&&&&&\run^{0,i}
	\ar@{-|>}[dlllll]
	\ar@{-|>}[dlll]\\
&&&&&&\tread^{0,2i}
	\ar@{-|>}[d]
	\ar@{--}@/^/[urrrrr]
	\ar@{~}[rr]&&
\tread^{0,2i+1}
	\ar@{-|>}[d]
	\ar@{--}@/^/[urrr]\\
&&&&&&\tfalse^{0,0}
	\ar@{-|>}[dddllllll]
	\ar@{-|>}[dddddr]
	\ar@{--}@/^/[u]
&&\tfalse^{0,0}
	\ar@{-|>}[d]
	\ar@{--}@/^/[u]\\
&&&&&&&&\tread^{0,2i}
	\ar@{-|>}[d]
	\ar@{--}@/^/[uuurrr]\\
&&&&&&&&\tfalse^{0,0}
	\ar@{-|>}[dlllllll]
	\ar@{--}@/^/[u]\\
\run^{+, \tuple{i,0}}
	\ar@{-|>}[d]&
\run^{+, \tuple{i,0}}
	\ar@{-|>}[d]\\
\done^{-,k}
	\ar@{-|>}[drrrrr]
	\ar@{--}@/^/[u]
	\ar@{-|>}[dddrrrrrrr]&
\done^{-,l}
	\ar@{-|>}[drrrrrrr]
	\ar@{--}@/^/[u]\\
&&&&&\twrite_{\ttrue}^{0, \tuple{0,k}}
	\ar@{~}[rr]
	\ar@{--}@/^/[uuuuuuurrrrrr]
	\ar@{-|>}[d]&&
\tread^{0, 2i+1}
	\ar@{-|>}[d]
	\ar@{--}@/^/[uuuuuuurrrr]&
\twrite_{\ttrue}^{0, \tuple{0,l}}
	\ar@{-|>}[d]
	\ar@{--}@/^/[uuuuuuurrr]\\
&&&&&\tok^{0,0}
	\ar@{--}@/^/[u]
	\ar@{-|>}[d]&&
\tfalse^{0,0}
	\ar@{--}@/^/[u]
	\ar@{-|>}[d]&
\tok^{0,0}
	\ar@{--}@/^/[u]\\
&&&&&\tread^{0,2i+1}
	\ar@{--}@/^/[uuuuuuuuurrrrrr]
	\ar@{-|>}[d]&&
\twrite_{\ttrue}^{0, \tuple{0, k}}
	\ar@{--}@/^/[uuuuuuuuurrrr]
	\ar@{-|>}[d]\\
&&&&&\ttrue^{0, 0}
	\ar@{--}@/^/[u]
	\ar@{-|>}[dl]&&
\tok^{0, 0}
	\ar@{--}@/^/[u]\\
&&&&\run^{+, \tuple{i, 0}}
	\ar@{-|>}[d]\\
&&&&\done^{-, m}
	\ar@{--}@/^/[u]
	\ar@{-|>}[drrrrrrr]\\
&&&&&&&&&&&\done^{0, \tuple{0,k,m,0}}\!\!\!\!
	\ar@{--}@/_/[uuuuuuuuuuuuu]
	\ar@{-|>}[drrr]\\
&&&&&&&&&&&&&&\done^{+,\tuple{0,k,m,0}}\!\!\!\!
	\ar@{--}@/_/[uuuuuuuuuuuuuuu]
}}
\]
which, after hiding, yields a strategy with sub-event structures such as:

\[
\scalebox{.9}{$
\xymatrix@R=10pt@C=0pt{
\intr{\lambda x^\com,\,y^\com.\,\newref\,r\,\tin\,(\ifpcf\,(!r)\,\bot\,(x;r:=\ttrue))
\parallel (\ifpcf\,(!r)\,y\,\bot)} =\hspace{-200pt}\\
&\!\!\!\!\!\!\!\!\!\!\!\!\!\!\!\!!\intr{\com}^\perp~~~~~~~~~~~~       \ar@{}[rrr]|\parallel&~~~~~~&~~~~~~&
\!\!\!\!\oc \intr{\com}^\perp\!\!\!\!    \ar@{}[rrr]|\parallel&~~~~~~&~~~~~~&
\oc \intr{\com}\\
&&&&&&&\run^{-,i}
        \ar@{-|>}[dllllll]
        \ar@{-|>}[dlllllll]\\
\run^{+,\tuple{i,0}}
        \ar@{~}[r]
        \ar@{-|>}[d]
        \ar@{--}@/^/[urrrrrrr]
&\run^{+,\tuple{i,0}}
        \ar@{-|>}[d]
        \ar@{--}@/^/[urrrrrr]\\
\done^{-,l}
        \ar@{--}@/^/[u]
&\done^{-,k}
        \ar@{--}@/^/[u]
        \ar@{-|>}[drrr]\\
&&&&\run^{+,\tuple{i,0}}
        \ar@{--}@/^/[uuurrr]
        \ar@{-|>}[d]\\
&&&&\done^{-,m}
        \ar@{-|>}[drrr]
        \ar@{--}@/^/[u]\\
&&&&&&&\done^{+,\tuple{0,k,m,0}}
        \ar@{--}@/_/[uuuuu]
}$}
\]

Here, there are several observations to make. 

Firstly, the copy index of the call to $y$ does not depend on $k$. This might seem
surprising: \red{the diagram suggests that} if Opponent plays two occurrences of
$\done^{-,k}$, once with $k=0$ and once with $k=1$, Player will play the
subsequent $\run^{+,\tuple{i,0}}$ twice, breaking local injectivity. 
\red{In fact, 
 recomputing the interaction with two occurrences of $\done^{-,k}$ one
realizes that the two occurrences of $\run^{+,\tuple{i,0}}$ do exist, but they
conflict with each other: this triggers new $\twrite_{\ttrue}^{0,\tuple{0,k}}$
events, which would be sequentialized in some order by the memory. The
$\tread^{0, 2i+1}$ would happen at some point during that sequentialization; each
of these possible occurrences of $\tread^{0, 2i+1}$ would lead to a call to $y$
-- so there would be multiple non-deterministic calls to $y$. This illustrates
that our symbolic representation of strategies is incomplete, and does not
specify in general their behaviour if Opponent replicates their moves (though it
is crucial in \cite{lics15} that this representation becomes complete for \emph{innocent} strategies).}

Secondly, we note that this term, in the Ghica-Murawski model of IPA \cite{gm}, would be interpreted by
the same strategy than that for sequential composition. Unlike their model, we keep some information
about non-deterministic branching; meaning that we do remember here that the term has a chance to
diverge. In the interpretation presented in this paper, we do not remember \emph{all} the information 
about divergences though. If one
was to simplify the term above to $\newref\,r\,\tin\,\lambda x^\com,\,y^\com.\,
x;r:=\ttrue \parallel \ifpcf~!r~y~\bot$, the branch where the read arrives too early \emph{w.r.t.}
the write would be hidden away by composition. The sole purpose of the superfluous read in our
example above is to create a race in memory before $x$, spawning two non-deterministic copies of the
execution of $x$. In one of them the computation is doomed, as the second thread is stuck in a loop.


\section{Conclusions}
\label{sec:conclusion}
In this paper, we have given the detailed development leading to our cartesian closed category
$\CHO$ of \emph{Concurrent Hyland-Ong} games, a setting that we illustrated with an interpretation
of IPA. The cartesian closed category $\CHO$ conservatively extends standard Hyland-Ong games, in
the sense that in our setting purely functional programs are interpreted as (copy-index aware versions
of) their tree of P-views -- but our setting also supports stateful, non-deterministic, or
concurrent languages, or any combination thereof.

The cornerstone of our construction is a compact closed category $\TCG$ of \emph{thin concurrent
games}, which extends Rideau and Winskel's category $\CG$ of games and strategies as event
structures \cite{lics11,cg1}. Note the interest of $\TCG$ is not restricted to the construction of
$\CHO$. It supports games that are much more general than those obtained from arenas. The future
will tell how this mathematical space is best exploited, but in subsequent work
we have already sometimes found it more convenient to build directly on $\TCG$
and on the AJM-style exponential $\ocajm$ rather than on $\CHO$. 

\red{Overall, we believe the framework is a very powerful setting for game
semantics, whose ramifications will take some time to explore. Because it is
conservative over traditional innocent game semantics but remembers the
non-deterministic branching points, it natively supports a notion of
non-deterministic innocence \cite{lics14,DBLP:phd/hal/Castellan17} (achieving
this in traditional game semantics has long remained an open problem, only solved
recently via reworking Hyland-Ong games using ideas from sheaf theory to remember
the non-deterministic branching points \cite{DBLP:conf/lics/TsukadaO15}).
For the same reason, it has been possible to
extend the present framework to give models of non-deterministic languages
adequate for any of \emph{may, must} and \emph{fair}-equivalence, whereas
traditional game semantics are mostly confined to \emph{angelic} non-determinism.
The framework extends, also transparently, with quantitative information: in
\cite{lics18,csl18} it has been extended with probabilities, along with a notion
of probabilistic innocence permitting a definability result and a collapse to the
probabilistic relational model -- such results were not within reach using the
traditional toolbox of game semantics.  A number of further extensions are under
active development. 

Beyond theoretical results, it is our hope that the truly concurrent nature of
this model will prove useful as a basis for algorithmic analysis and verification
of concurrent programs.}

\paragraph{Acknowledgments.} This work was partially supported by the LABEX
MILYON (ANR-10-LABX-0070), and by the ERC Advanced Grant ECSYM.


\bibliographystyle{alpha}
\bibliography{main}

\newpage
\appendix
\section{Examples and counter-examples}
\subsection{Necessity for uniformity witnesses}
\label{app:unif_wit}

In this first appendix, we give a few examples illustrating why uniformity is
achieved by having strategies carry uniformity witnesses,
rather than simply as a lifting property with respect to the symmetry in the
game.

\subsubsection{Necessity of witness.}
Our first example illustrates that simply requiring strategies on expanded arenas
to satisfy a lifting property with respect to the reindexing isos in the game (as
in AJM games) is unsound: it is too strict, and rejects some valid uniform
strategies. In particular, it becomes required when one wishes to express
uniformity for strategies with a non-deterministic branching behaviour. For
instance, consider the following strategy.

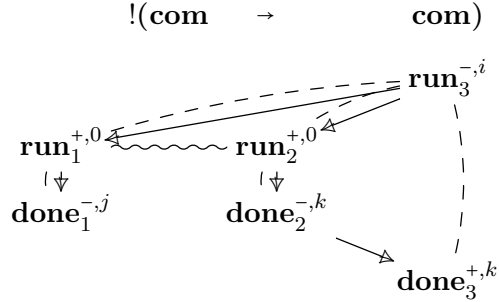
\[
\xymatrix@R=5pt@C=0pt{
&\oc(\com
	\ar@{}[rrr]|\to&&&&&
\com)\\
&&&&&&\run_3^{-,i}
	\ar@{-|>}[dllllll]
	\ar@{-|>}[dllll]\\
\run_1^{+,0}\ar@{--}@/^/[urrrrrr]
	\ar@{~}[rr]
	\ar@{-|>}[d]
	&&
\run_2^{+,0}
	\ar@{--}@/^/[urrrr]
	\ar@{-|>}[d]\\
\done_1^{-,j}
	\ar@{--}@/^/[u]&&
\done_2^{-,k}
	\ar@{--}@/^/[u]
	\ar@{-|>}[drrrr]\\
&&&&&&\done_3^{+,k}
	\ar@/_/@{--}[uuu]
}
\]
where subscripts are there just to distinguish occurrences of moves in the
discussion.

The strategy throws a non-deterministic coin. If it gets heads it calls its
argument, then diverges. On the other hand if it gets tails it calls for its
argument, then returns. This behaviour is easily definable in IPA, however a
naive attempt at defining uniformity rejects it. Indeed, the two configurations
$\{\run_3^{-,0}, \run_1^{+,0}, \done_1^{-,0}\}$ and $\{\run_3^{-,0},
\run_2^{+,0}, \done_2^{-,0}\}$ have the same image in the game, so in particular
they coincide up to reindexing iso. However, the latter can be extended with
$\{\done_3^{+,0}\}$ whereas the former cannot. This is because while the two
events $\run_1^{+,0}$ and $\run_2^{+,0}$ have the same image in the game (and
hence are symmetric), they correspond to entirely distinct events in the dynamic
behaviour of the program and should not be considered symmetric by the strategy.
Hence we need to equip strategies with \emph{uniformity witnesses} that record
which events are symmetric. 

\subsubsection{Non-uniqueness of witness.}
If a uniformity witness is needed, one may still hope to prove that it can be
recovered in a unique or canonical way, so that being uniform can still be
considered as a property rather than a piece of structure. Our next example shows
that it is not the case: uniform strategies can very well differ only via their
uniformity witnesses (this is best read after Section
\ref{sec:gameswithsymmetry}).

Consider the tcg $\A$ with events $\{\circleddash _0, \circleddash_1,
\oplus_0, \oplus_1\}$, all compatible, with trivial causality
and symmetry consisting in all polarity-preserving bijections.
Now consider the strategy (without symmetry) $ \sigma : S  \rightarrow  A$ described as follows:
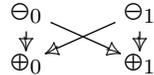
\[
  \xymatrix@R=5pt{
     \circleddash _0 \ar@{-|>}[d] \ar@{-|>}[dr] &  \circleddash _1 \ar@{-|>}[d] \ar@{-|>}[dl] \\
     \oplus _0 &  \oplus _1
  }
  \]

  By strong receptivity, any isomorphism family $\tilde{S}$ on $S$ making $\sigma$
  a $\sim$-strategy must   contain the two permutations $\id$ and $ \pi $
  of the set   $\{ \circleddash _0, \circleddash _1\}$. We now consider
  how a valid isomorphism family $\tilde{S}$ may allow these two
  permutations to extend (in $\tilde{S}$) to
  permutations on the   full set $\{\circleddash _0, \circleddash_1,
  \oplus_0, \oplus_1\}$. By thinness, the identity   bijection can
  only extend by the identity.    However, the thin axiom   does not
  restrict the possible extensions of $ \pi $. This means   that $ \pi $
  can extend \emph{e.g.} by $( \oplus _0, \oplus _0)$ or by   $( \oplus _0, \oplus
  _1)$. In fact, there are exactly two isomorphism families
  $\tilde{S}_1$ and $\tilde{S}_2$ on $S$ making $\sigma$ a
  $\sim$-strategy, with maximal extension of $\pi$ respectively given
  by $ \pi _1$ (left), and  $\pi _2$ (right):

  $$
  \xymatrix@R=0cm@C=0.1cm {
     \circleddash _0 &\longmapsto&  \circleddash _1 & \qquad&      \circleddash _0 &\longmapsto & \circleddash _1\\
     \circleddash _1 &\longmapsto  &\circleddash _0 & \qquad&      \circleddash _1 &\longmapsto & \circleddash _0\\
     \oplus _0 &\longmapsto & \oplus _1 & \qquad&      \oplus _0 &\longmapsto  &\oplus _0\\
     \oplus _1 &\longmapsto  &\oplus _0 & \qquad&      \oplus _1 &\longmapsto & \oplus _1\\
     &  \pi _1  \in  \tilde{S}_1 & & &  &\pi _2  \in  \tilde{S}_2
     }
     $$

This yields two $\sim$-strategies $\sigma_1 : \S_1 \to \A$ and
$\sigma_2 : \S_2 \to \A$, which differ only by how they
react to Opponent permuting their copy indices: either by also permuting 
($\sigma_1$) or by doing nothing $(\sigma_2)$. The two resulting
$\sim$-strategies are not weakly isomorphic or weakly equivalent
because there are no maps between $\S_1$ and $\S_2$.

\subsubsection{Operational interpretation.} Here we attempt to give an
operational reading on the difference between $\sigma_1$ and $\sigma_2$
above, by recasting the phenomenon in IPA (this is best read after
Section \ref{sec:interpretation}).
In the interpretation of IPA, these subtle differences between
uniformity witnesses can convey indirect information reminiscent from
the difference between causality due to the program order, and that due
to the dependencies in memory.

     Consider the two terms $M_1, M_2$ of type
     $(\com \rightarrow \com) \rightarrow \com \rightarrow \com$:
     $$
     \begin{aligned}
M_1 &=  \lambda fx.\,\newref\, r\,\tin\, f\, (\texttt{incr}\, r;
\mathtt{wait}\, (!r = 2).\, x; \bot) \\
M_2 &=  \lambda fx.\,\newref\, r\,\tin\, (f (\texttt{incr}\, r; \bot))
\parallel  (\texttt{wait}\, (!r = 2).\, (x  \parallel  x))\\
\end{aligned}$$ where $\mathtt{incr}\, r$ is a short-hand for
$r := !r + 1$ and $\mathtt{wait}\, b.\, M$ for
$Y\, ( \lambda l. \ifpcf\, b\, M\, l)$. For both terms, the strategy
includes the following pattern, where the variable $f$ calls its
argument twice:
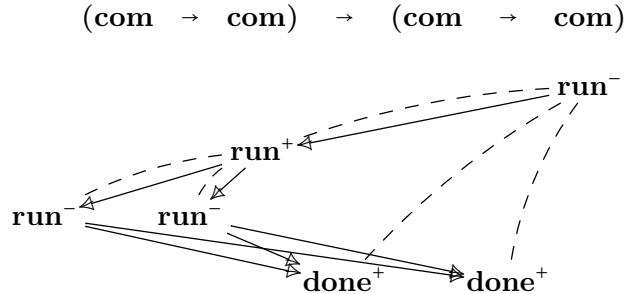
\[
\xymatrix@R=10pt@C=-5pt{
&(\com	\ar@{}[rr]|\to&&
\com)	\ar@{}[rr]|\to&&
(\com	\ar@{}[rr]|\to&&
\com)\\
&&&&&&&\run^{-}
	\ar@{-|>}[dllll]\\
&&&\run^{+}
	\ar@{--}@/^/[urrrr]
	\ar@{-|>}[dlll]
	\ar@{-|>}[dl]\\
\run^{-}\ar@{-|>}[drrrr]
	\ar@{-|>}[drrrrrr]
	\ar@{--}@/^/[urrr]&&
\run^{-}\ar@{-|>}[drr]
	\ar@{-|>}[drrrr]
	\ar@{--}@/^/[ur]\\
&&&&\done^{+}
	\ar@{--}@/^/[uuurrr]&&
\done^{+}
	\ar@{--}@/^/[uuur]
}
\]

We omit copy indices to reduce clutter. As soon as
$f$ calls its argument \emph{twice}, the two programs call $x$ twice.
The reader will recognize in the four lowest events the same pattern as in the
previous example -- and the isomorphism families for $\intr{M_1}$ and
$\intr{M_2}$ restricted to this pattern indeed behave respectively like
$\sigma_1$ and $\sigma_2$. 

It follows from the definition of the interpretation that this dynamics
of an index exchange causing an index exchange occurs in the purely
functional stage of the interpretation, as opposed to the accounting of
state. As a consequence, the dynamics of indices exchange follows the
causal links that originates from the program syntax tree, but ignores
those that come from communication through the memory. Hence
in $\intr{M_1}$ the Opponent exchange causes a Player exchange because
the occurrence of $x$ appears within the argument of $f$ \emph{in
the syntax tree}. In contrast, in $\intr{M_2}$ the Opponent exchange
causes no Player exchange as the dependency only flows through the memory.

%
%
%
%

\subsection{Absence of pullbacks in $\ESS$}
\label{app:no_pb}

The category of event structures with symmetry does not have pullbacks
in general. 
For that we first note that if a diagram has a pullback in $\ESS$,
then, forgetting symmetry, it is also a pullback in $\ES$. The reason for that is the following
proposition.

\begin{prop}\label{prop:forg}
The forgetful functor
$\ESS \to \ES$
which to any event structure with symmetry $\A = (A, \tilde{A})$ associates $A$, has a left
adjoint.
\end{prop}
\begin{proof}
The right adjoint associates, to any event structure $A$, the event structure with symmetry 
$(A, \refl_A)$, where 
\[
\refl_A = \{\{(a, a)\mid a\in x\}\mid x\in \conf{A}\}
\]
is the minimal symmetry on $A$. It is straightforward that this defines an adjunction.
\end{proof}

Hence the symmetry-forgetting functor is a right adjoint, and as such preserves pullbacks. Now, in
order to prove that $\ESS$ does not have pullbacks, we are going to construct a diagram in $\ESS$
whose pullback in $\ES$ has no possible isomorphism family.
%
%
Indeed, consider $A$ the following event structure:

$$
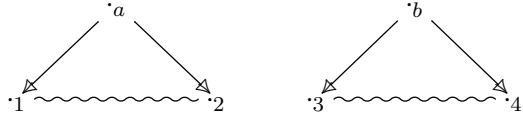
\xymatrix {
& \cdot_a \ar@{-|>}[dl]\ar@{-|>}[dr]& & & \cdot_b \ar@{-|>}[dl]\ar@{-|>}[dr]\\
\cdot_1\ar@{~}[rr] & & \cdot_2 & \cdot_3\ar@{~}[rr] & & \cdot_4 \\
}$$

Write $\A$ for $A$ equipped with the maximal isomorphism family: all
order-isomorphisms are in the family. Write $\A_1$
for the sub-event with symmetry where $\cdot_1$ can only be sent to
itself and to $\cdot_3$; and $\cdot_2$ can only be sent to itself and to $\cdot_4$. Similarly,
write $\A_2$ for that where $\cdot_1$ can only be sent to $\cdot_4$ and $\cdot_2$ to $\cdot_3$.
%

Now we have the following diagram:

$$
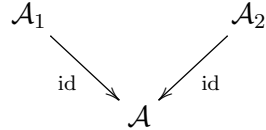
\xymatrix {
\ar[dr]_{\id} \mathcal {A}_1 & & \mathcal A_2 \ar[dl]^\id \\
& \mathcal A
}$$

Assume this diagram has a pullback $(\mathcal A_3, \Pi_1, \Pi_2)$. By Proposition \ref{prop:forg}
its underlying event structure is $A$ and the projection maps are both identities on objects. The isomorphism
$\{(\cdot_a, \cdot_b)\} : \{\cdot_a\} \isf_{\tilde{A_3}} \{\cdot_b\}$ must be in $\tilde{A_3}$ as
it is in both $\tilde{A_1}$ and $\tilde{A_2}$. However, its left hand side $\{\cdot_a\}$
can be extended with $\cdot_1$, so by the extension property we must have
\[
\{(\cdot_a, \cdot_b), (\cdot_1, \cdot_i)\} : \{\cdot_a, \cdot_1\} \isf_{\tilde{A_3}} \{\cdot_b, \cdot_i\}
\]
 with $i \in \{3, 4\}$. But by construction such an iso cannot be in
both $\tilde{A_1}$ and $\tilde{A_2}$, absurd.

\subsection{Weak equivalence is not a congruence}
\label{app:pentagram}

This key observation is one of the key facts guiding the design of our games.

Let $\A$, $\B$ and $\C$ be ``games'', \emph{i.e.} essps in the context of this
discussion. Consider $\sigma_1 : \S_1 \to \A^\perp \parallel \B$, $\sigma_2 :
\S_2 \to \A^\perp \parallel \B$, and $\tau : \T \to \B^\perp \parallel \C$ be
``strategies'', \emph{i.e.} maps of essps in the context of this discussion.
Assume further than $\sigma_1$ and $\sigma_2$ are weakly equivalent, \emph{i.e.}
that there are $f : \S_1 \to \S_2$ and $g : \S_2 \to \S_1$ such that $g\circ f
\sim \id_{\S_1}$, $f \circ g \sim \id_{\S_2}$, and the two obvious triangles
commute up to symmetry. It first came as a surprise to us that then, $\tau \odot
\sigma_1$ and $\tau \odot \sigma_2$ might \emph{not} be weakly equivalent.
Indeed, the extension property of isomorphism families ensures that two symmetric
configurations have ``bisimilar futures''. So it is natural to expect $\tau \odot
\sigma_1$ and $\tau \odot \sigma_2$ to behave similarly, and indeed they do, but
in a way less strict than that expressed by weak equivalence.

To be more precise, first consider the essp $\P$ (the ``pentagram''):
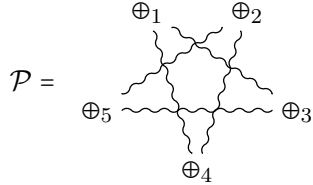
\[
\P = 
\raisebox{30pt}{
\xymatrix@R=1pt@C=0pt{
&\oplus_1
	\ar@{~}[ddddrrr]
	\ar@{~}[ddddddr]
&&\oplus_2
	\ar@{~}[ddddlll]
	\ar@{~}[ddddddl]\\\\\\\\
\oplus_5
	\ar@{~}[rrrr]
&&&&\oplus_3\\\\
&&\oplus_4
}}
\]

Its isomorphism family is the maximal one, \emph{i.e.} all bijections between configurations are in the
family. In $\P$, two events will eventually be played. It does not matter which ones, since
they are all symmetric -- the only thing that matters is the multiplicity. 

We consider $\P$ as a strategy on a game $\B$ with the same events as $\P$
($\{\oplus_1, \oplus_2, \oplus_3, \oplus_4, \oplus_5\}$), the maximal isomorphism
family, and no conflict. We write $\alpha_1$ for the obvious labeling
\[
\begin{array}{rcrcl}
\alpha_1 &:& \P &\to& \B\\
&& \oplus_i &\mapsto& \oplus_i
\end{array}
\]
which indeed informs a strategy on $\B$. 

We will also be interested in another strategy:
\[
\alpha_2 ~~ : ~~ \S ~~ \to ~~ \B
\]
where $\S$ has events $\{\oplus_1, \oplus_2\}$ and again maximal isomorphism family. The map $\alpha_2$ sends
$\oplus_i$ in $\S$ to $\oplus_i$ in $\B$.
The strategies $\alpha_1$ and $\alpha_2$ behave similarly, since both will eventually
play two events; and we do not care which ones since all possible choices are symmetric in the game.
Despite that, $\alpha_1$ and $\alpha_2$ are \emph{not} weakly equivalent. In fact, \emph{there is no
map from $\alpha_1$ to $\alpha_2$}: such a map would require us to build a map of event structures
from $P$ to $S$, but the reader can check that this would induce a $2$-coloring of $P$, which is not
bipartite.

We will now obtain $\alpha_1$ and $\alpha_2$ respectively as compositions $\tau \odot \sigma_1$ and
$\tau \odot \sigma_2$, for weakly equivalent $\sigma_1$ and $\sigma_2$. We introduce the game 
\[
\A = \xymatrix{\oplus_a\ar@{~}[r]&\oplus_b}
\]
with again the maximal isomorphism family. The strategy $\tau : \T \to \A^\perp \parallel \B$
selects $\alpha_1$ or $\alpha_2$ depending on Opponent's choice in $\A$.
Its events are represented below.
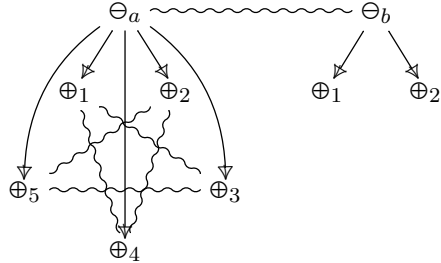
\[
\xymatrix@R=1pt@C=0pt{
&&\ominus_a
	\ar@{~}[rrrrrrr]
	\ar@{-|>}[dddl]
	\ar@{-|>}[dddr]
	\ar@/_1pc/@{-|>}[dddddddll]
	\ar@/^1pc/@{-|>}[dddddddrr]
	\ar@{-|>}[ddddddddd]&&&&&&&
\ominus_b
	\ar@{-|>}[dddl]
	\ar@{-|>}[dddr]\\\\\\
&\oplus_1
        \ar@{~}[ddddrrr]
        \ar@{~}[ddddddr]
&&\oplus_2
        \ar@{~}[ddddlll]
        \ar@{~}[ddddddl]&&&&&\oplus_1&&\oplus_2\\\\\\\\
\oplus_5
        \ar@{~}[rrrr]
&&&&\oplus_3\\\\
&&\oplus_4
}
\]

Its isomorphism family is, again, the maximal one: all order-isomorphisms between
configurations are valid symmetries. One can check that this satisfies indeed the
axioms for an isomorphism family; crucially the extension axiom uses the fact
that $\P$ and $\S$ are bisimilar (and that the symmetry on $\B$ is the maximal
one).

Finally, consider $\sigma_1, \sigma_2$ on $\A$, with $\sigma_1$ playing only
$\oplus_a$ and $\sigma_2$ playing only $\oplus_b$. They are clearly weakly
equivalent, since $\{(\oplus_a, \oplus_b)\}$ is in $\tilde{A}$. But by
construction we have $\tau \odot \sigma_1 \siso \alpha_1$ and $\tau \odot
\sigma_2 \siso \alpha_2$, which as we observed are not weakly equivalent.
%
%
%

Note that the games $\A$ and $\B$ are both tcgs; but crucially $\tau$ is not \emph{thin} (Definition
\ref{def:thin}). Indeed, for instance, the symmetry $\{(\ominus_a, \ominus_b)\}$ extends to both 
$\{(\ominus_a, \ominus_b), (\oplus_1, \oplus_1)\}$ and $\{(\ominus_a, \ominus_b), (\oplus_1,
\oplus_2)\}$, which is forbidden by Definition \ref{def:thin}. For \emph{thin}
strategies, positive extensions of the symmetry must be canonically chosen,
making it impossible that composite strategies as above are bisimilar but not
weakly equivalent.

\subsection{Failure of extension for copycat on general games}
\label{app:failextrace}

In the main text, we give (Definition \ref{def:symm_cc}) a candidate
$\CC_{\sym{A}}$ for the isomorphism family on copycat for any essp $\A$. The valid
symmetries on $\CC_\A$ are simply those order-isomorphisms between configurations of $\CC_A$ which
map to valid symmetries on $\A^\perp \parallel \A$. 

We prove in Proposition \ref{prop:cc_sim_strat} that this satisfies the extension
property of isomorphism families if the game is a tcg. This boiled down to Lemma
\ref{lem:tcg_racepres}, which shows that tcgs are \emph{race-preserving} : races in the isomorphism
family always originate to races in the game. As this phenomenon played an important role in the
design of the theory, we find it useful to include here an example demonstrating the fact that
without this race-preservation property (so if the games are plain event structures with polarities
and symmetry, rather than tcgs), the extension property fails in general for the isomorphism family on copycat.

Consider an essp $A = \{a^-, b^+\}$. We form $\oc_2 A$ with
events/causality/polarities/conflict those of $A\parallel A$ (we write $a^{-, i}$ for $(i, a)^-$), 
and isomorphism families the set of bijections between configurations included in the two maximal
ones:
\[
\begin{array}{l}
\{(a^{-, 1}, a^{-, 1}), (b^{+, 1}, b^{+, 1}), (a^{-, 2}, a^{-, 2}), (b^{+, 2}, b^{+, 2})\}\\
\{(a^{-, 1}, a^{-, 2}), (b^{+, 1}, b^{+, 2}), (a^{-, 2}, a^{-, 1}), (b^{+, 2}, b^{+, 1})\}
\end{array}
\]

So maximal symmetries either globally preserve the copy indices, or globally swap them. It is not
possible for a symmetry to, \emph{e.g.} send $a^{-, 1}$ to $a^{-, 1}$ and $b^{+, 1}$ to $b^{+, 2}$.
It is, in fact, a binary version of the $\ocajm$ operation of Definition
\ref{def:bang_ajm}, applied to $\A$.

From Definition \ref{def:symm_cc}, $\CC_{\oc_2 A}$ is equipped with a candidate
isomorphism family $\CC_{\tilde{\oc_2 A}}$. We now show that this however fails
the \emph{(Extension)} axiom of isomorphism families. From the definition, the
diagram below represents a valid symmetry in 
$\CC_{\tilde{\oc_2 A}}$.

\[
\xymatrix@R=5pt{
\oc_2 A^\perp\ar@{}[r]|\parallel&
\oc_2 A	\ar@{}[rrr]|{\isf_{\CC_{\tilde{\oc_2 A}}}}&&&
\oc_2 A^\perp\ar@{}[r]|\parallel&
\oc_2 A\\
b^{-, 1}\ar@{.}@/_/[rrrr]&
a^{-, 1}\ar@{.}@/_/[rrrr]&&&
b^{-, 1}&
a^{-, 2}
}
\]

The issue will come from the fact that the symmetry follows irreconciliable courses in the left and
the right components of $\CC_{\oc_2}$: in the left component it preserves copy indices, whereas in
the right component it swaps them. So it the left hand side of this symmetry extends as depicted
below

\[
\xymatrix@R=5pt{
\oc_2 A^\perp\ar@{}[r]|\parallel&
\oc_2 A \ar@{}[rrr]|{\isf_{\CC_{\tilde{\oc_2 A}}}}&&&
\oc_2 A^\perp\ar@{}[r]|\parallel&
\oc_2 A\\
b^{-, 1}\ar@{.}@/_/[rrrr]
	\ar@{-|>}[dr]&
a^{-, 1}\ar@{.}@/_/[rrrr]&&&
b^{-, 1}&
a^{-, 2}\\
&b^{+,1}
}
\]
the only matching extension on the right hand side is with $b^{+, 1}$ as well ($b^{+, 2}$ is not
possible as it would require to play $b^{-, 2}$ first), but $(b^{+, 1},
b^{+,1})$ is not a valid extension of the symmetry above, as for that it would need to swap the copy
index instead of preserving it.

\section{Postponed proofs}
\subsection{Pullbacks of dual $\sim$-receptive maps}
\label{app:pb_exist}

\pbsym*
\begin{proof}
  The (Groupoid) and (Restriction) axioms are direct consequences of
  the corresponding conditions for $\tilde S$ and $\tilde T$.

  \emph{(Extension).} Let
  $ \theta : w \isf _{\tilde{S}\wedge \tilde T} z$. Assume $w$ can be
  extended by an event $e \in S \wedge T$ to $w'$. Write
  $s = \Pi _1 e$ and $t = \Pi _2 e$, and assume \emph{e.g.} $\sigma s$ is positive in
  $A$. We then have:

\[
\xymatrix@R=15pt@C=15pt{
\Pi_1\,w' &  \sbij & (  \sigma  \wedge  \tau  )\,w' &  \sbij  & \Pi_2\,w' \\
& \ar@{-C}[ul]^s\ar@{}[d]|{\theta_S \rotatebox{90}{$\isf$}_{\tilde S}}
  \Pi_1\,w \ar@{}[r]|{\sbij}^ \sigma  & (  \sigma  \wedge  \tau  )\,w
  \ar@{}[r]|{\sbij}^ \tau  &
\Pi_2\,w \ar@{}[d]|{\theta_T \rotatebox{90}{$\isf$}_{\tilde T}} \ar@{-C}[ur]^t\\
& \Pi_1\,z \ar@{}[r]|{\sbij}^ \sigma  & (  \sigma  \wedge  \tau  )\,z
\ar@{}[r]|{\sbij}^ \tau  &
\Pi_2\,z \\
}
\]

We first use the extension property on $ \theta _S$ as
$ \Pi _1 w \longcov {s}$: $ \theta _S$ extends by $(s, s')$. Since
$ \sigma  \theta _S =  \tau  \theta _T$, this means that $ \tau  \theta _T$
extends by
$( \sigma s,  \sigma s')$ which is negative in $\A^\perp$. By
$\sim$-receptivity of $ \tau $, $ \theta _T$ extends by
$(t, t')$ with $ \tau t =  \sigma s$ and $ \tau t' =  \sigma s'$. The picture is
now:
\[
\xymatrix@R=15pt@C=15pt{
\ar@{}[ddd]|{ \rotatebox{90}{$\isf$}_{\tilde S}}\Pi_1\,w' &  \sbij  & (  \sigma
\wedge  \tau
)\,w' &  \sbij  & \Pi_2\,w' \ar@{}[ddd]|{ \rotatebox{90}{$\isf$}_{\tilde T}}\\
& \ar@{-C}[ul]^s\ar@{}[d]|{\theta_S \rotatebox{90}{$\isf$}_{\tilde S}}
  \Pi_1\,w \ar@{}[r]|{\sbij}^{ \sigma } & (  \sigma  \wedge  \tau  )\,w
  \ar@{}[r]|{\sbij}^ \tau  &
\Pi_2\,w \ar@{}[d]|{\theta_T \rotatebox{90}{$\isf$}_{\tilde T}} \ar@{-C}[ur]^t\\
& \ar@{--C}[dl]_{s'} \Pi_1\,z \ar@{}[r]|{\sbij}^ \sigma  & (  \sigma  \wedge
\tau  )\,z
\ar@{}[r]|{\sbij}^ \tau  & \Pi_2\,z \ar@{--C}[dr]^{t'}\\
z_1 & \sbij  &  \sigma z_1= \tau z_2 &  \sbij & z_2
}
\]

The obtained $\varphi : z_1 \sbij z_2$ is secured by construction, so
as observed in 
Definition \ref{def:secbij} its graph is ordered by $\leq_\varphi$ compatible
with $\leq_S$ and
$\leq_T$. Therefore restricting $\varphi$ to 
the causal history of $(s',t')$ yields $e' = [(s',t')]_{\varphi}$ a prime
secured bijection,
\emph{i.e.} an event $e' \in S \wedge T$ such that $z \longcov {e'} z'$.
Finally, $ \theta \cup \{(e, e')\} \in \tilde {S} \wedge \tilde T$
because $ \theta _S \cup\{(s, s')\}  \in  \tilde S$ and
$ \theta _T \cup\{(t, t')\}  \in  \tilde T$.
  
  If $ \sigma s$ is negative, the dual reasoning uses extension on $\tilde T$ and then
  $\sim$-receptivity of $ \sigma $.

  \emph{It is a pullback.} Clearly the maps
  $ \Pi _1 : S \wedge T \rightarrow S$ and
  $ \Pi _2 : S \wedge T \rightarrow T$ preserve symmetry: they map
  $ \theta $ to $ \theta _S$ and $ \theta _T$ respectively. We only need
  to check the universal property. Assume we have two morphisms of ess
  $ \varphi : \X \rightarrow \S$ and
  $ \psi : \X \rightarrow \T$ such that the square
  commutes:
  $$\xymatrix@R=15pt@C=15pt{
    & \X \ar@/_/[ddl]_{ \varphi }\ar@/^/[ddr]^{ \psi } \\
    & \ar[dl]\ar[dr]\S \wedge \T \\
    \ar[dr]^ \sigma  \tilde \S & & \T \ar[dl]_ \tau  \\
    & \A }$$
  Because $S \wedge T$ is a pullback in $\ES$ there is a map of event structures
  $ \langle \varphi , \psi \rangle : X \rightarrow S \wedge T$ making
  the two triangles commute, which is unique in $\ES$. This uniqueness
  lifts to $\ESS$ as the forgetful functor $\ESS \rightarrow \ES$ is
  faithful. To conclude we need only to prove that
  $ \langle \varphi , \psi \rangle $ preserves symmetry and is thus a
  morphism in $\ESS$. Let $ \theta : x \isf _{\tilde X} y$. It is transported to
  a 
  bijection $\tuple{\varphi, \psi}\,\theta : \tuple{\varphi, \psi}\,x \bij
  \tuple{\varphi,
  \psi}\,y$ such that $(\tuple{\varphi, \psi}\,\theta)_S = \varphi\,\theta$ and 
  $(\tuple{\varphi, \psi}\,\theta)_T = \psi\,\theta$, thus $\tuple{\varphi,
  \psi}\,\theta \in
  \tilde{S} \wedge \tilde{T}$ by definition.
\end{proof}

\subsection{Composition of $\sim$-receptivity and componentwise courtesy}
\label{app:comp_simrec_court}

To prove that, we first introduce the following more local characterisation for
$\sim$-receptivity.

\begin{lem}\label{lemma:charac_simr}
  Let ${\A}$ be a tcg and $ \sigma : {\S}  \rightarrow  {\A}$ be a map of ess.
  Then, $\sigma$ is $\sim$-receptive iff for all $x\in \conf{S}$ and
  $x\longcov{s_1^-}$,
  for all $\id_{\sigma\,x} \cup \{(\sigma\,s_1,a_2)\} \in \tilde{A}$, there
  exists a unique
  $s_2$ such that $\sigma\,s_2 = a_2$, and we have $\id_x \cup \{(s_1, s_2)\}
  \in \tilde{S}$.
\end{lem}
\begin{proof}
  \emph{only if.} Particular case of the definition of $\sim$-receptivity.

  \emph{if.} Assume $\theta :x_1 \isf _{{\tilde S}} x_2$, $x_1 \longcov{s_1^-}$
  and
$\sigma\,x_2 \longcov{a_2^-}$ such that $\sigma\,\theta \cup \{(\sigma\,s_1,
a_2)\} \in \tilde{A}$.
By (Extension), there is $s'_1$ such that $\theta \longcov{(s_1, s'_1)}$. Since
$\sigma$ is a map of
ess, we must have $\sigma\,\theta \cup \{(\sigma\,s_1, \sigma\,s'_1)\} \in
\tilde{A}$ as well. By
(Groupoid), it follows that $\id_{\sigma x_2} \cup \{(\sigma\,s'_1, a_2)\} \in
\tilde{A}$.
By hypothesis, we get a unique $s_2$ such that $\sigma\,s_2 = a_2$, satisfying
$\id_{x_2} \cup \{(s'_1, s_2)\} \in \tilde{S}$. And finally, by
(Groupoid) again, $\theta \cup \{(s_1, s_2)\} \in \tilde{S}$.
\end{proof}

Using that, we prove:

\compcourt*
\begin{proof}
As a preliminary to the proof, we note that thanks to $(A, B)$-courtesy of
$\sigma$ and $(B, C)$-courtesy of $\tau$, the immediate dependencies of negative
$p \in T\odot S$ in $T\circledast S$ \emph{have} to be visible as well, and must
map to the same component -- this is the key argument of the proof. Indeed assume
$p'\imc p$ in $T\circledast S$ with visible $p$ mapping to a negative event in
$A^\perp \parallel C$, for instance in $C$. Then by general properties of the
pullback in $\ES$, since $p$ maps to $C$, its immediate causal dependencies
$p'\imc p$ in $T\circledast S$ must be such that $\Pi_2\,p' \imc \Pi_2\,p$ in
$A\parallel T$ (consequence of \emph{e.g.} Lemma 2.7 of \cite{cg1} with
Proposition \ref{prop:rep_int}), but since $p$ maps to $C$ those must actually
both be in $T$, and since $\tau$ is $(B,C)$-courteous $p'$ must map to $C$ as
well, therefore it is visible. 

From that, it is clear that $\tau \odot \sigma$ is $(A, C)$-courteous. We now
show that it is $\sim$-receptive. We prove it via Lemma \ref{lemma:charac_simr}.
Take $z \in \conf{T \odot S}$, assume $z$ extends via some negative $p$, say in
$C$. The configuration $z$ has a witness $[z] \in \conf{T\circledast S}$, however
in general this witness might not extend with $p$, as it may need to perform some
invisible events prior to that.
In our case though, the preliminary above shows that this is not possible: the
immediate dependencies in $T\circledast S$ of $p$ are visible as well, and hence
in $z \subseteq [z]$. Now, if we also have that $(\tau \odot \sigma)\,z$ extends
with $c^-$ with $\id_{(\tau \odot \sigma)\,z} \cup \{((\tau \odot \sigma)\,p,
c)\} \in \tilde{A} \parallel \tilde{B} \parallel \tilde{C}$, then  
\[ 
\Pi_2\,[z] \,\,\,\,\,\,\, \longcov{\Pi_2\,p} \,\,\,\,\,\,\, \id_{\Pi_2\,[z]}
\cup
\{((A\parallel \tau)\,(\Pi_2\,p),
c)\} \,\,\,\, \in \,\,\,\, \tilde{A} \parallel \tilde{B} \parallel \tilde{C}
\]
so using $\sim$-receptivity of $A\parallel \tau$, we can uniquely lift $c$ to
$A\parallel T$, hence to $T\circledast S$ and $T\odot S$, and that lifting is by
construction compatible with $\tilde{T} \circledast \tilde{S}$ and $\tilde{T}
\odot \tilde{S}$.
\end{proof}

\subsection{Proofs for the copycat $\sim$-strategy}
\label{app:copycat}
This section contains proofs relative to the construction of the isomorphism
family for the copycat strategy. We start with a simple caracterisation of the
valid symmetries announced in the main text.

\symmcc*
\begin{proof}
Take $\theta = \theta_1 \parallel \theta_2 : x_1 \parallel x_2 \iso y_1
\parallel y_2$.

If $\theta$ is an order-iso, then take $a\in x_1 \cap x_2$. Assume
without loss of generality that $\pol_A(a) = +$, so that $(1,
a) \imc (2, a)$ in $\CC_A$. But then since $\theta$ is an order-iso,
it preserves immediate causal dependency, therefore $(1, \theta_1\,a)
\imc (2, \theta_2\,a)$. But since these two events are in different
components of $A^\perp \parallel A$, this necessarily means that
$\theta_1\,a = \theta_2\,a$ as required (using \emph{e.g.} the
characterisation of immediate causality of copycat in Lemma 3.3 of
\cite{cg1}).

Reciprocally, assume that for all $a\in x_1 \cap x_2, \theta_1\,a =
\theta_2\,a$. Using again
Lemma 3.3 of \cite{cg1}, it is immediate that $\theta$ preserves immediate
causal links. The same
reasoning applies to $\theta^{-1}$ (it is easy to show that the hypothesis is
stable under inverse), 
so it reflects immediate causal links as well; and is an order-iso.
\end{proof}

We now set to prove Proposition \ref{prop:cc_sim_strat}. Most verifications are
direct; the main issue being to show that $\CC_{\sym{A}}$ always satisfies the
axioms of isomorphism families. The first two axioms are immediate
consequences of the definition of $\CC_{\tilde{A}}$ in Definition
\ref{def:symm_cc}:

\begin{lem}\label{lemma:symm_cc_no_rf}
  For any tcg $\A$, the family $\CC_{\tilde A}$ satisfies the axioms
\emph{(Groupoid)} and   \emph{(Restriction)} of isomorphism families. 
\end{lem}

The main difficulty is to show the \emph{(Extension)} axiom.
And for good reasons: indeed, this axiom fails if $\A$ is an essp with
no further constraints (as illustrated in Appendix \ref{app:failextrace}). That
it holds when $\A$ is a tcg boils down to the property below.

\begin{lem}\label{lem:tcg_racepres}
Let $\A$ be a tcg. Then $\tilde{A}$ is \emph{race-preserving}, in the sense that
for any $\theta : x \isf_{\tilde{A}} y$, for any $\theta \subseteq^+ \theta_1 :
x_1
\isf_{\tilde{A}} y_1$ and $\theta \subseteq^- \theta_2 : x_2 \isf_{\tilde{A}}
y_2$, 
if $x_1$ and $x_2$ are compatible ($x_1 \cup x_2 \in \conf{A}$), then so are
$\theta_1$ and
$\theta_2$: $\theta_1 \cup \theta_2 \in \tilde{A}$ as well.
\end{lem}
\begin{proof}
We first prove that $\tilde A_+$ and $\tilde A_-$ are
    race-preserving.  Let
    $ \theta : x \isf _{\tilde A_+} y$ with a positive extension
    $ \theta_1 : x_1 \isf _{\tilde A_+} y_1$ and a negative
    extension $ \theta _2 : x_2 \isf _{\tilde A_+} y_2$, with 
$x_1 \cup x_2 \in \conf{A}$.

Using (Extension) of
    $\tilde A_+$ twice to $ \theta _1$ and $ \theta _2$, we get to the following
    picture:

$$\xymatrix@R=10pt@C=10pt{
 \theta _1' : x_1 \cup x_2  \isf_{\tilde{A}_+}  y'_1 && \theta'_2 : x_1 \cup x_2
\isf_{\tilde{A}_+}  y'_2 \\
\ar@{}[u]|{\rotatebox{90}{$\subseteq^-$}}  \theta _1 : x_1  \isf_{\tilde{A}_+}
y_1 & &
\ar@{}[u]|{\rotatebox{90}{$\subseteq^+$}}  \theta _2 : x_2  \isf_{\tilde{A}_+}
y_2 \\
& \theta  : x  \isf_{\tilde{A}_+}  y
\ar@{}[ul]|{\rotatebox{135}{$\subseteq^+$}}\ar@{}[ur]|{\rotatebox{45}{$\subseteq^-$}}
}$$
By the (Groupoid) axiom on $\tilde{A}_+$, we have $\id_y \subseteq \theta'_1
\circ
{\theta'_2}^{-1} : y'_2 \isf_{\tilde{A}_+} y'_1$. By (Restriction), we build
$\varphi = \theta'_1 \circ {\theta'_2}^{-1} \restrict y_2$.
By construction, we have $\id_y \subseteq^- \varphi \in \tilde{A}_+$, so
$\varphi = \id_{y_2}$ (as $\A_+^\perp$ is thin).
It follows that $\theta_2 \subseteq \theta'_1$, hence
$\theta'_1 = \theta_1 \cup \theta_2$ as required. A dual reasoning shows that
$\tilde{A}_-$ is
race-preserving as well.

Now, we deduce the result for $\tilde{A}$, using the decomposition of Lemma
\ref{lemma:decomp}.
Assume $ \theta  =  \theta ^- \circ  \theta ^+$ has extensions $\theta
\subseteq^+ \theta_1$ and 
$\theta \subseteq^- \theta_2$, with decompositions 
$ \theta _1^- \circ  \theta _1^+$ and $ \theta _2^- \circ  \theta _2^+$. By
monotonicity 
of the decomposition, we have $\theta^+ \subseteq^+ \theta_1^+$, $\theta^+
\subseteq^-
\theta_2^+$, $\theta^- \subseteq^+ \theta_1^-$ and $\theta^- \subseteq^-
\theta_2^-$.
By race-preservation of $\tilde{A}_+$ it follows first that $\theta_1^+ \cup
\theta_2^+ \in
\tilde{A}_+$, and then by race-preservation of $\tilde{A}_-$ it follows that
$\theta_1^- \cup
\theta_2^- \in \tilde{A}_-$. Thus $( \theta _1^- \cup  \theta _2^-) \circ (
\theta _1^+ \cup  \theta
_2^+) =  (\theta_1^- \circ \theta_1^+ ) \cup (\theta_2^- \circ \theta_1^-) =
\theta _1 \cup  \theta
_2  \in  \tilde A$.
\end{proof}

That $\sym{A}$ is race-preserving is actually a sufficient condition for the
\emph{(Extension)} axiom to hold on $\CC_{\sym{A}}$. With this we can finally
complete the proof.

\ccsimstrat*
\begin{proof}
By Lemma \ref{lemma:symm_cc_no_rf} it remains to prove
\emph{(Extension)}.
Let   $ \theta _1  \parallel   \theta _2 : x  \parallel  y  \isf
_{\CC_{\tilde A}} x'  \parallel  y'$. Assume \emph{e.g.} $x  \parallel
y \longcov{(2, a)}$.  There are two cases: 
  \begin{itemize}
  \item If $\pol_A(a) = -$, then by (Extension) for
    $\tilde A^\perp  \parallel  \tilde A$ we have 
    $ \theta _1  \parallel   \theta _2 \subseteq  \theta _1  \parallel   \theta _2'  \in  \tilde
    A^\perp  \parallel  \tilde A$ whose
    domain is $x  \parallel  y \cup \{a\}$. Its codomain is $x'  \parallel y' \cup \{ a'\}$.
    Since $\pol_A(a) = -$, we cannot have $a'\in x'$ -- indeed $x' \supseteq^+ x' \cap y'
    \subseteq^- y'$, so we would have $a' \in y'$ as well, absurd. So we have
    $x' \cap (y' \cup \{ a'\}) = x' \cap y' \subseteq^+ x'$, and 
    $x' \cap (y' \cup \{ a'\}) = x'\cap y' \subseteq^- y' \subseteq^- y' \cup \{ a'\}$, 
    which establishes that $x'  \parallel  (y' \cup \{ a'\}) \in \conf{\CC_A}$.

    Likewise we have $\theta_1 \cap \theta'_2 = \theta_1 \cap \theta_2$, hence we still have $\theta_1 \cap
    \theta'_2 \subseteq^+ \theta_1$ but also $\theta_1 \cap \theta'_2 \subseteq^- \theta_2
    \subseteq^- \theta'_2$, therefore $\theta_1 \parallel \theta'_2 \in \CC_{\tilde{A}}$.

  \item If $\pol_A(a) = +$ is positive then $a \in x$ as well. Thus, $[a) \subseteq x \cap y$. Therefore, we have
    $(x\cap y) \cup \{a\} \in \conf{A}$, and $(x \cap y) \cup \{a\} \subseteq x$. Define $\theta'_1 =
    \theta_1 \restrict (x \cap y) \cup \{a\}$. We have:
    \[
       \theta'_1 \supseteq^+ \theta_1 \cap \theta_2 \subseteq^- \theta_2
    \]
    By construction, the domains of $\theta'_1$ (which is $(x\cap y) \cup \{a\}$) and the domain of
    $\theta_2$ (which is $y$) are compatible, so by Lemma \ref{lem:tcg_racepres},
    $\theta'_2 = \theta'_1 \cup \theta_2 \in \tilde{A}$,
    and by construction its domain is $y \cup \{a\}$. To sum up, we have:
    \[
       \theta_1 \supseteq^+ \theta_1 \cap \theta'_2 \subseteq^- \theta'_2
    \]
    Hence $\theta_1 \parallel \theta'_2 \in \CC_{\tilde{A}}$ provides the required extension.
  \end{itemize}

We have established that $\CC_{\tilde{A}}$ is an isomorphism family. It is obvious that $\cc_A :
\CC_\A \to \A^\perp \parallel \A$ preserves symmetry. It remains to show that it is
$\sim$-receptive, for which we apply Lemma \ref{lemma:charac_simr}.
Assume $x \parallel y \in \conf{\CC_A}$ can be extended by $(2, a^-)$ in $\CC_A$ and
by $(2,b^-)$ in $A^\perp \parallel A$ (in which case it is immediate that
it is a valid extension in $\CC_A$ as well), such that:
\[
\id_x \parallel (\id_y \cup \{(a,
b)\}) \in \tilde{A}^\perp \parallel \tilde{A}
\]

We need to check that this is a valid extension in $\CC_{\tilde A}$ as well.
By the characterisation of Proposition \ref{prop:symm_cc}, we only have to check that 
$\id_x\,c = (\id_y \cup \{(a,b)\})\,c$ for each $c \in x \cap (y\cup \{a\})$, but in fact we must
have $c \in x \cap y$. Indeed, we cannot have $a\in x$, as by $x \supseteq^+ x \cap y \subseteq^- y$ 
and $\pol_A(a) = -$ that would imply $a \in y$ as well, absurd. So the verification is obvious.

Finally, that copycat is thin is an immediate consequence of $\A_-$ and
$\A_+^\perp$ being thin along with 
the characterisation of symmetries in copycat of Proposition \ref{prop:symm_cc}.
\end{proof}

\subsection{Positivisation of mediating maps}
\label{app:positivisation}

We start with the following lemma, which intuitively allows us to canonically
``transport'' a configuration along a negative symmetry.

\begin{lem}\label{lem:transp_negsym}
Let $\sigma : \S \to \A$ be a pre-$\sim$-strategy, $x\in \conf{S}$,
along with $\theta_- : \sigma x \isf_{\sym{A}_-} y$.

Then, there is a unique $\varphi : x \isf_{\tilde{S}} x'$ \emph{s.t.} $\sigma
\varphi = \theta_+ \circ \theta_- : \sigma x \isf_{\sym{A}} \sigma x'$ for some
$\theta_+ : y \isf_{\tilde{A}_+} \sigma x'$.
\end{lem}
\begin{proof}
\emph{Uniqueness.} If there are two such $\varphi_1 : x \isf_{\tilde{S}} x'_1$
and $\varphi_2 : : x \isf_{\tilde{S}} x'_2$, then from the hypotheses $\psi =
\varphi_2 \circ \varphi_1^{-1} : x'_1 \isf_{\tilde{S}} x'_2$ such that $\sigma
\psi \in \tilde{A}_+$; by Lemma \ref{lemma:pos_symm} it follows that $\psi$ is an
identity and $x'_1 = x'_2$, $\varphi_1 = \varphi_2$.

\emph{Existence.} Direct by induction of $x$ and $\theta_-$. For negative
extensions, it follows from the \emph{(Extension)} property for $\tilde{A}_+$ and
$\sim$-receptivity of $\sigma$. For positive extensions, it follows from the
\emph{(Extension)} axiom for $\tilde{S}$ and axiom (d) of tcgs on $\A$.
\end{proof}

\positivisation*
\begin{proof}
We show that for all $x\in \conf{S}$, there is a unique $\varphi_x : f x
\isf_{\tilde{S'}} x'$ such that $\sigma x \isf_{\tilde{A}_+} \sigma' x'$.

\emph{Uniqueness} is again a consequence of Lemma \ref{lemma:pos_symm}.

\emph{Existence.}
If $x\in \conf{S}$, then by hypothesis we know that the triangle induces 
\[
\theta_x : \sigma' (f x) \isf_{\tilde{A}} \sigma x
\]

By Lemma \ref{lemma:decomp}, $\theta_x$ decomposes as 
$\sigma' (f x) \stackrel{\theta_x^-}{\isf_{\tilde{A}_-}} y
\stackrel{\theta_x^+}{\isf_{\tilde{A}_+}} \sigma x$.
By Lemma \ref{lem:transp_negsym}, there is $\varphi_x : f x
\isf_{\tilde{S'}} x'$ such that $\sigma \varphi = {\theta'_x}^+ \circ
\theta_x^-$. We then have ${\theta'_x}^+ {\theta_x}^{-1} : \sigma x
\isf_{\tilde{A}_+} \sigma' x$ as required.

It is routine to show that the assignment from $x$ to $x'$ is monotonic,
preserves cardinality and unions, hence it is generated by a map of event
structures $f' : S \to S'$, which by construction preserves symmetry. By
construction, $f'$ satisfies the desired properties. The uniqueness of
$f'$ follows directly from Lemma \ref{lemma:pos_symm}.
\end{proof}

\subsection{Preservation of single-threadedness}
\label{app:single_threaded}

\compst*
We first prove that the interaction $T\circledast S$ satisfies the single-threadedness conditions.
More precisely, we prove by induction on $\varphi$ that for any secured bijection 
$\varphi : x_S \parallel x_C \sbij y_A \parallel y_T$
representing (via Proposition \ref{prop:rep_int}) a configuration of $T\circledast S$, then 
\[
\varphi = \biguplus_{1\leq i \leq n} \varphi_i
\]
where each $\varphi_i$ is a secured bijection with a unique minimal event. Indeed, 
assume $\varphi \longcov{(c,d)} \varphi'$ where $\varphi'$ fails this condition. Necessarily, either
$c = (1, s^+)$ or $d = (2, t^+)$, \emph{w.l.o.g.} assume the first. Then, the immediate 
predecessors of $(c, d)$ in $\leq_{\varphi'}$ must be $((1, s_1^-), d_1),
\dots, ((1, s_p^-), d_p)$ (using Lemma 2.7 of \cite{cg1}), with
$s_i \imc_S s$. By hypothesis, there are $1\leq i,j \leq p$ and distinct $1\leq k\neq l \leq n$
such that $((1, s_i^-),d_i) \in \varphi_k$ and $((1, s_j^-), d_j) \in \varphi_l$. But $\varphi_k$
(resp. $\varphi_l$) must contain an event synchronized with $\init(s_i)$ (resp. $\init(s_j)$).
Since $\sigma$ is single-threaded and $s_i, s_j \in [s]$ we have $\init(s_i) = \init(s_j)$, which
contradicts $\varphi_l \cap \varphi_k = \emptyset$.

Now, we go on to prove single-threadedness.

\emph{(1)}
Prime secured bijections have no non-trivial decomposition as above, therefore they 
have a unique minimal event.
This is true in particular for the \emph{visible} prime secured bijections. Condition \emph{(1)} of
single-threadedness follows then from the fact used in the proof of Lemma \ref{lem:comp_neg}
that a minimal event in the interaction of negative strategies is always visible.

\emph{(2)}
Finally, assume there is a minimal conflict $\varphi\!\mconflict~\psi$ in $T\odot S$ between visible
prime secured bijections. This means that there are non-necessarily visible prime secured bijections
$\varphi' \subseteq [\varphi]_{T \circledast S}$, $\psi' \subseteq [\psi]_{T \circledast S}$, such that $\varphi'\!\mconflict~\psi'$ in
$T\circledast S$. Writing $\varphi''$ (resp. $\psi''$) for $\varphi'$ (resp. $\psi'$) without its
top event, minimality of $\varphi'\!\mconflict~\psi'$ means that $\varphi'' \cup \psi''$ is a valid
secured bijection. Therefore, it decomposes:
\[
\varphi'' \cup \psi'' = \biguplus_{1\leq i \leq n} \varphi_i
\]
With each $\varphi_i$ a secured bijection having exactly one minimal event. If $n=1$, we are done since as remarked the
unique minimal event is necessarily visible. Otherwise, there are at least two $\varphi_i,
\varphi_j$ with distinct minimal events.

Then, using Lemma \ref{lem:pb_mconf}, $\varphi'\!\mconflict~\psi'$ implies that their
top elements have the form $((1, s_{\varphi'}), d_{\varphi'})$ and $((1, s_{\psi'}), d_{\psi'})$
with $s_{\varphi'} \mconflict_S s_{\psi'}$, or $(c_{\varphi'}, (2, t_{\varphi'}))$ and $(c_{\psi'}, (2,
t_{\psi'}))$ with $t_{\varphi'} \mconflict_T t_{\psi'}$, \emph{w.l.o.g.} say the first. By 
receptivity and courtesy of $\sigma$, we have $\pol(s_{\varphi'}) = \pol(s_{\psi'}) = +$.
Since $n\geq 2$, there are $(c_1, d_1) \imc_{\varphi} ((1, s^+_{\varphi'}), d_{\varphi'})$
with $(c_1, d_1) \in \varphi_i$ and $(c_2, d_2) \imc_{\psi} ((1, s^+_{\psi'}), d_{\psi'})$
with $(c_2, d_2) \in \varphi_j$.
By Lemma 2.7 of \cite{cg1}, as an immediate dependency of $((1, s^+_{\varphi'}),
d_{\varphi'})$, we have $(c_1, d_1) = ((1, s_1^-), d_1)$ with $s_1 \imc_S s_{\varphi'}$ 
(similarly, $(c_2, d_2) = ((1, s_2^-), d_2)$ with $s_2 \imc_S s_{\psi'}$).
But by single-threadedness of $\sigma$, $\init(s_{\varphi'}) = \init(s_{\psi'})$, so
there should be an event synchronized with $\init(s_1) = \init(s_2)$ both in $\varphi_i$ and
$\varphi_j$, absurd.



\newpage

\section{Indexes}

\printnomenclature

\printindex
\end{document}